**Theories of phosphorescence in organo-transition metal complexes – from relativistic effects to simple models and design principles for organic light-emitting diodes**

B. J. Powell*

Centre for Organic Photonics and Electronics, School of Mathematics and Physics, The University of Queensland, QLD 4072, Australia

*bjpowell@gmail.com

**Abstract**

We review theories of phosphorescence in cyclometalated complexes. We focus primarily on pseudooctahedrally coordinated $t_{2g}^6$ metals (e.g., $[Os(II)(bpy)_3]^{2+}$, $Ir(III)(ppy)_3$ and $Ir(III)(ptz)_3$) as, for reasons that are explored in detail, these show particularly strong phosphorescence. We discuss both first principles approaches and semi-empirical models, e.g., ligand field theory. We show that together these provide a clear understanding of the photophysics and in particular the lowest energy triplet excitation, $T_1$. In order to build a good model relativistic effects need to be included. The role of spin-orbit coupling is well-known, but scalar relativistic effects are also large – and are therefore also introduced and discussed. No expertise in special relativity or relativistic quantum mechanics is assumed and a pedagogical introduction to these subjects is given. It is shown that, once both scalar relativistic effects and spin-orbit coupling are included, time dependent density functional theory (TDDFT) provides quantitatively accurate predictions of the radiative decay rates of the substates of $T_1$ in phosphorescent organotranstion-metal complexes. We describe the pseudo-angular momentum model, and show that it reproduces the key experimental findings. For example, this model provides a simple explanation of the relative radiative rates of the substates of $T_1$, which differ by orders of magnitude. Special emphasis is placed on materials with potential applications as active materials in organic light-emitting diodes (OLEDs) and principles for the design of new complexes are identified on the basis of the insights provided by the theories reviewed. We discuss the remaining theoretical challenges, which include deepening our understanding of solvent effects and, vitally, understanding and predicting non-radiative decay rates.

# 1    Introduction

There has been long-standing fundamental interest in low-energy excitations of phosphorescent organo-transition metal complexes [1-4], some examples of which are given in Figure 1. This interest stemmed from the discovery that the emission properties of these complexes can be very different from typical organic molecules, particularly at low temperatures [5-27]. However, interest in the field has grown significantly as the number of potential applications for such complexes have increased. These include dye-sensitised solar cells [28-34], non-linear optics [34], photocatalysis [4, 34-37], biological imaging [38-40], chemical and biological sensing [34, 38, 41-43], photodynamic therapy [34], light-emitting electro-chemical cells [44-47] and OLEDs [2, 33, 48-55].

For many of these applications a detailed understanding of the lowest energy triplet excitation, $T_1$, would greatly facilitate the design of better materials. For example, in OLED applications the active organometallic complex is embedded in a host matrix, which transports the charge to the complex. When an excitation is formed on the complex it rapidly decays to the $T_1$ state. In phosphorescent organo-transition metal complexes strong spin-orbit coupling (SOC) enables fast intersystem crossing. This is estimated to occur in tens to hundreds of femtoseconds [56]. In contrast the lifetimes of the $T_1$ states, which are discussed in detail below, range from microseconds to milliseconds, i.e., many orders of magnitude slower than the intersystem crossing. Thus, even the generation of singlet excitations rapidly leads to the occupation of $T_1$ - a process known as triplet harvesting, cf. Figure 2. Therefore, the design of an active material for an OLED, in large part, amounts to controlling the key properties of $T_1$ such as its zero field splitting (ZFS), and the radiative and non-radiative decay rates of its three substates. Of course, this is far easier to write than to achieve via the chemical modification of a complex.

In non-relativistic quantum mechanics conservation of spin dictates that a triplet state may not decay radiatively to a singlet state. However, in relativistic quantum mechanics such processes are allowed due to SOC. Therefore, a proper description of phosphorescence will necessarily involve relativistic effects. In order to keep this review as self-contained as possible, in section 2 we provide a brief introduction to relativity and its role in chemistry. The topics covered here are not intended to be exhaustive, but rather to give the necessary introduction for non-specialists wanting to read the remainder of the review. Experts may wish to skip over this section. More comprehensive discussions of these subjects can be found elsewhere, for example in the excellent monograph by Dyall and Færgri [57].

Two major approaches have been taken to modelling the phosphorescence in organo-transition metal complexes: phenomenological or semi-empirical approaches, such as ligand field theory, and first principles approaches, primarily based on (relativistic) (time dependent) density functional theory.

Semi-empirical theories have fallen somewhat out of fashion in the theoretical chemistry community since the rise of computational chemistry. However, they have continued to be developed, particularly in the context of strongly correlated electron materials [58]. For example, models involving strong SOC have recently gathered much attention from the solid-state physics community in the context of the iridates [59, 60]. In section 5 we give a modern reformulation of some semi-empirical models of phosphorescent organo-transition metal complexes. This allows us to place several different models in a unified context and explain the consistency of their key predictions. In particular we stress that many of the key properties of the complexes we discuss are unavoidable consequences of their (approximate) (pseudo)trigonal and pseudooctahedral symmetries. These models also explain why (pseudo)trigonal pseudooctahedral $t_{2g}^6$ complexes are often strongly phosphorescent.

While the semi-empirical theories described in section 5 provide a powerful unified framework for understanding phosphorescence in organo-transition metal complexes, first principles approaches have the advantage of making more specific predictions for individual complexes. Therefore, in section 6, we review first principles calculations for phosphorescent organo-transition metal complexes, which have mainly been based on relativistic TDDFT. We see that such calculations are capable of predicting the observed ZFS and radiative decay rates of individual complexes within the typical spread between different experiments, e.g. in different solvents.

In order to give context to these theoretical results, we also give a very brief review of the most pertinent experimental literature in section 4, focusing on a few key examples. However, there are a number of very common pitfalls in comparing experiments with quantum chemistry. We discuss some of these in section 3. We conclude by discussing the outlook for the field and the major challenges still to be addressed in section 7, most importantly the understanding and prediction of non-radiative decay rates.

In his wonderful article "*The future of chemical physics*", Ahmed Zewail [61] wrote that

> "*Breakthroughs will continue to emerge when … pertinent concepts are generalized with the help of "simple, but not too simple" theories. Computations should be considered as tools, keeping in mind that large-scale computations without a "final" theoretical condensate (or better yet, a "simple equation") are like large-scale experiments which produce numerous results that do not boil down to a meaningful finding. From both experimental and theoretical studies, the ultimate goal is to provide an understanding of the function from knowledge of structure and dynamics.*"

In this review we will try to keep this advice in mind and focus on what has been learnt from calculations and experiments rather than the details of the calculations. Likewise we will focus on the key concepts of relativistic quantum mechanics as it applies to chemical physics, rather than the subtle issues related to its implementation, as we believe that for non-specialists this is what is important to understand. For practitioners of relativistic chemistry there is already a great wealth of literature describing the implementation of these methods, e.g. Ref. [57].

## 2 A brief introduction to special relativity, relativistic quantum mechanics, and relativistic effects in chemistry

### 2.1 Special relativity

While quantum mechanics is well known to chemists, special relativity does not form part of the core of the chemistry curriculum - so before discussing relativistic quantum mechanics, we begin with a brief discussion of special relativity. In the classical, or Newtonian, mechanics that one learns at school velocities are already relative [62]. Imagine you are sitting on a beach watching two boats moving in opposite directions at equal speeds, $s$. To a passenger looking out of a porthole below deck on one boat it will appear as if the second boat is moving with a speed of $2s$. This is known as 'Newtonian relativity' and the equations of motion are said to be 'Galilean invariant'. An inertial reference frame is any set of coordinates that is *not* accelerating. To transform coordinates between two inertial reference frames $S$ and $S'$ moving at a relative speed $v$, we use the 'Galilean' transformation, which, if we define the $x$-axis to be parallel to direction of the relative motion, can be written as

$$x' = x + vt,$$
$$y' = y,$$
$$z' = z,$$
$$t' = t,$$

and

$$u_x' = u_x + v,$$

where $x$, $y$, and $z$ are the Cartesian coordinates and $u_x$ is the x-component of velocity in frame $S$ at time $t$; primed quantities are the equivalent quantities in frame $S'$ at time $t$. It can be shown directly from Galilean invariance that if Newton's laws hold in any inertial frame then they hold in all inertial frames.

However, Galilean invariance allows for arbitrarily large velocities. Einstein argued [62-64], and it has been shown experimentally [65], that the speed of light, $c$, provides an upper speed limit for the universe. This contradicts the predictions of Newtonian relativity. For example, consider motion parallel to $v$ with speed $u_x = 0.6c$ and $v = 0.6c$, Galilean invariance predicts that $u_x' = 1.2c$. Thus, we must replace Galilean invariance with 'Lorentz invariance'. This is achieved via the Lorentz transformations, which, if we again take the $x$-axis to be parallel to direction of the relative motion, can be written as

$$x' = \gamma(x - vt),$$
$$y' = y,$$
$$z' = z,$$

and

$$t' = \gamma \left( t - \frac{vx}{c} \right),$$

where

$$\gamma = \frac{1}{\sqrt{1 - \left(\frac{v}{c}\right)^2}}$$

Note that, whereas in Newtonian physics all observers agree on a universal time, in special relativity time is different in different inertial frames – this is known as time dilation and often sloganized as "moving clocks run slow". Furthermore, for low velocities ($v \ll c$) Lorentz invariance reduces to Galilean invariance, as required to reproduce many experiments and everyday experience.

Lorentz invariance is the heart of special relativity and has many counterintuitive consequences [66-68]. One key point is that Galilean invariance implicitly assumes that there is a preferred reference frame of the universe (presumably the frame of the "fixed stars"). Lorentz invariance removes this preferred frame and treats all inertial frames equivalently [64]. Indeed, the Lorentz transformations can be derived from the assumption that the laws of physics should not depend on our choice of (inertial) reference frame. In particular, Maxwell's equations for electromagnetism are Lorentz invariant rather than Galilean invariant (indeed Lorentz discovered this before Einstein's work [69]).

It can be shown that in the special theory of relativity [62] the energy, $E$, of a particle is given by $E^2 = m^2c^4 + p^2c^2$, where $p$ is the momentum and $m$ is the rest mass. For massless particles, such as photons, $m = 0$ and $E = pc$, which, as, quantum mechanically, $E = \hbar\omega$ and $p = \hbar k$, where $\omega$ is frequency and $k$ is the wavenumber, is equivalent to $\omega = ck$, as we expect for photons.

For massive particles in their rest frame, i.e., the frame in which they do not move, $p = 0$ and we have perhaps the most famous equation in all of science, $E = mc^2$. At low velocities ($u \ll c$) we must get back Newtonian mechanics as there are countless experiments confirming Newton's laws in this regime. Thus, for particles moving at low velocities in our frame we require that $p = mu$, where $u = |\boldsymbol{u}|$, which is true non-relativistically. Taking a binomial expansion of the energy and keeping only the first term we find that

$$E = \frac{p^2}{2m},$$

for $u \ll c$ as expected from Newtonian mechanics.

For particles moving close to the speed of light we would like to be able to write $p = mu$. But, we also want to retain two of the most important ideas in physics: the conservation of energy and the conservation of momentum (as these conservation laws follow directly from the assumption that the laws of physics do not depend on time or position respectively [62]). We can allow all three simultaneously if we define the "relativistic mass" as

$$m^* = \gamma m.$$

This is known as relativistic mass enhancement and shows that objects moving fast become heavy.[1]

The upper speed limit of the universe has many important consequences. One of which is that it forbids instantaneous action at a distance. For example, in the Newtonian theory of gravity the sun and the earth attract each other instantaneously even though they are separated by $\sim 150$ Gm. More pressingly for chemical applications, Coulomb's law cannot be the correct description of the interactions between electrons or between electrons and the nucleus as, again, this assumes instantaneous action at a distance. The theory of quantum electrodynamics (QED) shows that electromagnetic forces are mediated by the exchange of (virtual) photons, which travel at the speed of light [70-72].

## 2.2 Special relativity and quantum mechanics

Westminster Abbey is the traditional venue of the coronation of English monarchs and a UNESCO world heritage site. This magnificent gothic church is decorated with a single equation[2]:

$$i\gamma \cdot \partial \psi = m\psi.$$
$$(1)$$

This is the Dirac equation (in an extremely compact notation), which is engraved on a commemorative plaque to Paul Dirac, who, by writing down this equation, unified quantum mechanics with the special theory of relativity. In so doing Dirac also predicted the existence of antimatter and gave a natural explanation of spin (intrinsic angular momentum), which is a somewhat artificial addition to non-relativistic quantum theory.

Before discussing the Dirac equation, let us briefly outline the tension between the Schrödinger equation and special relativity. Recall that, in the absence of an external potential, the time-dependent Schrödinger equation is

$$\left( i\hbar \frac{\partial}{\partial t} - \frac{\hbar^2 \boldsymbol{\nabla}^2}{2m} \right) \psi(\boldsymbol{r}, t) = 0,$$

where $\hbar$ is the reduced Planck constant, and $\psi(\boldsymbol{r}, t)$ is the wavefunction, which, in general, varies in both space, $\boldsymbol{r}$, and time, $t$. Note that the time-dependent Schrödinger equation is a second order differential equation with respect to space, but only first order with respect to time, *i.e.*, the spatial derivative appears as $\boldsymbol{\nabla}^2 = (\partial^2/\partial x^2, \partial^2/\partial y^2, \partial^2/\partial z^2)$ but the time derivative only appears as $\partial/\partial t$. As Lorentz transformations mix spatial and temporal coordinates, the Schrödinger equation cannot be Lorentz invariant and hence is inconsistent with special relativity.

---

[1] It is also possible to formulate special relativistic dynamics so that mass is not velocity dependent. To do this one must redefine momentum. While this is certainly an elegant approach we do not take it as using the relativistic mass will allow us to simplify some arguments below.

[2] Note that the $\gamma$ in equation (1) is not the same as the $\gamma$ in introduced in section 2.1, but rather the four-vector of Dirac matrices, which we will discuss further below.

As we have seen already, the Dirac equation (1) is often written in rather compact notation [70]. A more transparent notation is

$$\left[i\hbar\left(\gamma_0\frac{\partial}{c\partial t}-\gamma_1\frac{\partial}{\partial x}-\gamma_2\frac{\partial}{\partial y}-\gamma_3\frac{\partial}{\partial z}\right)-mc\right]\psi(\boldsymbol{r},t)=0,$$

where, the Dirac matrices are

$$\gamma_0=\begin{pmatrix}-1&0&0&0\\0&-1&0&0\\0&0&1&0\\0&0&0&1\end{pmatrix},$$

$$\gamma_1=\begin{pmatrix}0&0&0&1\\0&0&1&0\\0&-1&0&0\\-1&0&0&0\end{pmatrix},$$

$$\gamma_2=\begin{pmatrix}0&0&0&-i\\0&0&i&0\\0&i&0&0\\-i&0&0&0\end{pmatrix}$$

and

$$\gamma_3=\begin{pmatrix}0&0&1&0\\0&0&0&-1\\-1&0&0&0\\0&1&0&0\end{pmatrix}.$$

The Dirac equation is constructed to be Lorentz invariant, but a minimal check is that, unlike the Schrödinger equation, it is first order in both spatial and temporal derivatives. Clearly, in the interest of brevity, we have passed over many important subtleties. More detailed discussions are given in many textbooks, *e.g.*, [71, 73].

As the Dirac equation is a 4x4 matrix equation it must admit four solutions at every point in space-time. Furthermore, the scalar wavefunction familiar from non-relativistic quantum mechanics is replaced by a 4-vector wavefunction, known as a spinor. These four solutions have simple physical interpretations: two have positive energy and correspond to the two spin states of the electron; while two solutions have negative energy and correspond to the two spin states of positrons (antimatter electrons). To deal with these negative energy states Dirac invoked a sea of negative energy states filled up with electrons, similar to the valence band of a semiconductor. Positrons are then analogous to holes in a semiconductor [73]. However, the modern approach is to treat positrons as fundamental particles on an equal footing with the electron [71, 72, 74].

## 2.3   Relativistic effects in atomic physics and chemistry

It is helpful to separate relativistic effects into so-called scalar relativistic effects and spin-orbit coupling (SOC). Spin-orbit effects are more widely known and discussed than scalar effects. Nevertheless scalar effects are important to understand and cause important quantitative and qualitative changes in chemical problems, particularly when excited states are considered. Scalar relativity has important consequences for atomic orbitals - known as the direct and indirect effects. The former tends to stabilise atomic orbitals whereas the later tends to destabilise atomic orbitals. Below we discuss the origin of these effects and the competition between them. We then move on to discuss their consequences for chemistry, particularly the inert pair effect. Rather than giving rigorous derivations we will attempt to give simple physical arguments for these effects. Readers interested in more detailed derivations are referred to the extensive literature, e.g. Refs. [57, 75].

In order to make this discussion as simple as possible we use the semi-classical theory of the atom below. Clearly, this theory fails to describe many important experiments – but the predictions discussed below are qualitatively correct and give important physical insights. For an introduction to the semi-classical theory in a chemical context see [76, 77]; for a discussion of the modern applications of this approach to chemical problems see [78].

### 2.3.1  Scalar relativistic effects

We have noted above that, when a particle moves near the speed of light its mass is enhanced: $m^* = \gamma m$ (note that $\gamma \geq 1$, see above). Therefore the relativistic Bohr radius, $a_0^r$, is smaller than the non-relativistic Bohr radius, $a_0^{nr}$:

$$a_0^r = \frac{\hbar}{c\alpha Z}\frac{1}{m^*} = \frac{\hbar}{c\alpha Z}\frac{1}{\gamma m} < \frac{\hbar}{c\alpha Z}\frac{1}{m} = a_0^{nr},$$

where $\alpha = \frac{1}{4\pi\varepsilon_0}\frac{e^2}{\hbar c}$ is the fine structure constant and $Z$ is the atomic number. This means that fast moving electrons will have their average distance from the nucleus reduced. This increases the strength of their Coulomb interaction with the nucleus and hence increases their binding energy. This is known as the *direct effect* [79, 80]. The electrons that travel the fastest are those that spend the most time near the nucleus. The amount of time an electron spends near the nucleus is determined, to first order, by its angular momentum; this can easily be understood by considering the classical analogues of orbits of different angular momenta (see Figure 3).

As s-electrons have $l = 0$: they carry no angular momentum. Classical motion with no angular momentum is rectilinear, that is motion in a straight line back and forth through the nucleus. Thus, p-electrons ($l = 1$) have the smallest possible non-zero angular momentum; therefore the classical analogue of a p-electron's orbit is motion in a highly eccentric ellipse. As $l$ is increased the eccentricity of the ellipse is reduced. However, circular motion is only found in the limit $n \to \infty$ for $l = n - 1$, cf. Figure 3. Therefore the direct effect is strongest for s-electrons and weakens as $l$ increases. The principal quantum number, $n$, also has an effect on the size of the direct effect, with larger $n$ reducing the size of the direct effect (because the requirement for orthogonallity with other wavefunctions of the same $l$ suppresses the amplitude at the nucleus) [81].

The direct effect binds electrons more tightly to the nucleus. This means that electrons more effectively screen one another from the nucleus. Hence, electrons become more weakly bound to the nucleus, which is known as the *indirect effect* [79, 80]. Clearly, the indirect effect is largest for those electrons that spend most of their time far from the nucleus, *i.e.*, the size of the indirect effect increases as $l$ increases.

As the direct and indirect effects have opposite consequences for how strongly electrons are bound within atoms, there is clearly a competition between them. A rough rule of thumb is that overall scalar relativistic effects, which we understand as the sum of the direct and indirect effects, increase the net binding of s- and p-electrons to the nucleus; and decrease the net binding of d-, f- and higher $l$ electrons. The principal quantum number has a somewhat smaller effect on the relative importance of the direct and indirect effects, but, particularly for p-electrons can be important for determining the sign of the net change in binding due to scalar relativistic effects [80].

The inert pair effect [82], the tendency of the electrons in the outermost s-orbital of post-transition metals to resist oxidation, can be understood in the light of scalar relativistic effects. In particular the direct effect means that s-electrons are more tightly bound to the nucleus, which is one of the principal causes of the inert pair effect [83].

To make the above discussion more concrete it is worth considering a few examples. In Table 1 we compare the energies of the atomic orbitals of an isolated Ir atom calculated from density functional theory (DFT) in the non-relativistic approximation with the equivalent quantity calculated in the scalar relativistic approximation [84]. From a chemical perspective the key energy is that of the 5d orbital – which we see is destabilised by 0.28 eV – consistent with the above discussion. We note that, as the 5d orbitals are at the Fermi energy, the energy of these orbitals calculated from Kohn-Sham DFT is far more reliable than the energies of the other orbitals [85, 86].

It is well known that mercury is a liquid at room temperature. It was hypothesised long ago that this is a relativistic effect [83, 87-90] and recent calculations of the melting temperature support this claim [91]. To understand why, it is helpful to compare the electronic structure of mercury with that of cadmium. For a single atom the electronic structures are: Cd = [Kr] $4d^{10} 5s^2$ and Hg = [Xe] $4f^{14} 5d^{10} 6s^2$, *i.e.*, both atoms contain only filled shells. However, because of its large atomic number, there is a significant contraction of the 6s levels in Hg due to the direct effect, while the contraction of the 5s orbitals in Cd is far less pronounced. This means that the Cd 5s electrons form bonds far more readily than the 6s Hg electrons. This contributes significantly to the lower melting point of Hg. In this respect Hg is somewhat like a noble gas [83].

The colour of silver is typical of most metals, but gold has a yellow hue. This is also a relativistic effect [92]. The colour of metals is a rather complicated subject and we will not attempt to give a description of the colour of silver, other than to note that it is largely determined by the behaviour of surface plasmons [93, 94]. This is in part because in silver the lowest energy atomic transition, 4d → 5s, is in the ultraviolet and so does not contribute to the colour. However, in gold the 6s orbital is stabilised by the direct effect *and* the 5d orbital is destabilised by the indirect effect. This means that the 5d → 6s transition is in the blue. In the solid state the relevant transition is 5s → Fermi energy, but nevertheless a similar argument goes through and gold absorbs blue light whereas silver reflects it. Thus relativity leads, eventually, to the yellow colour of gold [92, 95].

### 2.3.2 Spin-orbit coupling (SOC)

SOC is intrinsically both quantum mechanical and relativistic and appears naturally in the Dirac equation. However, we can give a plausibility argument for the existence of SOC without applying the full machinery of the Dirac equation. A Lorentz transformation (partially) transforms electric fields, $\boldsymbol{E}$, into magnetic fields, $\boldsymbol{B}$, and *vice versa* [96]:

$$\boldsymbol{E} \to \gamma(\boldsymbol{E} + \boldsymbol{v} \times \boldsymbol{B}) - (\gamma - 1)(\boldsymbol{E} \cdot \hat{\boldsymbol{v}})\hat{\boldsymbol{v}}$$

$$\boldsymbol{B} \to \gamma\left(\boldsymbol{B} - \frac{\boldsymbol{v} \times \boldsymbol{E}}{c^2}\right) - (\gamma - 1)(\boldsymbol{B} \cdot \hat{\boldsymbol{v}})\hat{\boldsymbol{v}},$$

where $\hat{\boldsymbol{v}} = \boldsymbol{v}/|\boldsymbol{v}|$ is a unit vector in the direction of $\boldsymbol{v}$, the semi-classical velocity of the electron in the rest frame of the nucleus. Therefore, if an electron travels at a relativistic speed in the rest frame of the nucleus, then the field it sees in its own rest frame (due to the nucleus) will not be a purely electric because the Lorentz transformation partially transforms the electric field into a magnetic field. An electron in a magnetic field is subject to a Zeeman interaction, $\mu_0 \boldsymbol{B} \cdot \boldsymbol{S}$. It can be seen from the equations above that only the components of the electric field perpendicular to $\boldsymbol{v}$ are transformed into a magnetic field and therefore $\boldsymbol{B} \propto \boldsymbol{L}$, *cf*. Figure 3. Thus we have a SOC term in the Hamiltonian, $H_{SOC} \propto \boldsymbol{L} \cdot \boldsymbol{S}$.

Carrying through the above semi-classical analysis gives qualitatively correct coupling, but underestimates the strength of the SOC by a factor of two [57]. We can therefore understand some of the important properties of SOC from the semi-classical picture. Firstly, there is no SOC for electrons in s-orbitals. This is because the semi-classical motion is rectilinear motion back and forth through the nucleus (see Figure 3 and Refs. [76, 77, 97]). Therefore there is no component of the motion perpendicular to the field and hence no B-field in the rest frame of the electron and so no SOC. Similarly, the SOC is larger for large $l$ because the tangential component of the semi-classical motion increases proportional to $l$. Therefore, so does the B-field in the rest frame of the electron and so does the SOC.

Perhaps the most important consequence of SOC, for the discussion below, is that neither spin nor orbital angular moment are good quantum numbers once SOC is present. This means that, strictly, we can no longer talk about singlet or triplet excitations as they are mixed. Indeed in a molecule, unlike an atom, the electrons do not just move in the central potential of a single nucleus, so orbital angular momentum is not actually a constant of the motion even in the non-relativistic problem. In the non-relativistic self-consistent field approximation the molecular orbitals must transform according to an irreducible representation of the point group describing the symmetry of the molecule [98], with a structure of singlet and triplets superimposed on top of this. In a relativistic self-consistent field approximation the molecular spin-orbitals transform according to representations of the relevant point group. There is no additional spin structure overlaid on top of this. Beyond the self-consistent field approximation, similar statements hold for the full many-body eigenstates. That is, in the absence of SOC the full many-body states must transform according to an irreducible representation of the point group, but have an additional SU(2) spin-degeneracy overlade on top of this. With SOC the full many-body eigenstates must transform according to an irreducible representation of the point group, with no further spin degeneracy.

### 2.3.3 Phosphorescence

In the absence of SOC spin is a good quantum number, i.e., the total spin of the universe is conserved. Therefore, in non-relativistic quantum mechanics radiative transitions from triplet to singlet states (phosphorescence) are forbidden. However, once SOC is non-zero (which it always is in reality), phosphorescence is allowed. Nevertheless because SOC is typically weak compared to other relevant terms, e.g., the dipole coupling, phosphorescent decay rates are typically orders of magnitude slower than fluorescent decay rates.

For a system in an arbitrary state, $|\phi\rangle$, the rate of radiative decay to the ground state, $|S_0\rangle$, is given by [99],

$$k_r(\phi) = \frac{2(\Delta E)^3}{3\varepsilon_0 hc^3} \sum_{i \in \{x,y,z\}} |\langle S_0|\mu_i|\phi\rangle|^2$$

where $\Delta E$ is the energy of $|\phi\rangle$ relative to $|S_0\rangle$ (or, equivalently, the frequency of emitted light multiplied by ℏ), $\varepsilon_0$ is the permittivity of free space, $h$ is Plank's constant, and $\mu_i$ is the $i$th component of the electric dipole operator. In the absence of SOC $\langle S_0|\mu_i|T_m^k\rangle = 0$ where $|T_m^k\rangle$ is the kth substate of the mth triplet. For non-zero SOC the terms singlet and triplet are not well defined with respect to spin, although, as SOC is often weak, it is usually convenient to continue to use these labels. If we know the eigenstates both with and without SOC, it is helpful to write

$$k_r(\phi) = \frac{2(\Delta E_k)^3}{3\varepsilon_0 hc^3} \sum_{i \in \{x,y,z\}} \left| \sum_n \langle S_0|\mu_i|S_n\rangle\langle S_n|\phi\rangle \right|^2$$

(2)

where we have used the facts that $\sum_n(|S_n\rangle\langle S_n| + |T_n\rangle\langle T_n| + \cdots) = 1$ and that the operator $\mu_i$ conserves spin. Finally we note that, if SOC is weak enough, we can treat its effect on the triplet perturbatively, therefore to first order in $\hat{H}_{SOC}$

$$|\phi\rangle = |T_m^k\rangle + \sum_n \left( |S_n\rangle \frac{\langle S_n|\widehat{H}_{SOC}|T_m^k\rangle}{E(S_n) - E(T_m)} + \sum_q |T_n^q\rangle \frac{\langle T_n^q|\widehat{H}_{SOC}|T_m^k\rangle}{E(T_n) - E(T_m)} + \cdots \right),$$

and we have

$$k_r(T_m^k) = \frac{2(\Delta E_k)^3}{3\varepsilon_0 hc^3} \sum_{i \in \{x,y,z\}} \left| \sum_n \frac{\langle S_0|\mu_i|S_n\rangle\langle S_n|\widehat{H}_{SOC}|T_m^k\rangle}{E(S_n) - E(T_m)} \right|^2,$$

where, again, we have used the fact that $\mu_i$ conserves spin. Notice that we have now formally carried out a bilinear perturbation theory (i.e., one that is linear in both $\mu_i$ and $\widehat{H}_{SOC}$).

## 2.4   Relativity and density functional theory (DFT)

If one is interested in high accuracy methods in quantum chemistry, for sufficiently heavy atoms, relativistic effects must be included as they become large relative to other sources of error. Therefore, a great deal of effort has been expended on including relativistic effects in post-Hartree-Fock methods [57]. However, density functional methods require significantly less computational resources and this makes them appealing for large molecules and crystals, despite the lower accuracy of DFT. The complexes that we will be interested in below, cf. Figure 1, are sufficiently large that post Hartree-Fock methods are prohibitively expensive and so DFT calculations have been the main approach taken. Nevertheless, because of the heavy metals in phosphorescent organo-transition metal complexes, relativistic effects are important and need to be included.

DFT is based on the Hohenberg-Kohn theorem [100, 101], which proves, among other things, that the ground state energy of an electronic system is a functional[3] of the electronic density (given certain reasonable assumptions [101]). However, the Hohenberg-Kohn theorem is an existence theorem and does not tell us the form of this functional.

Practical implementations of DFT rely on the Kohn-Sham scheme [86, 101-103]. This approach introduces a set of equations - known as the Kohn-Sham equations - that look somewhat like a non-linear Schrödinger equation. Certain terms, which can be evaluated straightforwardly from a given density, are included explicitly in the Kohn-Sham equations. The remaining terms are described by the 'exchange-correlation functional'. The form of the exact exchange-correlation functional is not known. Indeed, it can be shown, given reasonable assumptions about the computational complexity of quantum mechanics, that there is no efficiently computable form of the exact exchange-correlation functional [104]. Nevertheless many approximate functionals have been introduced [74, 101, 103].

---

[3] Recall that a function is a mathematical machine that takes a number, $x$, as input and returns (another) number, $f(x)$, as output, e.g., $f(x) = \sin(x)$. A functional is another type of mathematical machine that takes a function, $f(x)$, as input and returns a number, $F[f(x)]$ as output, e.g., $F[f(x)] = \int_{-\infty}^{\infty} f(x)dx$.

To generalise DFT to the relativistic case [105] a number of issues have to be overcome. Perhaps the most important is that, as we have seen above, electricity and magnetism are no longer separate phenomena in a relativistic framework [96]. The simple physical reason for this is Lorentz invariance. Because no information can be transmitted faster than the speed of light, Lorentz invariance forbids instantaneous action at a distance. In non-relativistic theory the Coulomb force between electrons and nuclei and amongst electrons is taken to act instantaneously. Treating the interaction in a relativistic theory one is led to quantum electrodynamics [71, 72, 74], which in its most complete form treats both light and matter quantum mechanically and relativistically. It is possible to derive a DFT in the full field theoretic case [75, 106-109]. However, this remains too computationally expensive to apply to systems of the size we will discuss below. Therefore a common approximation is to abandon the requirement of strict Lorentz invariance and chose to work in a particular frame. For molecules and solids the natural choice of reference frame is the Born-Oppenheimer frame, *i.e.*, the frame in which the nuclear potential is a static Columbic potential once the Born-Oppenheimer approximation is made [57].

The relativistic Hohenberg-Kohn theorem [106, 107, 110] proves that the ground state energy of a Lorentz invariant system is a unique functional of the ground state four-current. (The density is mixed with the current by Lorentz transformations; the four-current is the natural generalisation of current and density in Lorentz invariant systems.) However, analogously to non-relativistic theory, practical implementations require the solution of the Dirac-Kohn-Sham equations [57, 105],

$$\begin{pmatrix} V & c\boldsymbol{\sigma}\cdot\boldsymbol{p} \\ c\boldsymbol{\sigma}\cdot\boldsymbol{p} & V \end{pmatrix} \begin{pmatrix} \psi_L \\ \psi_S \end{pmatrix} = E \begin{pmatrix} \psi_L \\ \psi_S \end{pmatrix}$$

where $V = V_{nuc} + V_{Hartree} + V_{xc}$, $V_{nuc}$ is the nuclear potential, $V_{Hartree}$ is the Hartree potential, $V_{xc}$ is the exchange-correlation potential and $\psi_L$ and $\psi_S$ are two-spinors known as the large and small components of the spinor respectively. These names arise because for the matter solutions the large component is much larger than the small component (somewhat confusingly the reverse is true for the antimatter solutions). As we will only be concerned with the matter solutions we can simplify Dirac-Kohn-Sham equations by eliminating the small component, which yields

$$\left[ T + V - E + \frac{V-E}{2mc^2+E-V}T - \frac{c^2}{(2mc^2+E-V)^2}(\{\boldsymbol{p}V\}\cdot\boldsymbol{p} + \hbar\boldsymbol{\sigma}\cdot\{\boldsymbol{\nabla}V\}\times\boldsymbol{p}) \right] \psi_L$$
$$= 0$$
(3)

where $T = p^2/2m$ is the non-relativistic kinetic energy operator and we have used the result that, quantum mechanically, $\boldsymbol{p} = -i\hbar\boldsymbol{\nabla}$. This two-component relativistic theory is a dramatic simplification as now one need only deal with a two-spinor. The loss of information about the antimatter solutions is not a significant drawback for chemical applications. The first two terms in equation (3) are non-relativistic (independent of $c$) and the last term in equation (3) is the spin-orbit coupling – all other terms describe scalar relativistic effects.

For a single electron orbiting a (point-like) nucleus of atomic number $Z$

$$V = -\frac{e^2}{4\pi\varepsilon_0}\frac{Z}{r},$$

and one has

$$\boldsymbol{\nabla}V = \frac{e^2}{4\pi\varepsilon_0}\frac{Z}{r^3}\boldsymbol{r}.$$

So the SOC is proportional to

$$\{\boldsymbol{\nabla}V\} \times \boldsymbol{p} \cdot \boldsymbol{\sigma} \propto \frac{\hbar}{2}\boldsymbol{r} \times \boldsymbol{p} \cdot \boldsymbol{\sigma} = \boldsymbol{L} \cdot \boldsymbol{S},$$

where, in the final equality, we have used two well-known relationships: $\boldsymbol{S} = \frac{1}{2}\hbar\boldsymbol{\sigma}$ and $\boldsymbol{L} = \boldsymbol{r} \times \boldsymbol{p}$.

A further common approximation is simply to neglect spin-orbit coupling. The two components of $\psi_L$ then decouple and we have a one-component wavefunction, which contains only scalar relativistic effects. Solving the one-component scalar relativistic Hamiltonian is no more computationally expensive than solving the Schrödinger equation. As spin-orbit effects are often small they can then be added perturbatively to the one-component solutions. In phosphorescent organo-transition metal complexes, including spin-orbit coupling at second order has been found to reproduce the solutions of the two and four component theories extremely accurately [111, 112].

The discussion above is clearly very incomplete, but, hopefully, gives a broad overview of the key concepts. However, let us just mention that it is $\psi_L + \psi_S$ rather than $\psi_L$ that is normalised. This means that the formalism above is not quite ready to be applied to real chemical problems – although this is easily dealt with [57] and does not change the qualitative discussion above.

## 3  Some common errors made when comparing computational chemistry to experiment

An easy trap to fall into is to compare the interpretations of experiments with the interpretations of theoretical research. This can lead to problems that can be avoided if one instead compares the actual results of calculations with the actual results of experiments. Crucially however, this requires the calculation of the actual property the experiment actually measures.

For example, the solution of the (Dirac)-Kohn-Sham equations leads to a set of one electron orbitals. Except for the highest occupied orbital [85, 86], the energies of the orbitals do not correspond to any physical observable. However, the (Dirac)-Kohn-Sham equation is reminiscent of the (Dirac)-Hartree-Fock equation and it is therefore tempting to interpret the (Dirac)-Kohn-Sham eigenstates as molecular orbitals. This temptation is rarely resisted. This amounts to treating DFT as a self-consistent field approximation.

However, whether one considers Kohn-Sham or self-consistent field molecular orbitals it is still fallacious to conflate molecular orbitals (which are artefacts of particular theoretical approximation schemes) with anything in the real world, particularly with the results of experiments.

It should be stressed that in both relativistic and non-relativistic quantum mechanics all observables must be the expectation values of Hermitian operators [81]. The true eigenstates of molecules involve all of the electrons; they are many-electron wavefunctions. To put it more formally, the true eigenstates are highly entangled and involve multiple Slater determinants. Molecular orbitals are single electron like. Therefore, one cannot write the wavefuntion as a product state of molecular orbitals or construct an operator, the expectation value of which is a molecular orbital. Thus, there is no sense in which molecular orbitals can be said to exist. There have been high profile claims to have measured the properties of molecular orbitals with angle resolved photoemission (ARPES) [113] and scanning tunnelling microscopy [114]. While these are experimental *tours de force*, they do not represent measurements of molecular orbitals, but rather the one electron spectral density. While this may have a passing resemblance to the density of a molecular orbital in some systems it is, formally, quite different.

Nevertheless, in organic electronics the interpretation of experiments in terms of molecular orbitals is widespread. This leads to many problems – both conceptual and practical [115, 116]. For example, (i) the highest occupied molecular orbital (HOMO) is often taken to be equivalent to the ionisation energy (IE); (ii) the lowest unoccupied molecular orbital (LUMO) is frequently assumed to be synonymous with electron affinity (EA); and (iii) the HOMO-LUMO gap is often taken to be the same as the optical excitation energy. None of these assumptions are valid when there are electronic correlations or once the environment is accounted for [115, 116].

It is easy to understand, on a purely physical basis, why electronic correlations mean that the electron affinity is not the same as molecular orbital energies. When an electron is added to the system, one does not only have to account for the energy of that electron; that electron now interacts with all of the other electrons in the molecule. Similarly, when an electron is removed one must also adjust for all the interactions that are no longer present. These effects can be very large in organic molecules [117, 118]. Moreover, the geometry of a molecule will typically relax once an electron is added/removed, further changing the energy.

In practice, further complications arise because the quantities of interest, such as the IE and EA, depend strongly on the environment surrounding the molecule. This means that the electrochemical measurements of the oxidation or reduction potential in solution will typically differ from measurements of the solid state IE by, say, ultraviolet photoemission spectroscopy (UPS) and the solid state EA using inverse photoemission spectroscopy (IPES). Furthermore, these differences are typically on the scale of changes one wants to control to optimise material design [119], for example, the offset in the energy between a donor-acceptor pair required for exciton dissociation in bulk heterojunction solar cells. Therefore, any theoretical approach should include a description of the environment sufficient to capture these effects. Typically, the materials used in organic electronic devices are amorphous, which suggests that some type of averaging procedure should be carried out to give an effective model of the solid state environment.

There have been a number of recent attempts to predict the IEs and EAs of molecules used in organic electronics using DFT [115, 120-123]. The accurate calculation of IEs and EAs from the Kohn-Sham eigenvalues is a longstanding problem in DFT and is often referred to as the 'band gap problem' [124]. A classic example is the band structure of Si. The eigenvalues of the Kohn-Sham equations do reproduce, qualitatively, what is known experimentally about the band structure of Si, such as the indirect gap and the locations of the valence band maxima and conduction band minima; but, the band gap so calculated is about 50 % larger than that measured experimentally [124]. Similar problems are found when estimating the IEs and EAs and optical gaps of molecular systems from the Kohn-Sham eigenvalues [125, 126]. Below, we will use the term 'band gap problem' to refer to both the intrinsic difficulties with DFT and the problems arising from the neglect of correlation, i.e., of treating molecular orbital energies as predictions of ionisation energies and electron affinities.

While the band gap problem is very serious in the solid state, for molecular systems the calculation of IEs and EAs need not face this issue. DFT gives very accurate numbers for the total ground state energies of molecular systems, both in principle and in practice. IEs and EAs are just differences in the total ground state energies of molecular systems:

$$I = E_0(+1) - E_0(0)$$
$$A = E_0(0) - E_0(-1),$$

where, $I$ is the IE, $A$ is the EA, and $E_0(q)$ is the ground state energy of the molecule or (system in question) with charge $q$, for example, $E_0(-1)$ is the ground state energy of the molecule with one *additional* electron. It has been shown that, if this approach is taken and combined with relatively low cost approaches to modelling the environment, then molecular IEs/EAs can be calculated to within the accuracy of typical measurements on organic electronic systems [115].

Time dependent density functional theory (TDDFT) [127-129] is an important advance on DFT as it allows for the calculation of excited state energies within the density functional formalism. Therefore TDDFT can be used to calculate, e.g., optical excitation energies of molecules used as the active components in organic photovoltaic and OLED devices [130]. In linear response TDDFT the excited states are straightforwardly written as linear superpositions of the transitions between the (Dirac)-Kohn-Sham orbitals [128]. This allows one to make connection with the intuitive, but overly simplified, pictures of molecular orbital theory. This can be a significant aid to understanding the relevant physical contributions to excited states.

## 4 Some key experimental facts

We will not attempt to give a comprehensive review of the measured properties of phosphorescent organo-transition metal complexes as the literature is vast (certainly in comparison to the theoretical literature) and already contains a number of excellent reviews, for example Refs. [1-3, 50, 131-134]. Rather, in this section, we will highlight a few results that are germane to the following discussion of theoretical work. We will also discuss other experiments in sections 5 and 6 as appropriate.

Because we are mostly interested in subtle relativistic effects, such as the ZFS and radiative decay rates of $T_1$, the most relevant experiments are performed at low temperatures – typically well below 77 K. In this temperature range experiments are performed in glassy matrices, solid solutions or Shpol'skii matrices [2, 135].

## 4.1 M(bpy)$_3$

A number of related complexes of the form given in the section heading, including [Ru(bpy)$_3$]$^{2+}$, [Rh(bpy)$_3$]$^{3+}$, [Os(bpy)$_3$]$^{3+}$, and [Ir(bpy)$_3$]$^{2+}$ (**1**), share a similar phenomenology. These complexes have been intensively studied for several decades because their emission, absorption and excitation spectra are very different to those of simple organic molecules [5-27]. This is largely due to relativistic effects: strong SOC means that these complexes phosphoresce and at sufficiently low temperature the ZFS is large enough to mean that populations of the different substates of $T_1$ have different populations. Several excellent reviews have already been written [1-3, 132-134]. Therefore, here we just briefly summarise that phenomenology, without a detailed discussion of the experiments that lead to this picture.

- Radiative emission comes overwhelmingly from the lowest energy triplet excitation, $T_1$, and its vibrational satellites [1]. It is convenient to label the three substates I, II and III, with I being the lowest energy substate and III the highest energy. We will adopt this nomenclature henceforth.
- When the three substates of $T_1$ can be individually resolved, either in a magnetic field or otherwise, I has the longest lifetime and III has the shortest lifetime (i.e., $\tau_I > \tau_{II} > \tau_{III}$), cf. Table 2.
- The ZFS of $T_1$ and the excited state lifetimes both differ by orders of magnitude across the family of complexes. Strong ZFS is correlated with short lived excitations. Therefore the excitations in those complexes with small ZFS and long lived triplet excitations have been classified as ligand centred (LC); whereas in complexes with large ZFS and relatively short lived triplets the excited states have been assigned as metal to ligand charge transfer (MLCT). For example, in [Rh(bpy)$_3$]$^{3+}$ the ZFS between I and III is very similar to that in isolated bpy; but the ZFS is three orders of magnitude larger in [Os(bpy)$_3$]$^{3+}$. Obviously some of this change stems from the larger charge on the Os nucleus, nevertheless the ZFS in [Ru(bpy)$_3$]$^{2+}$ is already two orders of magnitude larger than that in [Rh(bpy)$_3$]$^{3+}$.
- The ZFS between substates I and II is considerably smaller than that between substates II and III, cf. Table 2. For example, for [Ru(bpy)$_3$]$^{3+}$ in [Zn(bpy)$_3$](ClO$_4$)$_2$ the former is measured to be 8.7 cm$^{-1}$ whereas the latter is 52 cm$^{-1}$ [1].

It is also interesting to note that there has been a long-running debate [1, 136, 137] over the nature of the excited state in these complexes. This has focused on whether the excited state is delocalised over all three ligands or localised to a single ligand.

## 4.2 Ir(ppy)$_3$

Ir(ppy)$_3$ (**2**) became the subject of intense research when it was shown that it was possible to fabricate highly efficient OLEDs from Ir(ppy)$_3$ doped into a 4,4'-bis-(9-carbazolyl)-biphenyl (CBP) host matrix [138]. Our aim is not to focus on these applications (see, e.g., [49] for a recent review of these) but rather to understand the nature of the phosphorescence. Therefore, we focus here on the nature of T$_1$.

Yersin's group have given a very detailed experimental characterisation of the photophysics of the T$_1$ excitation in Ir(ppy)$_3$ [139-142]. As they have recently given a detailed review of this work [2], we will not discuss the experimental details and will again limit ourselves to recounting the main conclusions. The low temperature emission and excitation spectra are reproduced in Figure 4a [139]. Three electronic excitations, labelled I, II and III, are observed, which Hofbeck and Yersin [139] identified as corresponding to the three substates of T$_1$. No other excitations are observed within 500 cm$^{-1}$ of these states. Figure 4b shows how the emission spectra evolves as the temperature is increased. At 1.5 K a Boltzmann population analysis suggests that only substate I will have a macroscopic occupation. Therefore at 1.5 K the emission is dominated by radiative decay from this state and its vibrational satellites. As the temperature is raised substates II and III become thermally populated and contribute significantly to the emission, resulting in increased spectral weight at higher energies.

The energies of the substates of T$_1$ are summarised in Figure 5. It is interesting to note, particularly if we are interested in testing theoretical predictions against experiment, that the ZFS measured in CH$_2$Cl$_2$ differ by about a factor of two from those measured in poly(methyl methacrylate), PMMA (see Table 4 of Ref. [2]). The (total) excited state lifetimes of the excited substates of T$_1$ of Ir(ppy)$_3$ in these two solvents also differ by a factor of ~2 (see Table 4 of Ref. [2]). Thus one should be cautious of trying to compare theoretical predictions to experimental numbers at better than this accuracy unless the specific solvent used in the measurements has been taken into account in the calculations.

Hofbeck and Yersin [2, 139] measured the temperature dependence of the photoluminescent quantum yield (PLQY) and the emission decay times of Ir(ppy)$_3$ in PMMA and, from these measurements and the measured ZFS, they extracted the radiative and non-radiative decay rates of Ir(ppy)$_3$ in PMMA, which are also shown in Figure 5. Note that, as we say for M(bpy)$_3$, $\tau_I > \tau_{II} > \tau_{III}$, and here also $k_R^I > k_R^{II} > k_R^{III}$ – indeed this is the case in all complexes for which there is data (cf. Table 2). This allowed them to calculate a separate PLQY for each substate. For all three states the PLQY is very similar ~90 %, which suggests that the same factors determine both the relative radiative and non-radiative decay rates of the substates. We are not aware of a detailed explanation of why this should be the case. However, one presumes that the strength of SOC plays a key role in determining the non-radiative decay rate (as well as the radiative decay rate) and so gives rise to the proportionality.

### 4.3 Ir(ptz)$_3$

Perhaps the key challenge for full colour OLED displays is the development of highly-efficient deep blue emitters [2, 141, 143]. In the context of phosphorescent emitters, a particularly interesting complex is Ir(ptz)$_3$ (**3**). This complex is a highly efficient pale blue emitter with a PLQY of ~66 %, measured in toluene at room temperature [54]. Fluorination at either or both of the *ortho* and *para* positions (see Figure 1) drives the emission to a deeper blue. However, this fluorination also results in a rapid decrease in the PLQY (see Table 3). For all of these compounds the excited state lifetimes are a few microseconds and therefore are consistent with phosphorescent decay.

The high resolution spectroscopic studies, which led to the results discussed in sections 4.1 and 4.2, have not been carried out for Ir(ptz)$_3$ or its fluorinated derivatives (**4-6**), to date. This limits our understanding to the T$_1$ state. Nevertheless, significant insight can be gained from magnetic circular dichroism (MCD) experiments, which have been performed by Smith *et al.* for both Ir(ptz)$_3$ [84] and its fluorinated analogues [144]. These spectra are reproduced in Figure 6. In all spectra a strong MCD A-term (cf. Figure 7) occurs around the first absorption band.

One would suspect that the low-energy excitations of Ir(ptz)$_3$ will be similar to those of Ir(ppy)$_3$, discussed above. Therefore, the clear resolution of MCD A-terms in all four complexes is an interesting result. MCD A-terms are due to an excited state degeneracy (cf. Figure 7). However, the degree of degeneracy is only established within the linewidth of the feature. The linewidths in the MCD spectra of Figure 6 are comparable with the observed splitting in Ir(ppy)$_3$ [139]. The clear equal and opposite signed peaks observed in Figure 6 indicate that any lowering of the complexes' C$_3$ symmetry in the excited state does not split the E levels of the T$_1$ manifold enough to destroy the derivative shaped $\Delta\varepsilon_M$ expected for an A-term from a degenerate E state.

Smith *et al.* [84, 144] also observed clear mirror image symmetries between the lowest energy feature observed in absorption and the highest energy feature observed in emission in all four complexes. These mirror images reveal relatively small Stokes shifts (~220 cm$^{-1}$), consistent with the observed lowest energy absorption feature also being responsible for the emission. At temperatures ≥ 10 K most of the emission is coming from the two (nearly) degenerate E states - the same states that carry the absorption intensity. However, by 2 K the emission changes dramatically suggesting that the E levels are depopulated into a lower energy substate with much weaker oscillator strength (and hence radiative decay rate) [144, 145].

The experiments described above are consistent with the same electronic structure as has been observed in M(bpy)$_3$ and Ir(ppy)$_3$, i.e., very little oscillator strength associated with substate I and significantly more oscillator strength associated with substates II and III. However, the important caveat remains that while it is clear that there is a ZFS between substate I and substates II and III, no ZFS has yet been observed between substates II and III.

## 4.4 Conclusions

Thus we have seen that the properties of the low-energy excitations of Ir(ppy)₃ and Ir(ptz)₃ are similar to those of the M(bpy)₃ complexes. Indeed the same phenomenology is seen in a large number of tris-bidentate Ir(III) complexes, including many heteroleptic complexes. In particular Table 2 shows that in a vast range of complexes $\tau_I < \tau_{II} < \tau_{III}$ and $\Delta_{I,II} < \Delta_{II,III}$. Where there is data we also find $k_R^I > k_R^{II} > k_R^{III}$. Therefore, a proper theoretical understanding of these complexes should be able to describe these common properties and the clear differences among the complexes on the basis of simple chemical and physical understanding as well as via detailed computation [61].

## 5    The pseudo-angular momentum model

### 5.1    Warm up: the quantum mechanics of triangles and triskeles and the mapping to an $l = 1$ angular-momentum Hamiltonian

As the compounds we will discuss (cf. Figure 1 and Table 2) have at least approximate trigonal symmetry it will be helpful to review some of the properties of triangular systems. To do this we will use C₃ symmetry as our primary example so really we are describing a triskele rather than a triangle, although one could equally frame the following discussion in terms of any of the trigonal groups, C₃ᵥ, C₃ₕ, D₃, D₃ₕ, D₃ₐ, or even S₆, with only minor changes in nomenclature. The character table for C₃ is shown in Table 4. In particular note that in the C₃ point group, the $E$ representation is not really a two-dimensional representation, as one would usually expect. The $E$ representation is only two-dimensional if the Hamiltonian is also time-reversal symmetric, $T$.

To understand this $C_3 \times T$ symmetry it is helpful to solve the Hückel (or tight-binding) model [58, 146] on three sites (these might be atoms, ligands or other chemical moieties) on the arms of a triskele. The Hamiltonian for this model is, in matrix form,

$$H = \begin{pmatrix} \alpha & \beta & \beta \\ \beta & \alpha & \beta \\ \beta & \beta & \alpha \end{pmatrix},$$

(4)

where each row (column) refers to one of the sites, $\alpha$ is the energy of an electron on an isolated site and $\beta$ is the quantum mechanical amplitude for an electron to hop from one site to another; for a covalent bond $\beta < 0$. We have neglected spin at this stage as this just introduces a trivial two-fold degeneracy to all states. As, by the assumed C₃ symmetry, all sites are equivalent, we may, without loss of generality, set $\alpha = 0$. The solution of this problem is trivial, one finds that there are three eigenstates with energies $E_a = 2\beta$ and $E_{e^+} = E_{e^-} = -\beta$ [58, 146], where we have labelled the eigenstates by the symmetry labels of the C₃ point group. Clearly, as expected, the a-orbital is non-degenerate and there are two-fold degenerate e-orbitals.

The wavefunction of the a orbital is

$$\psi_a = \frac{1}{\sqrt{3}}(\phi_1 + \phi_2 + \phi_3),$$

where $\phi_i$ is the orbital on the $i^{\text{th}}$ site. Because the e orbitals are degenerate any linear superposition of the wavefunctions is a valid eigenstate, therefore we have a choice of how to write these molecular orbitals. A common choice is the real representation:

$$\psi_{e\theta}^{(\mathbb{R})} = \frac{1}{\sqrt{6}}(2\phi_1 - \phi_2 - \phi_3),$$
$$\psi_{e\epsilon}^{(\mathbb{R})} = \frac{1}{\sqrt{2}}(\phi_2 - \phi_3).$$

However, another interesting choice is

$$\psi_{e+} = \frac{-1}{\sqrt{3}}(\phi_1 + e^{i2\pi/3}\phi_2 + e^{-i2\pi/3}\phi_3),$$
$$\psi_{e-} = \frac{1}{\sqrt{3}}(\phi_1 + e^{-i2\pi/3}\phi_2 + e^{i2\pi/3}\phi_3).$$

We now choose to work in a spherical polar coordinate system. To define the coordinate system (cf. Figure 8) we choose (i) the centre of the triangle as the origin, (ii) the distance from centre of the triangle to a vertex to be of unit length (or equivalently the length of a side of the triangle to be of length $\sqrt{3}$), and (iii) site 1 to be at angle $\theta = 0$. We can now write all three molecular orbitals in the form of Bloch wavefunctions [93, 98] on the unit circle:

$$\Psi_{L^z} = \text{sgn}^{L^z}(-L^z) \sum_{i=1}^{3} e^{iL^z\theta_i/\hbar}\phi_i,$$

where $L^z$ is the $z$-component of the angular momentum of the state $\Psi_L$; here the $z$-axis is taken to be perpendicular to the plane of the triangle (cf. Figure 8) and the initial phase factor, $\text{sgn}^{L^z}(-L^z)$, is required to ensure that the states transform as required under time reversal symmetry[4]. It is clear, upon noting that $\theta_i = 2\pi(i-1)/3$, that $\psi_a = \Psi_{L^z}$ for $L^z = 0$, $\psi_{e+} = -\Psi_{L^z}$ for $L^z = \hbar$, and $\psi_{e-} = \Psi_{L^z}$ for $L^z = -\hbar$.

The Hamiltonian discussed above has time reversal symmetry: that is, if we make the mapping $T: t \mapsto -t$, where $t$ is time, the spectrum, and indeed all physical observables, are unchanged. Angular momentum, $\boldsymbol{L}$, is odd under time reversal, *i.e.*, $T\boldsymbol{L}T^{-1} = -\boldsymbol{L}$, this is easy to see for the classical angular momentum as $\boldsymbol{L} = \boldsymbol{r} \times \boldsymbol{p} = \boldsymbol{r} \times m\boldsymbol{v} = \boldsymbol{r} \times m(d\boldsymbol{r}/dt)$. This implies that the orbitals $\phi_{e+}$ and $\phi_{e-}$ must be degenerate so long as the Hamiltonian is time reversal symmetric as they map to one another under time reversal: $T\phi_{e+} = \phi_{e-}$; $T\phi_{e-} = \phi_{e+}$. This is just an instantiation of Kramer's theorem [147-149], whereby a similar argument will hold for any $C_3$ symmetric system with time reversal symmetry.

---

[4]Recall that $\text{sgn}(x) = \begin{cases} 1 & \text{if } x > 0 \\ 0 & \text{if } x = 0. \\ -1 & \text{if } x < 0 \end{cases}$

If we had an octahedral system, rather than one with triangular symmetry, one would find that the A and E states combined to form a representation of $T_{2g}$. One consequence is that the $L^z = 0$ state is degenerate with the $L^z = \pm 1$ states in an octahedral environment. Thus, if we start with an octahedral symmetry and add a small (i.e., perturbative) trigonal term to the Hamiltonian that lowers the symmetry to, say, $C_3$ then we expect the A and E states to be close in energy. Physically, such a perturbation might correspond to a small distortion of the molecule or a subtle change in the chemistry. We can then rewrite the Hamiltonian of the trigonal system [Eq. (4)] in terms of the (pseudo)-angular momentum:

$$H = \Delta(L^z)^2$$

where $\Delta = -3\beta$ and we have neglected a constant term ($\alpha + 2\beta$), which we are free to do without loss of generality. Importantly this model has the same energy spectrum and eigenstates as Eq. (4). We will make extensive use of this result below.

However, a magnetic field breaks time reversal symmetry. In terms of the model, if there is a net magnetic flux through the triangle this can be represented by a complex phase on $\beta$. If one solves the triangular Hückel model for complex $\beta$ one finds that $\phi_{e+}$ and $\phi_{e-}$ are no longer degenerate. A magnetic field breaks time reversal symmetry; this invalidates Kramer's theorem and means that the $e_+$ and $e_-$ states will no longer be degenerate. Therefore, quite generally, a magnetic field will lift the degeneracy of the E states of a $C_3$ symmetric complex.

The above discussion has introduced the idea of representing the solutions of a Hückel model via an angular momentum. However, we need not limit this description to non-interacting models, such as the Hückel model. Indeed, provided the relevant many-body states have the requisite symmetry one can represent arbitrarily strongly correlated states in the above pseudo-angular momentum language. This makes the pseudo-angular momentum approach extremely powerful.

## 5.2 Pseudooctahedral $t_{2g}^6$ complexes

We now want to start building up some simple models of phosphorescent organo-transition metal complexes – we do so following Ref. [150]. To do so we begin by noting that if the metallic ion forms six bonds then it sees an approximately octahedral environment, henceforth referred to as pseudooctahedral symmetry. We will argue below that it is helpful to start from the point-of-view that the complex is basically octahedral and then to consider the changes in the photophysics due to the departures from octahedral symmetry.

In an octahedral environment the d orbitals are split into a three-fold degenerate $t_{2g}$ set of orbitals and a two-fold degenerate $e_g$ pair of orbitals. The latter are virtual in the ground and low-lying excited states and will be disregarded in the pseudo-angular momentum model. We stress that the models discussed below are not specific to the Ir(III) complexes that we focus on and serve equally well for other pseudooctahedral complexes were the electronic state of the central transition metal is $t_{2g}^6$ in the ground state. As foreshadowed in section 5.1 we may choose to write the three t$_{2g}$ states in terms of the $L^z$ eigenvalues of an $L$=1 system [151]. In terms of the more usual notation these states are[5] [151]

$$d_{L^z=1} = \frac{-1}{\sqrt{3}}\left(d_{xy} + e^{-i2\pi/3}d_{xz} + e^{i2\pi/3}d_{yz}\right)$$

$$d_{L^z=0} = \frac{1}{\sqrt{3}}\left(d_{xy} + d_{xz} + d_{yz}\right)$$

$$d_{L^z=-1} = \frac{1}{\sqrt{3}}\left(d_{xy} + e^{i2\pi/3}d_{xz} + e^{-i2\pi/3}d_{yz}\right).$$

We now note that one can make different combinations of the ligand π and π* orbitals to form combinations of the various representations of O$_h$. Symmetry dictates that while the metal-t$_{2g}$ orbitals can mix with the t$_{2g}$ combinations of the ligand orbitals they cannot mix with the other representations [98, 149].

Let us now consider a hypothetical octahedral $t_{2g}^6$ complex where the HOMOs have t$_{2g}$ symmetry and have strongly mixed metal-ligand character. We will label the three t$_{2g}$ HOMOs by the $z$-component of their angular momentum, $L_H^z$, where we take the $z$-axis to be parallel to one of the C$_3$ axes of the octahedral symmetry. We will now posit that our hypothetical complex has a LUMO with t$_{1u}$ symmetry.[6] This orbital can only mix with atomic p-orbitals on the metal so, assuming (reasonably) that these are well separated in energy from the LUMO, any such mixing will be minimal. Nevertheless, because we have a three-fold representation, we again find three degenerate states, which we can label by the $z$-component of their angular momentum, $L_L^z$.

In order to construct a model of the lowest energy excitations of our complex we limit our ambition to describing states with one electron in the t$_{2g}$-HOMOs and one electron in the t$_{1u}$-LUMOs. We begin by considering an isolated ligand. It is well known that there is an exchange interaction between a hole in a π-orbital and an electron in a π*-orbital [152, 153]; this lowers the energy of triplet excitations relative to singlet excitations. Therefore, to the extent that the t$_{2g}$ HOMO has π character we expect our model of the low energy states to contain a term, $J\boldsymbol{S}_H \cdot \boldsymbol{S}_L$, where $\boldsymbol{S}_H$ is the spin of the hole in the HOMOs and $\boldsymbol{S}_L$ is the spin of the electron in the LUMOs and $J < 0$.

---

[5] The linear combination of d-orbitals above is appropriate for a coordinate system with the $z$-axis along one of the C$_3$ axes of the complex. In octahedral systems it is more conventional to take all three Cartesian axes to lie along the C$_4$ axes. In the latter coordinate system the appropriate linear combinations of the d-orbitals for the pseudo-angular momentum states are $d_{L^z=\pm1} = \frac{\mp1}{\sqrt{2}}\left(id_{xz} \pm d_{yz}\right)$ and $d_{L^z=0} = d_{xz}$.

[6] Note that the argument that follows would not change in any substantive way if we chose another three-dimensional representation, i.e., t$_{1g}$ or t$_{2u}$, instead. However, we contend that t$_{1u}$ is the relevant representation for the complexes in Figure 1, see section 5.3 for details.

Now, to the extent that the HOMOs have significant metallic character, we will have strong spin-orbit coupling: $\lambda \boldsymbol{L}_H \cdot \boldsymbol{S}_H$. However, we neglect the much smaller SOC for the $\pi$ system, therefore we do not expect to have significant SOC for the LUMO.

Thus we arrive at our model Hamiltonian for the low energy excitations of an octahedral complex:

$$H_{O_h} = J\boldsymbol{S}_H \cdot \boldsymbol{S}_L + \lambda \boldsymbol{L}_H \cdot \boldsymbol{S}_H.$$

(5)

We will henceforth refer to this model and its generalisations as pseudo-angular momentum models. We stress that the values of $J$ and $\lambda$ will depend strongly on the extent of mixing between the $t_{2g}$-metal orbitals with the $t_{2g}$-$\pi$ orbitals in the HOMOs. For example, if the HOMO wavefunctions are

$$h_{t_{2g}}^{Lz} = d_{t_{2g}}^{Lz} \cos \theta + \pi_{t_{2g}}^{Lz} \sin \theta,$$

(6)

where $d_{t_{2g}}^{Lz}$ and $\pi_{t_{2g}}^{Lz}$ are the metal and ligand orbitals respectively and $\theta$ describes the extent of mixing and that might be estimated by comparison to experiment or first principles theory, then one would estimate that $J \simeq J_\pi \sin^2 \theta$ and $\lambda \simeq \lambda_d \cos^2 \theta$ where $J_\pi$ is the exchange interaction on the isolated ligand and $\lambda_d$ is the SOC coupling on the isolated metal [153, 154]. It is interesting to make estimates for these parameters: for 2-phenylpyridyl (ppy) it has been estimated [153] based on the absorption spectra, emission spectra, and emission lifetimes [155] that $J_\pi \sim 2$ eV; and for an isolated Ir ion $\lambda_d \sim 0.43$ eV [156, 157]. Estimates for other ligands and complexes give similar values.

A few points to note about $H_{O_h}$ (Equation (5)) are: (i) the ground state, $S_0$, is not described by the pseudo-angular momentum model – it is assumed to be well separated from the low energy excitations by a gap that is large compared to gaps between the excitations; (ii) the energy is independent of $\boldsymbol{L}_L$, this makes the problem somewhat simpler to solve; (iii) $H_{O_h}$ commutes with $\boldsymbol{I} = \boldsymbol{L}_H + \boldsymbol{S}_H + \boldsymbol{S}_L$ therefore $I^2$ and $I^z$ are conserved quantities (as, trivially, are $L_L^2 = 2$ and $L_L^z$).

It is straightforward to solve $H_{O_h}$ exactly (numerically) and we plot the spectrum of excitations with $L_L^z = 0$ in Figure 9 – because $\boldsymbol{L}_L$ is decoupled from the other angular momenta it can be immediately seen that the other solutions simply triple the degeneracies of all states. For $\lambda = 0$ one finds that all the singlet excitations have energy $E_S^{(0)} = +3J/4$ and all the triplet excitations have energy $E_T^{(0)} = -J/4$. For $\lambda \neq 0$ this degeneracy is lifted and the nine "triplet" states are split into a non-degernate state, a three-fold dengerate manifold and a five-fold degenerate manifold. We can understand this degeneracy as the SOC couples the $S = 1$ (where $\boldsymbol{S} = \boldsymbol{S}_H + \boldsymbol{S}_L$) spin degree of freedom to the $L_H = 1$ angular momentum degree of freedom. Recall that $3 \otimes 3 = 1 \oplus 3 \oplus 5$ i.e., coupling two triplets yields a singlet, a triplet and a quintuplet. Therefore we identify state I as having quantum numbers $I = 0$, $I^z = 0$; states II-IV as having $I = 1$, $I^z = -1, 0, 1$; and states V-IX as having $I = 2$, $I^z = -2, -1, 0, 1, 2$, where we have numbered that states in order of their energy.

Considering the singlet states we have $1 \otimes 3 = 3$ i.e., coupling a singlet to a triplet yields only a triplet. Thus, in the presence of SOC the singlet levels yield states with quantum numbers $I = 1, I_z = -1, 0, 1$. Therefore, SOC only hybridises the singlet states with the $I = 1$ triplet states. An important consequence of this is that the wavefunction of the lowest energy excited state, I, has no singlet weight and therefore the I→0 transition, where 0 is the ground state (cf. Figure 5), remains forbidden even in the presence of SOC. This is a direct consequence of the conservation of total (pseudo plus spin) angular momentum, $I$.

We stress that the calculated spectrum of the pseudoangular-momentum model for true octahedral symmetry (shown in Figure 9) is not at all like the measured spectra (discussed in section 4).

### 5.2.1 Why are pseudooctahedral complexes important?

So far we have simply specialised to pseudooctahedral complexes without any particular justification. It is interesting to note that many of the most prominent phosphorescent complexes used in, for example, OLED applications are pseudooctahedral, cf. Figure 1. It is natural to ask why. As described above, in an octahedral complex the relevant pseudooctahedral d-orbitals form a three-fold degenerate ($t_{2g}$) manifold. Key processes in SOC involve flipping a spin and simultaneously moving an electron (or hole) from one d-orbital to another.[7] Such processes are supressed if the d-orbitals are significantly separated energetically. For example, if $\lambda$ is small compared to the other energy scales in the problem, second order perturbation theory predicts that the shift in energy of the $n$th state is

$$\Delta E_n^{(2)} = \sum_{m \neq n} \frac{|\langle \phi_m | H_{SOC} | \phi_n \rangle|^2}{E_m - E_n},$$

where, the $\phi_n$ are the wavefunctions of the nth state in the absence of SOC, $H_{SOC}$ is the SOC terms in the Hamiltonian, and the $E_n$ are the energies of the nth state in the absence of SOC.[8]

In a pseudooctahedral complex it is helpful to write the ligand field, $V(r)$, in the form

$$V(r) = V_{O_h}(r) + V_G(r)$$

(7)

where $V_{O_h}(r)$ contains all of the terms with octahedral symmetry, $G$ is the point group of the complex, e.g., D₃, and $V_G(r)$ are the terms that break the octahedral symmetry. When we describe a complex as pseudooctahedral the implication is that $V_G(r)$ is small compared to $V_{O_h}(r)$. However, in this review we will not be overly concerned with the precise formal definition of this.

---

[7] Recall that $\boldsymbol{L} \cdot \boldsymbol{S} = L^z S^z + \frac{1}{2}(L^+ S^- + L^- S^+)$ where the raising operators, $L^+$ and $S^+$ increase the orbital angular momentum and spin by one respectively and the lowering operators, $L^-$ and $S^-$ decrease the orbital angular momentum and spin by one respectively.

[8] Note that, for simplicity, we have neglected the possibility of degeneracy in the unperturbed Hamiltonian. Although this is likely to be important in most practical situations, it is not germane to the point under discussion here; see, e.g., Ref [54] for a discussion of degenerate perturbation theory.

Therefore in a pseudooctahedral complex the $t_{2g}$ d-orbitals are split by the $V_G(r)$ terms in the ligand field, which we will discuss at length below. Compare this to, for example, a square planar ($D_{4h}$) complex. Here the d-orbitals are split by the large (square planar) terms in the ligand field. Therefore, one expects that the separation of the d-orbitals will be much greater in a square planar complex than in a pseudooctahedral complex. This means that, even if $\lambda$ is the same size the effects of SOC will be significantly weaker in a square planar complex than a similar pseudooctahedral complex. Yersin *et al.* [2] have shown that this expectation is borne out experimentally. Therefore we choose to focus on pseudooctahedral complexes in this review.

### 5.3    $D_3$ complexes, e.g., $M(bpy)_3^{3+}$

When introducing the pseudo-angular momentum model of pseudo-octahedral complexes above we assumed a $t_{2g}$ HOMO and a $t_{1u}$ LUMO with relatively little justification. However, we will now show that this assumption is quite natural for bidentate ligands, which will dominate the discussion below. 2,2'-bipyridine (bpy) is an ideal complex to take as our prototype for this discussion. As bpy has a mirror plane perpendicular to the ligand that bisects the chelate angle[9] all $\pi$ and $\pi^*$ ligand orbitals can be classified as either symmetric or antisymmetric with respect to this reflection (cf. Figure 10) [158, 159]. These reflection symmetries are also present in the case of true octahedral symmetry. In the octahedrally symmetric case combinations of the symmetric $\pi$ or $\pi^*$ ligand orbitals will form a representation of $t_{2g}$ and combinations of the antisymmetric $\pi$ or $\pi^*$ ligand orbitals will form a representation of $t_{1u}$.

Furthermore if we consider the limit of the bond joining the two pyridine groups in bpy becoming weak, then the bpy HOMO will be approximately $\pi_{bpy} = \left[\pi_{py}^{\cdot}(1) + \pi_{py}^{\cdot}(2)\right]/\sqrt{2}$, where $\pi_{py}^{\cdot}$ is the singly occupied molecular orbital (SOMO) of an isolated pyridine radical and the numbers in parentheses distinguish the two different pyridine groups in bpy. Note that $\pi_{bpy}$ is symmetric under reflection through the plane perpendicular to the molecule and bisecting the chelate angle, which interchanges the labels 1 and 2. Similarly, the LUMO of the bpy will be approximately $\pi_{bpy}^* = \left[\pi_{py}^{\cdot}(1) - \pi_{py}^{\cdot}(2)\right]/\sqrt{2}$, which is odd under reflection through the same plane. Therefore, provided the orbitals do not reorder as the strength of the bond is returned to its real value, we expect the ligand HOMO to be symmetric under reflection and the ligand LUMO to be antisymmetric under reflection. Thus in the full complex one expects a $t_{1u}$ LUMO and a $t_{2g}$ HOMO, as assumed in the derivation of $H_{O_h}$ (Equation (5)). The above arguments are confirmed by both first principles [160] and semi-empirical [161] calculations.

---

[9] There is also a $C_2$ axis that bisects the chelate angle – one could equally frame this discussion in terms of the symmetry under this rotation, but the asymmetry of the $\pi$-orbitals under refection through the plane of the ligand makes this a little confusing.

However, as bpy is bidentate $M(bpy)_3$ complexes, such as $Fe(bpy)_3^{2+}$, $Ru(bpy)_3^{2+}$, $Os(bpy)_3^{2+}$ and $Ir(bpy)_3^{3+}$ (**1**), have $D_3$ rather than $O_h$ symmetry. This lifts the three-fold degeneracy of the $t_{2g}$ and $t_{1u}$ orbitals. Consulting the relevant character tables [98] one finds that $t_{2g} \rightarrow a_1 + e$ and $t_{1u} \rightarrow a_2 + e$. However, provided the $D_3$ terms remain small compared to the $O_h$ ligand field, the splitting of the HOMOs and LUMOs will be small compared to their separation from other orbitals. As we saw in our discussion of the triskele in section 5.1 the e states have angular momentum $L^z = \pm 1$, whereas the $a_1$ and $a_2$ states have $L = 0$. Therefore, we can add this trigonal splitting to the pseudo-angular momentum model Hamiltonian by introducing two parameters $\Gamma$ and $\Delta$ to yield

$$
\begin{aligned}
H_{D_3} &= H_{O_h} + \Gamma(L_L^z)^2 + \Delta(L_H^z)^2 \\
&= J\boldsymbol{S}_H \cdot \boldsymbol{S}_L + \lambda \boldsymbol{L}_H \cdot \boldsymbol{S}_H + \Gamma(L_L^z)^2 + \Delta(L_H^z)^2,
\end{aligned}
\tag{8}
$$

where $L_L^z$ and $L_H^z$ are the z-components of the LUMO and HOMO angular momenta respectively. An important question is what are the signs of $\Delta$ and $\Gamma$? We could find this from experiment or a first principles calculation (and indeed we will discuss these below), but it would be better to find this from our understanding of the symmetry of the complex. Because they have different symmetries, the two one-dimensional representations ($a_1$ and $a_2$) cannot interact. But the two pairs of two-fold degenerate states belong to the same representation (e). Therefore they will in general interact [159]. This will tend to stabilise the occupied e states and destabilise the virtual e levels. Thus we expect the frontier orbitals of $M(bpy)_3$ complexes to show the ordering sketched in Figure 11. It is therefore immediately clear that $\Gamma > 0$. $\Delta$ is a little more subtle as we must recall that this is the energy for putting a hole in an e state and holes "roll uphill". Therefore $\Delta$ is also positive. Comparing with scalar relativistic DFT calculations we estimate that for $Ir(ppy)_3$ $\Delta \sim 140$ meV and $\Gamma \sim 90$ meV [111]. Estimates for other complexes give similar values.

It is important to observe that $H_{D_3}$ is the same as a model of $Fe(bpy)_3^{2+}$, $Ru(bpy)_3^{2+}$ and $Os(bpy)_3^{2+}$ studied by Kober and Meyer [162], albeit in a slightly different nomenclature. However, the derivation above is somewhat more sophisticated and places the model in a more general context. Furthermore, Kober and Meyer argued that $\Gamma$ is negative [162]. However, both the simple symmetry arguments of a single bpy ligand (discussed above) and DFT calculations (discussed at length below) suggest that $\Gamma$ is positive. This means that Kober and Meyer's calculations gave very different results from those discussed below, see also [157].

We can again solve the $D_3$ pseudo-angular momentum model defined by Equation (8) (numerically) exactly and typical results are shown in Figure 12. We will see below that, in contrast to the purely octahedral pseudo-angular momentum model, the calculated spectra are now in good agreement with the measured spectra – discussed in section 4, see particularly Table 2.

The trigonal terms break the SU(2) symmetry of the octahedral model and therefore lift the three- and five-fold degeneracies. The calculated spectra are now like those calculated from first-principles for relevant complexes. For example, if trigonal symmetry is enforced for, e.g., $Os(bpy)_3^{2+}$, $Ir(ppy)_3$, $Ir(ptz)_3$ TDDFT calculations predict that SOC splits $T_1$ into a non-degenerate state (I) and, at slightly higher energies, a pair of degenerate states (II and III) [111, 136, 160, 163].

We saw above that in the octahedral model radiative decay from the lowest energy excited state (I→0) is forbidden by the conservation of $I$. Because $H_t$ breaks the SU(2) symmetry of the octahedral model $\boldsymbol{I}^2$ no longer commutes with $H$, nevertheless $I^z$ and $L_L^z$ remains a good quantum numbers for the trigonal model. Furthermore, the Hamiltonian is time reversal symmetric, therefore the parity of an eigenstate under time reversal, $\mathcal{T} = \pm 1$, is also a good quantum number. Note however, that $I^z$ does not commute with time reversal so it is not, in general, possible to form states that are simultaneously eigenstates of both. However, one may define states that are simultaneous eigenstates of the $\mathcal{T}$ and $\mathcal{I}^z = (-1)^{I^z}$. Therefore, we take these as our quantum numbers. To understand this it is helpful to work in an explicit basis, which we detail in Table 5.

Substate I is found [150] to be composed of the basis state $|T_{\parallel}\rangle$ admixed with $|T_{z^2}\rangle$ and has quantum numbers $\mathcal{I}^z = \mathcal{T} = +1$, $L_L^z = 0$ whereas states II and III are a degenerate pair with $\mathcal{I}^z = -1$, $\mathcal{T} = \pm 1$, $L_L^z = 0$, cf. Figure 12, whose largest contributions come from $|T_x\rangle$ and $|T_y\rangle$). The singlet states with the same quantum numbers contribute to substates II and III, but all of the singlets are forbidden from mixing with substate I by the combination of time reversal symmetry and the conservation of $\mathcal{I}^z$. Hence, the I→0 transition remains forbidden in the trigonal pseudo-angular momentum model. This is a direct consequence of the underlying octahedral symmetry of the complex [150].

If one considers the energies of the substates of $T_1$ one finds [150] that, provided the SOC is non-zero, the non-degenerate state (which is predominately $T_1^0$) is lower in energy than the pair of degenerate substates (which are predominately $T_1^{-1}$ and $T_1^1$) for any positive values of $J$, $\Gamma$ and $\Delta$. We have argued above that we expect all of these parameters to be positive in complexes such as $M(\text{bpy})_3$.

In general the radiative decay rates of the eigenstates are given by Eq. (2). In the absence of SOC the triplets are forbidden from decaying radiatively due to the conservation of spin. Therefore, as substate I remains a pure triplet even when SOC is included it cannot decay radiatively even in the presence of SOC. One can see this quite straightforwardly at first order in perturbation theory. Nevertheless in the pseudo-angular momentum model the result that the substate I is forbidden from decaying radiatively is a consequence of the conservation of the total angular momentum, $I$. Therefore this is an exact result and does not depend on the applicability of perturbation theory or on the smallness of SOC ($\lambda$).

Therefore, we have shown that in a D$_3$ pseudooctahedral complex SOC splits $T_1$ into a non-degenerate state, which we identify as I, from which radiative decay is forbidden and, at slightly higher energies, a degenerate pair of states, which we identify as II and III, that do phosphoresce.

## 5.4   C$_3$ complexes, e.g., Ir(ppy)$_3$ and Ir(ptz)$_3$

We now move on to consider $C_3$ pseudooctahedral complexes. To do this we will start from the pseudo-angular momentum model of pseudooctahedral $D_3$ complexes described in section 5.3 and treat the change from $D_3$ to $C_3$ as a perturbation. This is eminently reasonable if we consider that the change in chemical structure inherent in going from Ir(bpy)$_3^{3+}$ (**1**) to Ir(ppy)$_3$ (**2**) amounts to replacing N$^+$ with C. Of course, if we move to bidenate ligands that have less and less mirror symmetry perpendicular to the plane of the ligand this approximation may eventually break down. But we contend that this does not occur in the complexes shown in Figure 1 and Table 2.

The above arguments are equivalent to writing the ligand field (cf. Equation (7)) as

$$V(r) = V_{O_h}(r) + V_{D_3}(r) + V_{C_3}(r)$$

where $V_{D_3}(r)$ contains the terms in the ligand field without $O_h$ symmetry but with $D_3$ symmetry and $V_{C_3}(r)$ contains the terms in the ligand field without $O_h$ or $D_3$ symmetry.

Moving from $D_3$ to $C_3$ removes the distinction between a$_1$ and a$_2$ levels [98]. Thus, the t$_{2g}$ levels of the octahedral HOMO are split into an a orbital and a pair of e orbitals; as are the t$_{1u}$ levels of the octahedral LUMO. Therefore the HOMO and LUMO of a $C_3$ complex may interact, which decreases $\Gamma$ and $\Delta$. Nevertheless, so long as the $C_3$ terms of the ligand field are small compared to the $D_3$ terms of the ligand field one will still find that both $\Gamma$ and $\Delta$ will be positive, cf. Figure 11.

Therefore, for $C_3$ complexes we again expect $J$, $\Gamma$ and $\Delta$ positive. Thus the solution of the $C_3$ pseudo-angular momentum model is very similar to that for $D_3$ complexes. In particular, one again finds [150] that SOC splits $T_1$ into a non-degenerate non-radiative state (I) and, at slightly higher energies, a degenerate pair of states (II and III) that do phosphoresce.

## 5.5   $C_{2v}$, $C_2$ and $C_1$ complexes: broken trigonal symmetry via excited state localisation or in heteroleptic complexes

It is interesting to consider what happens in complexes with even lower symmetry. This is not just relevant because of the interest in complexes with intrinsically low symmetry, such as heteroleptic complexes, but also because of the longstanding idea that excited states may be localised on one ligand [152, 164-166]. This would naturally lead to a lowering of the symmetry of the complex due to vibronic coupling. If we view the complex as basically octahedral, we can view this as a Jahn-Teller distortion [150]. Furthermore, in many heteroleptic complexes the chemistry of the complex may dictate that the low energy excitations are localised to a particular ligand. In contrast the $O_h$, $D_3$ and $C_3$ pseudo-angular momentum models described above assumed that the excited states are delocalised over all three ligands.

Whatever the origin of the breaking of trigonal symmetry it will, in general, lift the degeneracy of the E representation orbitals of the $D_3$/$C_3$ pseudo-angular momentum models (i.e., HOMO-1 and LUMO+1). In terms of the pseudo-angular momentum model these effects are represented by adding the terms

$$H_\gamma = \gamma \left[ \left( L_L^x \right)^2 - \left( L_L^y \right)^2 \right]$$

and

$$H_\delta = \delta \left[ \left( L_H^x \right)^2 - \left( L_H^y \right)^2 \right]$$

to the D$_3$ Hamiltonian. (Note that the values of $\Gamma$, $\Delta$ and, probably to a lesser degree, $J$ may also be altered by excited state localisation or ligand substitution.) This yields the following Hamiltonian for complexes with broken trigonal symmetry:

$$H_{bt} = J\boldsymbol{S}_H \cdot \boldsymbol{S}_L + \lambda \boldsymbol{L}_H \cdot \boldsymbol{S}_H + \Gamma (L_L^z)^2 + \Delta (L_H^z)^2$$
$$+ \gamma \left[ \left( L_L^x \right)^2 - \left( L_L^y \right)^2 \right] + \delta \left[ (L_H^x)^2 - \left( L_H^y \right)^2 \right].$$

Note that for variation of $\Delta$ and $\delta$ ($\Gamma$ and $\gamma$) allows one to represent arbitrary relative energies of the three HOMOs (LUMOs) so this Hamiltonian allows for the description of complexes with heteroleptic ligands as well as localised excitations.

Firstly we note that $I^z$ does not commute with H$_{bt}$. However, $(L_H^x)^2 - \left( L_H^y \right)^2 = \frac{1}{2}[(L_H^+)^2 + (L_H^-)^2]$, where the ladder operators are given by $L_H^\pm = L_H^x \pm iL_H^y$, therefore I$^z$ is conserved modulo two. Thus, $\mathcal{J}^z$ is conserved even for a trigonal system that has undergone a Jahn-Teller distortion, cf. Table 5. Similarly $L_L^z$ is conserved modulo two, which gives rise to the quantum number $\mathfrak{L}^z = (-1)^{L_L^z}$. Therefore, unless $\gamma$ and/or $\delta$ are large enough to drive level crossings involving the substates of T$_1$, the consequences of the conservation of $\mathcal{J}^z$ and $\mathcal{T}$ in the broken trigonal symmetry pseudo-angular momentum model will be the similar to those for the D$_3$ pseudo-angular momentum model, discussed at length in section 5.3. In particular substate I is forbidden from mixing with any of the singlet states, regardless of the strength of SOC, due to the conservation of $\mathcal{J}^z$ and $\mathcal{T}$ [150].

In Figure 13 we plot the exact (numerical) solution of pseudo-angular momentum model for broken trigonal symmetry. In this figure we take $\Delta = J/2$ and $\lambda = J/5$, but the conclusions that follow are robust to wide variations in these parameters [150]. The most interesting part of this solution is that even modest breaking of the trigonal symmetry in the ligand system drives significant changes in the ZFS and in particular drives us to $\Delta_{I,II} < \Delta_{II,III}$, as is seen experimentally for the complexes in Table 2. Therefore, we conclude that the data in Table 2, combined with the theory described above, shows that the excited state is localised in all of these complexes.

In order to understand the ZFS of T$_1$ more fully it is helpful to parameterise the substates I-III in terms of the 'spin Hamiltonian' that is widely used in the discussion of ZFS – particularly in electron paramagnetic resonance experiments:

$$H_{spin} = D(S^z)^2 + E[(S^x)^2 - (S^y)^2].$$

(9)

In Figure 14 we plot the parameter $D$ and $E$ (not to be confused with the energy, $E$) calculated from the pseudo-angular momentum model for broken trigonal symmetry. One sees that: (i) $D$ is always positive and (ii) $D > E$ for all parameters. Firstly, this underlines that the $T_1^0$ substate is always the lowest energy substate (I). Secondly this shows that even weak breaking of trigonal symmetry has a great effect on the ZFS.

### 5.5.1 Radiative decay rates

We plot the radiative decay rate (calculated via Eq. (2)) in Figure 15. State I is dark – as expected from the conservation laws derived above. Furthermore, once the Jahn-Teller distortion becomes significant one finds that the radiative decay from state II is significantly slower than the radiative decay from state III. This is in precisely what is observed in experiments [1, 2, 139] on pseudo-octahedral d$^6$ complexes (cf. Table 2).

It is straightforward to understand both the changes in energy and the radiative rates of states II and III. The trigonal perturbation lowers the energy of (stabilises) states that are antibonding between the $I^z = \pm 1$ orbitals, e.g., $|S_x\rangle$ and $|T_x\rangle$, and raises the energy of (destabilises) those that are bonding between the $I^z = \pm 1$ orbitals, e.g., $|S_y\rangle$ and $|T_y\rangle$. It is clear from Table 5 that whereas $|S_x\rangle$ and $|T_y\rangle$ are even under time reversal $|T_x\rangle$ and $|S_y\rangle$ are odd. Thus, SOC mixes $|S_x\rangle$ with $|T_y\rangle$ and $|S_y\rangle$ with $|T_x\rangle$. Hence the trigonal distortion increases the energy difference between the triplet and singlet basis states that contribute to state II (i.e., $|S_x\rangle$ and $|T_y\rangle$ for $\delta > 0$); whereas trigonal symmetry reduces the energy difference between the triplet and singlet basis states that contribute to state III ($|S_y\rangle$ and $|T_x\rangle$ for $\delta > 0$). Thus the symmetry of the model dictates that $k_R^I < k_R^{II} < k_R^{III}$, as is observed experimentally [1, 2, 139], see Table 2.

### 5.5.2 From D$_3$ to C$_{2v}$ or C$_2$, e.g., $M$(bpy)$_3$

A number of authors have explored ligand field models of an excitation localised to a single ligand [153, 154, 160, 164, 167-169]. It is interesting to compare these with the pseudo-angular momentum model for broken trigonal symmetry, discussed in the preceding section.

Let us begin by taking $M$(bpy)$_3$ as our prototypical system, here we follow the discussion in Ref. [160], which is based on the work of Refs. [159, 164, 167, 168]. A single ligand M-bpy complex would have C$_{2v}$ symmetry, however, in the real $M$(bpy)$_3$ complexes the other two ligands lower the symmetry to C$_2$. We will nevertheless treat the system assuming it has pseudo-C$_{2v}$ symmetry. Note that one can repeat the following analysis for the C$_2$ case straightforwardly, by removing the numerical subscripts in the labels of the representations and allowing for some (weak) mixing between a$_1$ and a$_2$ and between b$_1$ and b$_2$, which is forbidden in the C$_{2v}$ analysis.

In C$_{2v}$ symmetric pseudooctahedral complexes the t$_{2g}$ orbitals are split into three non-degenerate levels, labelled a$_1$, a$_2$ and b$_1$. If the ligand on which the excitation is localised is taken to lie in the yz plane with the z axis along the C$_2$ axis of the bpy then these correspond to the $d_{x^2-y^2}$, $d_{xy}$ and $d_{xz}$ orbitals respectively. Again (cf. section 5.3) we apply the trick of considering the limit of a weak bond between the two pyridine groups, which shows that the highest energy $\pi$-orbital is $a_2$ whereas the lowest energy $\pi^*$-orbital is $b_1$. This is consistent with the findings of both first principles [160] and semi-empirical [161] calculations. Therefore, one expects that interaction with the ligand will stabilise the metallic $b_1$ state and destabilise the metallic $a_2$ state, cf. Figure 16, which defines the labels for the molecular orbitals of the complex. We can therefore characterise the three singlet excitations in the low-energy sector of the model:[10] the lowest energy $^1B_2(a_2 \rightarrow b_1^*)$, the intermediate energy excitation is $^1B_1(a_1 \rightarrow b_1^*)$, and highest energy singlet is $^1A_1(b_1 \rightarrow b_1^*)$.

However, Nozaki [160] argued that the exchange interaction is sufficiently strong for the $^3B_1(a_1 \rightarrow b_1^*)$ excitation to render this the lowest triplet excitation. The three spin substates of triplet excitations transform according to the $b_2$, $b_1$ and $a_2$ representations of $C_{2v}$ corresponding to the x, y and z projections respectively. So, when SOC is included, the three substates transform as $A_1$, $A_2$ and $B_1$ respectively. The $A_1$ and $B_1$ substates can mix with singlets with the same symmetry, but there is no $A_2$ singlet in the model and so this substate remains pure triplet and is therefore spin forbidden. Working through the details of the model one finds that the energies, $E_\Gamma$, of these states will be in the order $E_{A_2} < E_{B_1} < E_{A_1}$ and the oscillator strengths, $f_\Gamma$, will be ordered $0 = f_{A_2} < f_{B_1} < f_{A_1}$.

Thus the conclusion of the $C_{2v}$ model is, again, that the lowest energy triplet is split by SOC and that the lowest energy substate (I) has no oscillator strength. If we add the corrections on moving to $C_2$ symmetry this would allow for some singlet character, and hence some oscillator strength, to be gained by the lowest energy substate as the $A_2$ and $A_1$ substates are now both A and are therefore allowed to interact. But, as these symmetry breaking perturbations are expected to be weak, one still expects that substate I will have the smallest oscillator strength and substate III will have the greatest oscillator strength. Therefore the results of this model consistent with the pseudo-angular momentum model for broken trigonal symmetry.

### 5.5.3  From $C_3$ to $C_1$, e.g., Ir(ppy)$_3$

Finally, we consider the character of excitations localised to a single ligand of a $C_3$ symmetric complex, such as Ir(ppy)$_3$ (**2**). Formally, the complex now has $C_1$ symmetry, i.e., no symmetry at all. Thus, one might expect that group theory has little to say about the properties of the complex. Nevertheless, if, following the arguments in section 5.4, we treat Ir(ppy)$_3$ as [Ir(bpy)$_3$]$^{3+}$ (**1**) with a small perturbation due to replacing one N$^+$ per ligand with C, we can treat an excited state localised to one ligand of Ir(ppy)$_3$ as having pseudo-$C_{2v}$ symmetry. Therefore, this model predicts that with the $T_1$ state of Ir(ppy)$_3$ the substate I will have the smallest oscillator strength and the substate III will have the greatest oscillator strength,

---

[10] Here we use the notation $^{2S+1}\Gamma(\gamma \rightarrow \gamma^*)$ to indicate a transition of spin $S$ and representation $\Gamma$ that consists of moving an electron from an occupied molecular orbital of representation $\gamma$ to an unoccupied molecular orbital of representation $\gamma^*$.

consistent with experiment and the pseudo-angular momentum model for broken trigonal symmetry.

## 5.6   Correlation effects

So far we have only included electronic correlations (configuration interaction) insofar as we have included the exchange interaction, $J$. It is interesting to ask what role other correlations play. Jacko *et al.* have investigated this question in the context of a model similar to those described above (but without SOC) for excitations localised to one ligand [153, 154, 169]. Three important conclusions can be reached from this work.

Firstly, electronic correlations have a large impact on the parameters in the models discussed above, particularly $\Gamma$ and $\Delta$. Indeed Nozaki *et al.* [136, 160] have argued that this means that DFT is not accurate to calculate $\Gamma$ and $\Delta$.

Secondly, the correlations in the low-energy triplet and singlet excitations can be very different [169]. This can lead to important differences in the physical properties of these excitations, for example the degree of MLCT. We will see below that, even in absence of SOC, first principles calculations predict that the singlet excitations are largely single determinant (uncorrelated) but that there is strong configuration interaction between the low-energy triplet states, see, e.g., [84, 111, 144].

Thirdly, Jacko *et al.* [169] solved their model exactly (full CI) and were able to show that the solutions in the single excitation configuration interaction (CIS) approximation reproduce these results extremely well. As CIS is of similar quality to the approximations made in, for example, linear response TDDFT [169, 170], this gives one hope that TDDFT may be accurate enough to capture the important correlation effects in the complexes discussed in this review.

It is important to note that in the pseudo-angular momentum model correlations only act to change the values of the parameters. Therefore the main conclusions of this model are independent of the degree to electron correlations.

## 6   Density functional approaches

The majority of density functional calculations described below take one of two approaches to including scalar relativistic effects: either including them directly by solving the one-component Dirac-Kohn-Sham equations, cf. section 2.4, or by employing relativistic pseudopotentials, see, e.g., [171-173]. There have also been a few benchmarking solutions of the two- and four-component Dirac-Kohn-Sham equations. In the pseudopotential approach one does not treat the core electrons explicitly, rather one replaces the core electrons and the atomic nucleus by an effective core potential or pseudopotential. Relativistic pseudopotentials adjust this potential so as to reproduce the energy shifts that result from *scalar* relativistic effects, such as the direct and indirect effects [173].

Relativistic pseudopotential (TD)DFT calculations can be of similar accuracy to scalar relativistic (TD)DFT calculations, although there are a number of subtle issues at play [173]. In general scalar relativistic calculations are no more expensive than non-relativistic calculations. However, the fact that core electrons are not explicitly included means that relativistic pseudopotentials, like non-relativistic pseudopotentials, do result in a speed up over all electron calculations. The frozen core approximation therefore results in a similar speed-up for scalar relativistic calculations. A major advantage of performing scalar relativistic calculations is that this allows for a straightforward, unbiased 'apples to apples' comparison with non-relativistic calculations. This allows one to understand the role of scalar relativistic effects in a material.

After either a scalar relativistic or relativistic pseudopotential calculation SOC can be included perturbatively. This is typically done at second order. It is not immediately clear that the perturbative approach should be expected to be accurate for the complexes discussed in this review. For example, the spin-orbit coupling constant for Ir is $\sim 4500$ cm$^{-1} \sim 0.55$ eV. Although this is small compared to the $T_1 \rightarrow S_0$ and $S_1 \rightarrow S_0$ transitions (which are several eV), the energy gaps between singlet and triplet excited states are often comparable to or smaller than this, cf. Figure 24, which shows many energy differences of just a few tens of meV between singlet and triplet excitations. Therefore, it is not clear that the perturbation series converges. It is therefore natural to ask how well such an approach compares with more computationally expensive approaches. However, the lowest triplet states are typically well separated, in energy, from the lowest singlet states, at the scalar relativistic level of theory, cf. Figure 24. This gives one hope that at least the low energy excitations may be correctly described by a perturbation theory in the SOC.

To address the accuracy of perturbative approaches to SOC in these complexes Smith *et al.* [111] calculated the excitation spectra of Ir(ppy)$_3$ using both the two-component relativistic formalism and the one-component formalism with spin-orbit coupling included perturbatively. Two component calculations are significantly more computationally expensive. Even with a large core and a modest double zeta plus polarization (DZP) basis set, on Smith *et al.*'s computational architecture (see [111] for details) one SCF iteration took $\sim 2$ hours for the one-component calculation on a single node but $\sim 16$ hours for the two-component calculation. This makes the two-component calculations of a wide spectrum of excited states impractical except for benchmarking purposes and prevents one from improving such calculations by working with higher quality basis sets. However, Mori *et al.* [174] emphasised that the non-perturbative nature of two component calculations means that accurate results are obtained for the low-energy states even when only these states are retained in the calculation. This means that, for (the common) case where we only consider a small number of excited states, such as just $T_1$, full two-component calculations may indeed be practical.

However, Smith *et al.* found that the scalar relativistic theory with SOC included perturbatively does an excellent job of reproducing the two-component calculations (see Figure 17 and Figure 18). The most significant differences are that some states found to have very low oscillator strengths in the two-component theory have somewhat larger (although still small) oscillator strengths in the one-component calculations. These states all have extremely small singlet contributions, which accounts for their weak oscillator strengths. Therefore these errors can be straightforwardly understood as resulting from the perturbative treatment of SOC. The SOC is responsible for mixing singlet and triplet excitations, therefore small *absolute* errors in the degree of singlet-triplet mixing (particularly when the singlet contribution is very small) can lead to large *relative* errors in the oscillator strength. Fortunately, the states most prone to such errors are precisely the states that are least important spectroscopically. Jansson *et al.* [112] found similar agreement between perturbative treatment of SOC and a four-component relativistic treatment.

Furthermore, Smith *et al.* [111] found that increasing the size of the basis set and reducing the size of the core produced much more important changes to the calculated spectra. The changes due to basis sets are far more significant than the differences between the one- and two-component calculations. This suggests that, when seeking the best quality results for phosphorescent organo-transition metal complexes for a given computational cost, one is best served by working in a high quality basis within the one-component theory rather than a small basis or larger core in the two-component theory.

## 6.1 $M$(bpy)$_3$

Nozaki *et al.* studied a series of $M$(bpy)$_3$ complexes with relativistic pseudopotentials both with [160] and without [136] SOC included perturbatively. They also included vibrational broadening of the emission spectra of these complexes via the Huang-Rhys factor [136] and modelled solvent effects via the Onsager model [136, 160, 175]. Nozaki *et al.* studied [Zn(bpy)$_3$]$^{2+}$, [Ru(bpy)$_3$]$^{2+}$, [Os(bpy)$_3$]$^{2+}$, [Rh(bpy)$_3$]$^{3+}$ and [Ir(bpy)$_3$]$^{3+}$ (**1**) and found that in the ground state (S$_0$) the complex is D$_3$ symmetric, consistent with crystallographic measurements [176-179]. Mineev *et al.* [180] have also studied [Ir(bpy)$_3$]$^{3+}$ using relativistic pseudopotentials and including SOC perturbatively and obtained results consistent with those of Nozaki *et al.*

Ground state DFT calculations with relativistic pseudopotentials [160, 180] find that in $M$(bpy)$_3$ there are six frontier orbitals that give rise to the low energy photophysics: two HOMO-1 levels (e), the HOMO (a$_1$), the LUMO (a$_2$), two LUMO+1 levels (e). This is consistent with the HOMO and HOMO-1 being derived from t$_{2g}$ orbitals split by the D$_3$ ligand field and the LUMO and LUMO+1 being derived from t$_{1u}$ orbitals split by the D$_3$ ligand field. Consistent with this the DFT calculations suggest that the HOMO and HOMO-1 are antibonding combinations of metal and ligand orbitals whereas the LUMO and LUMO+1 are predominantly ligand orbitals with only small contributions from the Ir atomic orbitals. However, Nozaki *et al.* [136] note that the trigonal splitting, parameterised by Δ and Γ, cf. Equation (8), are difficult to estimate from (TD)DFT because they are strongly dependent on correlation effects [153, 154, 169], cf. section 5.6. This, in turn, implies that the level of theory used in first principles calculations, e.g., the choice of functional, will be vital for the correct description of the excited states.

Nozaki *et al.* [136, 160] predicted that the symmetry of the Zn, Rh and Ir compounds is lowered to C$_2$ in the T$_1$ state due to structural relaxation consistent with excited state localisation. However, in these compounds the T$_1$ state is predicted to be strongly LC. Consistent with the predictions of the pseudo-angular momentum model (section 5.2 and Ref. [150]), Nozaki *et al.* also found that the effects of SOC are weak in these complexes. This in turn leads to a slow radiative rate from the $T_1$ state even for Ir(bpy)$_3^{3+}$, where one would naïvely expect strong SOC to be associated with the heavy Ir atom. This is quantitatively and qualitatively consistent with what is observed experimentally [181].

The excited state geometries of [Ru(bpy)$_3$]$^{2+}$ and [Os(bpy)$_3$]$^{2+}$ appear to be a somewhat more subtle issue. Nozaki *et al.* predicted that the excited state geometries of these complexes are strongly affected by the polarisability of the solvent - with a $D_3$ structure found *in vacuo* and in less polarisable solvents, but a $C_2$ geometry realised in highly polarisable solvents (see Figure 19). Nozaki *et al.* [136] made a detailed comparison of the predicted and measured phosphorescence spectra of Ru(bpy)$_3^{2+}$ in various environments and found that the experimental spectra were best reproduced by a $C_2$ geometry in acetonitrile (298 K) and glassy butyronitrile (77 K), but that a $D_3$ geometry best reproduced the data from crystals of [Ru(bpy)$_3^{2+}$](PF$_6$)$_2$. Similar comparisons for Os(bpy)$_3^{2+}$ suggest that this complex also takes a $C_2$ conformation in acetonitrile (298 K) and a $D_3$ geometry when doped into crystals of [Ru(bpy)$_3^{2+}$](PF$_6$)$_2$; but that Os(bpy)$_3^{2+}$ takes a $D_3$ structure in glassy butyronitrile (77 K).

In contrast to Zn(bpy)$_3^{2+}$, Rh(bpy)$_3^{3+}$ and Ir(bpy)$_3^{3+}$, Nozaki [160] predicted that $T_1$ has strong MLCT character in Ru(bpy)$_3^{2+}$ and Os(bpy)$_3^{2+}$. This is likely the origin of the greater importance of solvent effects in the Ru and Os complexes. The solvent reorganisation energy is much larger when $T_1$ is localised to a single ligand than when it is delocalised (because the $S_0$ geometry is $D_3$; cf. Figure 19). But, the dipole moment is small in the $D_3$-$T_1$ geometry and much larger in the $C_2$-$T_1$ geometry (Figure 19). This suggests that polar solvents allow for conformational change in the excited state geometry because the large dipole moment is stablised in a polar environment and therefore more than compensates for the energetic cost of the distorted conformation.

Because of the strong MLCT character of $T_1$ in Ru(bpy)$_3^{2+}$ and Os(bpy)$_3^{2+}$ SOC plays a significant role, e.g., inducing large ZFS and a significant radiative decay rate from $T_1$. For all of the $M$(bpy)$_3$ complexes studied by Nozaki *et al.* [136, 160] the calculated excitation spectra in the $D_3$ [136] and $C_2$ [160] geometries are entirely consistent with the pseudo-angular momentum models discussed in sections 5.3 and 5.5 respectively. In particular, the order of the (representations of the) low-energy excitations and the relative radiative decay rates of the substates of $T_1$ are correctly reproduced by the pseudo-angular momentum model. However, in the first principles calculations substate I does attain some radiative rate (consistent with experiment) – although this remains much slower than those of II and III. Presumably this is achieved via coupling to higher energy singlets [144] that are not included in the pseudo-angular momentum model; however Nozaki did not investigate this explicitly.

## 6.2   Ir(ppy)$_3$

Ir(ppy)$_3$ (**2**) is a key active material for OLED displays and the poster-child for Ir(III) phosphorescent emitters. As such, many different approaches have been used to study these materials. This means that, beyond its clear scientific importance, Ir(ppy)$_3$ provides an important benchmark of theoretical methods.

Hay's 2002 paper [130] is a landmark in the theory of phosphorescent organo-transition metal complexes. Hay performed DFT and TDDFT calculations for Ir(ppy)$_3$ and related complexes using the B3LYP hybrid functional [182] and the LANL2DZ relativistic pseudopotential for the Ir atom [172], but did not include SOC. This was one of the first papers to suggest that this, relatively computationally inexpensive, level of theory could do a reasonable job of predicting the structural and optoelectronic properties of such complexes. Hay predominately focused on ground state geometries, and for Ir(ppy)$_3$ he found that the ground state geometry has C$_3$ symmetry. This prediction is consistent with both x-ray crystallography [183, 184] and gas phase electron diffraction [184].

In a C$_3$ geometry, DFT calculations find six key frontier orbitals, see Figure 20. The HOMO and LUMO are (non-degenerate) a orbitals whereas the HOMO-1 and LUMO+1 are the related (two-fold degenerate) e orbitals. The three occupied frontier orbitals (*i.e.*, the HOMO and the two HOMO-1 levels) are strong mixtures of Ir-5d and ppy-$\pi$ orbital, around 50 % from each. The three virtual frontier orbitals (*i.e.*, the LUMO and the two LUMO+1 levels) are essentially pure ppy-$\pi$* orbitals. Quantitatively similar results have been derived from scalar relativistic calculations [111]. This is entirely consistent with the C$_3$ pseudo-angular momentum model discussion in section 5.4, cf. Figure 11.

TDDFT calculations are ultimately more interesting than DFT calculations as they allow for actual predictions of optoelectronic properties [115, 116], cf. section 3. The low-energy excitation spectrum consists of excitations that are dominated by transitions between the frontier molecular orbitals. As the occupied frontier molecular orbitals have ~50 % Ir-5d character and the virtual frontier molecular orbitals have a negligible Ir-5d character, Hay argued that the low-energy excitations are approximately 50 % MLCT and 50 % ligand centred (LC). Again, quantitatively similar results are found in scalar relativistic calculations [111].

Nozaki [160] made a number of important contributions to the modelling of Ir(ppy)$_3$. Firstly, he optimised the geometry of Ir(ppy)$_3$ in the excited T$_1$ state. He found that in the T$_1$ electronic state the C$_3$ symmetry of the S$_0$ geometry is broken, which he attributed to a Jahn-Teller mechanism [136, 185]. Consistent with the pseudo-angular momentum model [150].

Secondly, Nozaki included SOC. To do this he started with TDDFT calculations using a relativistic pseudopotential (LANL2DZ) for the Ir atom, i.e., the same approach Hay employed. Nozaki then included one-centre spin-orbit coupling perturbatively. The one-centre spin-orbit coupling approximation includes mixing between, say, the [1]MLCT and [3]MLCT excitations, but neglects the coupling between say, [3,1]MLCT and [1,3]LC as these involve multi-centre spin-orbit coupling [167, 168].

Nozaki found that the spin-orbit coupling strongly mixes singlet and triplet excitations [160]. Nozaki argued that this means that assigning features in the optical spectrum as [1]LC, [3]MLCT and so forth has little meaning and is an inadequate description of the photophysics. For example: (i) the states responsible for the broad band in the absorption at around 26 000 cm$^{-1}$, which is usually assigned as $^1$MLCT, were found to result from states with a significant ($\sim$30 %) triplet character, and (ii) the state with the largest oscillator strength in the lowest band, which has been assigned to $^3$MLCT, were calculated to have >60 % singlet character. Nevertheless, whenever a perturbative treatment of SOC is valid describing individual transitions as triplet or singlet is meaningful and helpful. Therefore, we will take advantage of this nomenclature below.

Jansson *et al.* [163], who also included scalar effects via a relativistic pseudopotential on the Ir atom and included SOC perturbatively, further investigated the properties of Ir(ppy)$_3$ in the T$_1$ geometry. Comparing the molecular orbitals they find in the S$_0$ (Figure 21) and T$_1$ geometries (Figure 22) confirms that the distortion of the T$_1$ geometry occurs because both of the open shell orbitals are localised on a single ligand. This necessarily requires breaking the C$_3$ symmetry of the molecule at which point vibronic coupling will drive a conformational change. This symmetry breaking is consistent with the pseudo-angular momentum model discussed in section 5.5.3. In the S$_0$ geometry Jansson *et al.* find that the lowest energy substate is non-degenerate (A) and the other two substates are degenerate (E) to within a small amount of numerical noise. This is consistent with the pseudo-angular momentum model (see section 5.4 and Ref. [150]) and other density functional calculations. In the T$_1$ geometry Jansson *et al.* find that $\Delta_{I,II} > \Delta_{II,III}$ – which is inconsistent with both the pseudo-angular momentum model and, more importantly, experiment (see, particularly, Table 2). Therefore these calculations do not correctly reproduce the ZFS.[11]

Jansson *et al.* found that S$_5$ is the singlet excitation most strongly mixed with T$_1$ by SOC and that the S$_5\rightarrow$S$_0$ transition mainly emits light polarised in the *xy*-plane (where the *z*-axis is taken to be parallel to the C$_3$ symmetry axis). The S$_5\rightarrow$S$_0$ transition has E symmetry [111]. This explains why it does not emit z-polarised light. Specifically the electric dipole operator $\boldsymbol{\mu}\sim(x, y, z)$, where $x$, $y$ and $z$ and the Cartesian coordinates [98]. $(x, y)$ form a representation of E, whereas $z$ is a representation of A. Therefore an E symmetry transition, such as S$_5\rightarrow$S$_0$, may only emit light polarised in the *x*-*y* plane.

In the S$_0$ geometry T$_1$ is zero field split into a lower energy A substate and a pair of E substates. The A state (called $T_1^z$ in Jansson *et al.*) is found to have much lower radiative rate than the E states [111, 163]. The S$_5$ state has E symmetry and is predominately HOMO-1$\rightarrow$LUMO+1 [111], consistent with the pseudo-angular momentum model discussed in section 5.4. Therefore, it can mix effectively, via SOC, with the E sublevels of T$_1$ but not the A sublevel. This explains why the A sublevel has a much lower radiative rate. This is precisely the same argument that leads to the small oscillator strength for this state in the pseudo-angular momentum model, cf. section 5.

---

[11] Note however, the Jansson *et al.* use the Cartesian basis for the substates of T$_1$ rather than the eigenbasis. In a C$_3$ geometry the Cartesian basis with the z-axis parallel to the C$_3$ axis is a representation of a $\oplus$ e – so this is not important. But, in this could be responsible for the failure of the calculations to reproduce the experimentally observed ZFS in the T$_1$ geometry.

In the $T_1$ geometry the A and E sublevels of $T_1$ mix because the $C_3$ symmetry is broken. Nevertheless, the substate I still has less oscillator strength than the two substates at higher energies and there remains little $z$-polarised light emitted from the $T_1$ manifold [163], consistent with the approximate $C_3$ pseudo-angular momentum model [150], cf. section 5.5.

## 6.3    Fluorination of Ir(ptz)$_3$

### 6.3.1    The parent complex

X-ray crystallography [54] reveals that, like Ir(ppy)$_3$ (**2**), Ir(ptz)$_3$ (**3**), has an approximate $C_3$ symmetry. Ground state DFT calculations also find a $C_3$ geometry, which is in good agreement with the measured structure [111]. Figure 23 shows the frontier molecular orbitals reported by Smith *et al.* [111]. As for Ir(ppy)$_3$ (cf. Figure 20), the HOMO and LUMO transform according to the A representation and are therefore non-degenerate, whereas the HOMO-1 and LUMO+1 transform according to the E representation. Thus, the electronic structure calculations for both complexes yield trigonal splitting parameters $\Gamma, \Delta > 0$, cf. Equation (8).

Smith *et al.* [84, 111] have carried out detailed electronic structure calculations for Ir(ptz)$_3$ and investigated the different predictions made when relativity is included at different levels of approximation. This allowed them to understand how different relativistic effects affect the properties of Ir(III) complexes. These different levels of theory were then compared to low-temperature (10 K) absorption measurements and low-temperature, high-field (5 T) magnetic circular dichroism (MCD) spectra. Low temperature measurements pose a far stronger test of theory than room temperature spectroscopy does, because of the significant reduction in the broadening of peaks in the spectra as the temperature is lowered (cf. Figure 4 of Ref. [84]; the effect is even more pronounced in the MCD spectra). Furthermore, MCD experiments are particularly suited to understanding $C_3$ complexes because a magnetic field breaks time reversal symmetry and therefore lifts the degeneracy of the E representation (see section 5.1). Therefore MCD spectra strongly distinguish between A and E symmetry excitations, which the absorption spectra alone does not.

Figure 24 compares Smith *et al.*'s measured absorption and MCD spectra with the excitation spectra calculated at three levels of theory: non-relativistic TDDFT, scalar relativistic TDDFT and scalar relativistic TDDFT with spin-orbit coupling included perturbatively.  When comparing calculations one should recall that, although the absorption is strongest around 3.6 eV, this region is not relevant for technological applications of this complex as the active material in OLEDs as this is a far higher energy than the lowest energy transition, which is responsible for emission. (Also recall that the visible region of the spectrum is 1.7 to 3.1 eV.) Therefore, the primary requirement of a useful theory is an accurate description of the lowest energy excitations, around 2.8-3.0 eV, that are responsible for the emission.

It is immediately clear that the non-relativistic calculation does a very poor job of describing the low-energy photophysics – indeed there are no excitations with energy below 3.0 eV and no excitations with any oscillator strength below 3.4 eV.

The scalar relativistic calculation is a significant improvement. This calculation predicts that there are three triplets (one A and a degenerate pair of E's) in the low energy regime. Furthermore, there are singlets with larger oscillator strengths at significantly lower energies in the scalar relativistic calculation. This is vitally important for a correct description of the mixing of singlets and triplets via spin-orbital coupling, which is essential for phosphorescence.

The dramatic differences between the non-relativistic and the scalar relativistic calculations are straightforward to understand. The lowest energy triplet, $T_1$, is predominately a mixture of HOMO→LUMO and HOMO-1→LUMO+1 (69 % and 21 % respectively in the scalar relativistic calculation [111]). Similarly the lowest energy E triplet, $T_2$, is a mixture of, predominately, HOMO→LUMO+1 and HOMO-1→LUMO. The singlets with large oscillator strengths, $S_3$ and $S_5$, are predominately HOMO-1→LUMO and HOMO→LUMO+1 respectively. An examination of the frontier molecular orbitals in Figure 23 shows that these transition have a strong MLCT character; a population analysis shows that these transitions are ~50 % MLCT [84]. From the discussion in section 2.3.1 one expects that the most important scalar relativistic correction will be an increase in the energy of the Ir-5$d$ orbitals due to the indirect effect and that the direct and indirect effects will largely cancel for the π-electrons in the ligands. Scalar relativistic calculations for an isolated Ir atom predict that there is a 0.28 eV destabilisation of the 5$d$-orbital, cf. Table 1 and Ref. [84]. As the HOMO and HOMO-1 have a significant contribution from the Ir-5$d$ orbitals, one expects a significant shift in the energies of these orbitals when scalar relativistic effects are included. On the other hand, as the LUMO and LUMO+1 are predominately ligand π*-orbitals one does not expect scalar relativistic effects to significantly change their energies. This scenario is entirely consistent with the DFT results, see Table 6. Once these shifts are fed forward to the TDDFT calculations the shifts in HOMO and HOMO-1 energies accounts for the shifts in the energies of both the singlet and triplet excitations of around 0.2-0.3 eV [84].

Nevertheless, there remain significant problems with the scalar relativistic TDDFT calculations. Firstly, there are a number of clear features in the MCD and strong absorption in the 2.9-3.1 eV energy window, whereas the scalar relativistic calculation does not predict any states in this energy range. Secondly, if somewhat pedantically, at this level of theory all the triplet transitions are formally forbidden, whereas clearly they are allowed experimentally.

When SOC is included the predicted spectrum is again changed significantly. Most dramatically the zero field splitting of the triplets is non-negligible on this energy scale and results in a significant increase in the number of lines on the graph. These are not additional transitions, but are due to a reduction of the (spin) degeneracy as we no longer have three-fold degenerate triplet states. Indeed singlet and triplets are no longer well defined, but by tracing back the excitations to linear combinations of the scalar relativistic excitations the degree of singlet contribution can be defined – this is represented by the colours of the lines in Figure 24.

The agreement between theory and experiment is extremely good once SOC is included. Firstly, although the lowest energy excitation has A symmetry, the underlying triplet does not mix appreciably with any singlet state (as expected in the pseudo-angular momentum model [150]) and so radiative emission from this excitation is strongly forbidden and does not contribute to either the absorption or MCD spectra at 10 K. This is entirely consistent with the observed MCD A term coincident with the onset of absorption as the second lowest energy excitations are a degenerate pair states with E symmetry, which will be split by an applied magnetic field. Secondly, the zero field splitting produces a series of low energy states which can be used to give a detailed explanation of the MCD spectrum [111]. Thus the major qualitative issues with the scalar relativistic TDDFT are rectified once SOC is included and, indeed, good qualitative agreement with experiment is achieved.

The remaining qualitative differences between theory and experiment are consistent with what one would expect from TDDFT calculations. While it is entirely possible that some improvement may be possible by optimising things like the basis set and exchange-correlation functional, which we will not focus on in this review, major improvement would probably require an alternative method. Of course, higher accuracy methods are difficult to apply to such large molecules as these in the absence of significant discontinuities in computing technology [186]. However, one might also consider interactions with the environment as the effects of these could be larger than correlation effects [115, 169].

All of the result discussed above are consistent with the prediction of the pseudo-angular momentum model of $C_3$ complexes discussed in section 5.4 and Ref. [150].

### 6.3.2 Fluorinated analogues

Relativistic TDDFT calculations with SOC included have been shown [144] to accurately predict the changes in the low-temperature absorption spectra and the low-temperature high-field MCD spectra when $Ir(ptz)_3$ (**3**) is fluorinated in the ortho and para positions of the triazolyl ring (**4-6**), which are labelled *X* and *Y* respectively in Figure 1. The key experimental properties of these complexes are summarised in Table 3. It is clear that fluorination results in a rapid decrease in PLQY. However, closer examination shows that there are two effects at play. Fluorination at *X* leads to a decrease of the radiative rate by a factor of 2 or 3, whereas fluorination at *Y* leads to an increase of the non-radiative rate by an order of magnitude. One would like to understand both of these effects to facilitate progress in designing efficient blue emitters. The calculation of the non-radiative rate is an extremely challenging problem (see section 7.1.4). Therefore we will focus mainly on the radiative rate, however we will make a few observations about the non-radiative rate towards the end of this section.

Crystallographic studies [54] find that the facial isomer has $C_3$ symmetry. There is no experimental evidence to date of distortions in the excited state: in particular MCD experiments [144] see clear A-terms, indicating that the degeneracy of the E states is not lifted to within the width of the transition. Therefore we will conduct the discussion below on the basis of $C_3$ symmetry – even if this symmetry is broken in the excited state the labels will still be reasonably well defined.

The calculated low-energy excitation spectra of $Ir(ptz)_3$ and its fluorinated analogues are shown in Figure 25. It is clear that the details of these excitations change when the complex is fluorinated. However, simply examining this figure does not provide significant insight into how this might result in changes in the radiative or non-radiative rates. To gain such insight it is helpful to unpack these calculations and understand how the excited states are formed from singlets and triplets by turning on SOC. Then one can compare the TDDFT singlets and triplets with the molecular orbitals found in DFT calculations to gain an intuitive understanding of the changes in the photophysics arising from fluorination.

For each complex the lowest energy excitation (labelled 1A) has a negligibly small oscillator strength, as we also saw for $M$(bpy)$_3$ (**1**) and Ir(ppy)$_3$ (**2**) and expect from the pseudo-angular momentum models discuss in section 5. The 1A and 2E states are separated from states 3-6 and it is therefore natural to identify the states 1A and 2E with the zero field split T$_1$ states, whereas states 3-6 (two A and two E) can be identified with the zero field split T$_2$ states. Analysis of the perturbative spin-orbit coupling calculations confirms this assignment and gives a precise quantification of it [144].

Unpacking the linear response TDDFT one finds that in all four complexes T$_1$ is mostly (>62 %) built from the HOMO→LUMO transition and T$_2$ is predominately (>47 %) HOMO→LUMO+1 [144]. We stress that these numbers are far smaller than 100 %. Therefore any discussion (including the following) based on molecular orbitals cannot be viewed as quantitative. Nevertheless the analysis of the triplet excitations in terms of molecular orbital suggests that the energy difference between the T$_1$ and T$_2$ excitations is correlated with Γ, the energy difference between the LUMO and LUMO+1. This is indeed found to be the case (cf. Figure 25 and Table 7). Furthermore, we see that fluorination at the $Y$ position has little effect on the LUMO-LUMO+1 gap whereas fluorination at the $X$ position strongly effects the LUMO-LUMO+1 gap.

Fluorine is strongly electronegative. Thus fluorination enhances the asymmetry between the triazolyl and phenyl groups. This is emphasised by the partial charge analysis in Table 8. Note that this effect is realised over all the orbitals and therefore is not particularly noticeable in the distributions of the frontier orbitals [144]. Nevertheless the effect is clear in the energies of the frontier orbitals (cf. Table 7).

It is interesting to note that the change in partial charges caused by fluorination at $X$ and $Y$ are largely independent. That is, the partial charges are given by

$$Q_{frag}(F,H) = Q_{frag}(H,H) + \Delta Q_X,$$
$$Q_{frag}(H,F) = Q_{frag}(H,H) + \Delta Q_Y,$$
$$Q_{frag}(F,F) = Q_{frag}(H,H) + \Delta Q_X + \Delta Q_Y,$$

where $Q_{frag}(X,Y)$ is the partial charge on the fragment, $frag$ = Ir, triazolyl or phenyl, in the complex with atom $X$ in the ortho position and atom $Y$ in the para position, see labels in Figure 1. These equations can be combined to yield the sum rule

$$Q_{frag}(H,H) + Q_{frag}(F,F) = Q_{frag}(F,H) + Q_{frag}(H,F).$$

(10)

Indeed a similar linear response to fluorination has been demonstrated in a wide range of experimentally measured quantities as well as calculated properties [144]. Therefore, such sum rules may be general rules to aid the design of new phosphorescent complexes.

The discussion above considered only triplet excitations. If we are to give a description of the radiative rate we must also consider the singlet excitations, as it is only once SOC has mixed singlets with triplets that phosphorescence can occur. At room temperature most of the emission comes from the state 2E. We saw above that the largest contribution to this state comes from $T_1$, but, in terms of phosphorescence the mixing with $S_3$ is vitally important. $S_3$ only contributes a small weight to 2E ($\sim$5 %) [144], but it has a large oscillator strength compared to the other low lying singlet excitations.

The mixing of singlets and triplets by SOC is reduced as the energy gap between the states is increased – as can readily be derived from perturbation theory in the SOC. In a related series of complexes one expects the radiative rate to depend on the energy gap between a triplet and the singlet that it mixes with [154, 187-189]. Li *et al.* [187] pointed out that the rate depends on the inverse square of this gap at the lowest order in perturbation theory. However, recently Jacko *et al.* [153, 154, 169] have shown that a second inverse square relationship arises because of the hybridisation between metal and ligand orbitals. Thus overall the radiative rate should exhibit a quadratic dependence on the inverse of the energy gap between a triplet and the singlet that it mixes with [169], which is consistent with TDDFT calculations, cf. Figure 26.

The calculated energy gap $S_3$-$T_1$ is strongly dependent on fluorination at the *Y* position: this gap is $\sim$10 % larger in the two complexes with *Y*=F than the complexes where *Y*=H. However, fluorination at the *X* position has little effect on the $S_3$-$T_1$ gap. The $S_3$ transition is largely HOMO-1→LUMO. Therefore the $S_3$-$T_1$ gap is correlated with the trigonal HOMO-HOMO-1 gap, $\Delta$ [144]. As with the other energy gaps discussed above this can be understood from the different effects of drawing charge towards the *X* and *Y* positions.

With these trends noted it is interesting to ask why substitutions at the *X* and *Y* positions have such different effects on the non-radiative rates. However, we stress that no calculations have explicitly examined this issue to date. Nevertheless, Table 8 shows that fluorination at *Y* position causes large changes in the distribution of charge on the molecule (corresponding to a change in the partial charge on a fragment of order 0.1 electrons in the $S_0$ state). However, the change in the partial charges due to fluorination at the *X* position is two orders of magnitude smaller. If we assume that non-radiative decay involves interaction with the solvent (or solid state environment) [175, 190-193], then changes in the charge distribution, in either the ground or excited states would be expected to lead to dramatic changes in the non-radiative decay rate. Thus, the large changes in the partial charges under fluorination at the *Y* position are consistent with the concomitant large increase of the radiative decay rate. Similarly, the correlation between the small changes in the radiative decay rate and the small changes in the partial charges under fluorination at the *X* position is to be expected.

Furthermore, it is clear from Figure 25 and Table 7 that fluorination at the $X$ position causes more significant redistribution of the low energy spectral weight: decreasing the $T_1$-$T_2$ energy difference by reducing the trigonal LUMO+1-LUMO splitting, $\Gamma$, and decreasing the zero field splitting of $T_2$. This reduces the probability of excitations equilibrating into the main emissive state, 2E, which it is tempting to speculate, may be related to the dramatic increase in the non-radiative rate.

Li *et al.* [194] have investigated two fluorinated analogues of Ir(ppy)$_3$, namely Ir(F$_3$ppy)$_3$ (**7**) and Ir(F$_4$ppy)$_3$ (**8**). They found that TDDFT with relativistic pseudopotentials and SOC included perturbatively reproduces that broad trend (found experimentally [139, 140, 195]) that fluorination reduces the radiative rate. However, the subtle differences in the radiative rates of Ir(F$_3$ppy)$_3$ and Ir(F$_4$ppy)$_3$ were not reproduced; experimentally [195] the radiative rate of Ir(F$_3$ppy)$_3$ is about 20 % *smaller* than that of Ir(F$_4$ppy)$_3$ (at 298 K in degassed CH$_2$Cl$_2$) whereas Li *et al.* predicted that the radiative rate of Ir(F$_3$ppy)$_3$ is about 20 % *larger* than that of Ir(F$_4$ppy)$_3$. Unfortunately Li *et al.* did not report the energies of the singlet/triplet excitations prior to the inclusion of SOC, so we are unable to repeat the analysis described above for Ir(ptz)$_3$ on the basis of their calculations.

## 6.4   Heteroleptic complexes

De Angelis *et al.* [180, 196, 197] have investigated a range of heteroleptic complexes using relativistic pseudopotentials and including SOC perturbatively. The complexes studied include several of the form Ir(ppy)$_2X$, where $X$ is another bidentate ligand (often bpy or a derivative of bpy) and the closely related Ir(df-ppy)$_2$(dma-bpy)$^+$ (**11**). There are many important differences between homoleptic and heteroleptic complexes; these arise primarily because in many homoleptic complexes the HOMO and LUMO are localised on different ligands even in the $S_0$ geometry. This leads to changes in the low energy excitations.

Nevertheless, as we expect from the discussion in section 5.5 many of the key qualitative conclusions drawn from the homoleptic complexes discussed above continue to hold for these heteroleptic complexes. For example, Minaev *et al.* [180] find that the HOMO levels are formed from antibonding combinations of metal and ligand orbitals whereas the LUMO has negligible metal character. Furthermore, as in the homoleptic complexes, SOC mixes $T_1$ most strongly not with $S_1$, but with slightly higher energy excitation such as $S_3$ or $S_5$. This is as one would expect from the selection rules discussed in the context of the pseudo-angular momentum models derived in section 5.

Younker and Dobbs [198] calculated the emission frequencies, ZFS and radiative rates for a number of phosphorescent Ir(III) complexes. They found that properties calculated at the $S_0$ geometry were in better quantitative (absolute value) and qualitative (degree of correlation) agreement with experiment than calculations at the $T_1$ geometry. A naïve interpretation of this result would suggest that the triplet state is sufficiently short lived that the nuclei do not have time to relax. However, Younker and Dobbs rejected this conclusion. Jansson *et al.* [163] found that, in the $T_1$ electronic state, the nuclear potential energy surface is both anharmonic and soft.

Therefore, they argued that the relevant geometry may be somewhere between the predicted $S_0$ and $T_1$ geometries.

Mori *et al.* [174] carried out relativistic TDDFT calculations for twenty three phosphorescent organometallic complexes of potential interest for OLED applications. They showed that both scalar relativistic TDDFT with SOC included perturbatively and two-component TDDFT reproduce the trends seen in the total zero-field splitting ($\Delta_{I,III}$) and phosphorescent radiative lifetime across this set of complexes and correctly predict the values of these properties. For example, these calculations correctly reproduce the expected correlation between a large ZFS and a high radiative rate, as both are caused by strong SOC, as is observed experimentally [2].

Minaev *et al.* [199] investigated an interesting series: $Ir(ppy)_2(pic)$ (**13**), N984 (**14**), N984a (**15**). This corresponds to making a substitution at the ortho position, from a proton, to an amino group to a dimethylamino group (see Figure 1). They were able to reproduce the observed trend in radiative rates across this series. In particular Minaev *et al.* were able to give a clear explanation of how the changes in the triplet states arise from the changes of chemical composition. The replacement of one of the ppy ligands in $Ir(ppy)_3$ (**2**) breaks the $C_3$ symmetry, which lifts the two-fold degeneracy of the HOMO-1 and LUMO+1. Minaev *et al.* found that this leads to the localisation of the frontier molecular orbitals onto a single ligand (or a single ligand and the Ir atom), see Figure 27 – similar to the effect of the distortion in the $T_1$ geometry of $Ir(ppy)_3$. Note that the presence of the pic ligand means that the two ppy ligands are no longer symmetry equivalent; therefore Minaev *et al.* distinguish them by referring to the ligand opposite the Ir-O bond as ppy'. In N984a and N984 the LUMO is based on ppy', the LUMO+1 is based on ppy, and the LUMO+2 is based on pic. However, in $Ir(ppy)_2(pic)$ the pic orbital is dramatically lowered in energy and becomes the LUMO, as shown in Figure 28.

The changes in the orbital energy structure lead to important changes in low lying triplet excitations. In N984a $T_1$ is predominately $HOMO_{ppy+ppy'+d}{\rightarrow}LUMO_{ppy'}$ and $T_2$ is predominately $HOMO_{ppy+ppy'+d}{\rightarrow}LUMO+1_{ppy}$ [199], where we have added subscripts to emphasise the location on the molecular orbital (cf. Figure 27). But in $Ir(ppy)_2(pic)$ the lowering of the energy of the pic orbital leads to a strong mixing of the transitions between the frontier MOs with $T_1$ being $0.41(HOMO_{ppy+ppy'+d}{\rightarrow}LUMO_{pic}) + 0.53(HOMO_{ppy+ppy'+d}{\rightarrow}LUMO+1_{ppy'})$ and $T_2$ being $0.56(HOMO_{ppy+ppy'+d}{\rightarrow}LUMO_{pic}) - 0.42(HOMO_{ppy+ppy'+d}{\rightarrow}LUMO+1_{ppy'})$. This strong mixing means that the two levels are almost degenerate. Minaev *et al.* speculated that this leads to the lower phosphorescence quantum yield in $Ir(ppy)_2(pic)$ than N984 by reducing the rate of intersystem crossing and increasing the rate of non-radiative $T_1{\rightarrow}S_0$ transitions by increasing the matrix element $\langle S_0|H_{SOC}|T_1\rangle$.

Li *et al.* [200] studied the series $Ir(piq)_x(ppy)_{3-x}$ (**16**), with relativistic pseudopotential TDDFT and SOC included perturbatively. They found that this method gives an accurate prediction of the differences in the radiative rates observed in this series. However, they focused primarily on higher energy excitations, which are not relevant to the present discussion.

Thus we see that the predictions of first principles calculations for these homoleptic complexes are also entirely consistent with the predictions of the family of pseudo-angular momentum models developed in section 5.

## 6.5   Degree of metal-to-ligand charge transfer

Yersin *et al.* have emphasised that there is a strong correlation between the observed ZFS and the observed emission decay time across many phosphorescent organo-transition metal complexes, see, in particular, Figure 8 of Ref. [2]. As discussed above, both the ZFS and the radiative decay rate are reproduced by relativistic (TD)DFT calculations, cf. particularly Ref. [174]. We note also that this trend is what would be expected from the pseudo-angular momentum model described in section 5, cf. particularly Equation (6) and the surrounding discussion, when $\lambda$ is small compared to *both* $\Delta$ and $J$. However, away from this limit the ZFS is not monotonic in $\lambda$, cf. Figure 29. Therefore, one would expect this trend to break down for large MLCT, contrary to the assumption of Yersin *et al*.

On the basis of these and other experimental results, Yersin *et al.* [2] have assigned the $T_1$ excitations in those complexes with ZFS of just a few wavenumbers as predominately LC, those with large ZFS as predominately MLCT and those with intermediate ZFS as mixed LC/MLCT. The (TD)DFT calculations are *inconsistent* with this assignment. For example Ir(ppy)$_3$ (**2**) has the largest ZFS splitting of any of the complexes Yersin *et al.* considered and yet the $T_1$ is found to be a roughly equal mixture of LC and MLCT. Indeed none of the complexes discussed above is found to have more than about 50 % MLCT character to its $T_1$ excitation. This is important for, e.g., OLED applications as the exchange interaction comes predominately from the LC component of excitation, cf. equation (6). Thus in order to have triplets that are significantly lower in energy than the singlets, and hence efficient triplet harvesting, one must have a significant LC character in the low lying excitations. This does not explain why one does not find other complexes with larger MLCT components in $T_1$. But perhaps it is simply that the lack of efficient triplet harvesting in such complexes means that their phosphorescence has not been studied in such detail and therefore were not included in the Yersin *et al.*'s analysis.

Therefore while both the first principles and pseudo-angular momentum model calculations are consistent with the *results* of experiment, both are inconsistent with the previous *interpretation* of these experiments. This underlines the general point made in section 3, that one should compare theory to experimental results rather than interpretations!

## 7   Conclusions, outlook and future challenges

It is clear that the results of the family of pseudo-angular momentum models developed in section 5 are in excellent accord with experiment, cf. particularly Table 2. Thus we conclude that the basic conceptual framework for understanding pseudooctahedral $t_{2g}^6$ complexes is provided by the pseudo-angular momentum models.

Furthermore, the properties of the low-energy states predicted by (TD)DFT calculations reviewed in section 6 agree well with both the pseudo-angular momentum model Hamiltonians and experiment. Certainly the prediction of the radiative decay rates of the substates of $T_1$ from first principles are accurate enough to provide important guidance in the design of new complexes once scalar relativistic effects and SOC are included, although the ZFS (particularly the relative sizes of $\Delta_{I,II}$ and $\Delta_{II,III}$) remains a little more challenging. Furthermore, these methods are available in a number of readily available codes - including ADF [201-206], Dalton [207, 208], Dirac [209], NWChem [210, 211], ORCA [212-219] Molcas [220, 221], Molpro [222, 223] and Gamess [224, 225]. So there is no reason for such calculations not to become *de rigueur* within the field.

Nevertheless there is more to be done and we now briefly survey some of the major challenges that remain to be overcome.

## 7.1 Modelling challenges

An oft stated goal in computational and theoretical materials science is the design of new materials. There are a number of major challenges that need to be overcome to achieve this in phosphorescent organo-transition metal complexes. We now briefly discuss some of these. However, before doing so we note that there are a number of different ways in which one might achieve this. For example, a first principles solution might involve screening large numbers of candidate systems and predicting their properties at some desired level of accuracy. Some intriguing programs have taken early steps in this direction in closely related fields [226-228]. Alternatively, the kind of simple models described in section 5 provide conceptual insights that can lead to new design principles. Again there are many examples of this, including some we have discussed above [144, 169]. We believe that both of these approaches have important roles to play in the rational design of new complexes for specific applications. Indeed the interplay between these approaches may be the area where the most progress can be made.

### 7.1.1 Correlations

Matsushita *et al.* [229] studied Ir(ppy)$_3$ (**2**) via the multi-configuration self-consistent field (MCSCF) approximation with SOC included perturbatively. They studied the $T_1$ geometry, but constrained the optimisation to enforce $C_3$ symmetry. This appears to lead to some significant problems with the details of excitation spectrum – in particular, the excitation spectrum is inconsistent with experiment and TDDFT calculations that allow for a $C_1$ geometry of the $T_1$ state (see section 6.2). Nevertheless these calculations suggest that the states relevant to phosphorescence in Ir(ppy)$_3$ only involve single excitations (see, in particular, Table 1 of Ref. [229]).

Jacko *et al.* [153, 154, 169] have investigated correlations in the context of the models discussed in section 5. They have shown that these have profound effects: direct Coulomb interactions alter the values of the trigonal ligand field parameters ($\Gamma$ and $\Delta$); and the exchange interaction has a profound effect on the degree of MLCT in excited states. Likewise, Nozaki *et al.* [136, 160] found that the choice of exchange-correlation functional strongly affected the degree of MLCT.

The vast majority of the calculations discussed above are based on TDDFT and use the B3LYP functional. Younker and Dobbs [198] have compared a generalised gradient approximation (GGA) functional (BP86) with a hybrid functional (B3LYP) and found that the later does better at predicting the ZFS and phosphorescent radiative rate in a range of Ir(III) complexes. (However, they suggested that geometries optimised with BP86 and electronic properties found with B3LYP give the strongest correlation with experiment.) Younker and Dobbs also found that the choice of basis set has a significant impact on the calculated emission rate. There has also been some effort in using other functionals to describe relativistic effects in organometallic complexes, see e.g., Wang and Ziegler [230]. We are not aware of a systematic study of the effect of the choice of functional on relativistic effects in organometallic complexes. However, Świderek and Paneth [231] compared a range of functionals with configuration interaction singles (CIS) and concluded that hybrid functionals performed best – consistent with the findings of Younker and Dobbs. This is consistent with more general findings for (TD)DFT - it is well known that TDDFT descriptions of charge-transfer excited states are significantly improved by inclusion of Hartree-Fock exact exchange [86, 232-239].

### 7.1.2 Quantitative prediction of the spin Hamiltonian parameters

It is seen experimentally in many complexes that the ZFS between substates I and II of $T_1$ ($\Delta_{I,II}$) is significantly less than the ZFS between substates II and III of $T_1$ ($\Delta_{II,III}$), cf. Figure 5 and Table 2. This is not predicted by TDDFT calculations in the $T_1$ geometry for localised excitations [160, 163]. At first sight this seems to be inconsistent with the fact that, even in the $T_1$ structure the complexes are very close to having trigonal symmetry. In a trigonal geometry TDDFT calculations [84, 111, 144] find that the lowest energy substate is A and the higher energy substates are E and therefore degenerate. However, the pseudo-angular momentum model for broken trigonal symmetry naturally predicts that even weak breaking of the trigonal symmetry drives the system into the regime $\Delta_{I,II} \ll \Delta_{II,III}$, cf. Figure 13 and Figure 14. An important challenge for first principles calculations is to reproduce this result and correctly predict the parameters $D$ and $E$ in the spin Hamiltonian, Equation (9). To do so will require modelling of specific solvents as there is a significant solvent dependent shift of the ZFS, cf. sections 4.2 and 7.1.3, and Refs. [2, 139].

### 7.1.3 Solvent effects

Several authors have investigated continuum models of solvents. For example, Mori *et al.* [174] employed a continuum model of solvation effects [240, 241] in their study of phosphorescent organometallic complexes (discussed in section 6.4). They found that the solvent tends to decrease both the ZFS and radiative lifetime as the solvent tends to stabilise more polarised states such as MLCT states. This, in turn, enhances the degree to which triplets mix with, say, $^1$MLCT states. Mori *et al.* used dichloromethane as the solvent in all of their calculations and we are not aware of a systematic theoretical study of the effects of different solvents. However, Younker and Dobbs [198] found that the predictions of the radiative rate were significantly worse when they included solvent effects via the same method. This type of study has thus far been limited to effect of solvents on the radiative properties of the complexes.

Another important role of the solvent (or rather the glassy environment of a device) is to transport charges to the active materials. Marcus-Hush theory [242-245] tells us that the reorganisation energies of the acceptor and donor in a charge transfer process should be properly matched to the electronic energies to optimise the rate of charge transport. Some nascent efforts have been made to apply this to the design of organic electronic devices [246-249] but there is certainly room for this to be applied more widely. This is a cooperative effect and therefore demonstrates that the emissive material and the host should be optimised *together* to achieve the best results.

### 7.1.4  Non-radiative decay rates

We have seen above that the calculation of radiative decay rates for phosphorescent organometallic complexes is largely a solved problem. Certainly, predictions for most materials are accurate to better than an order-of-magnitude. This is probably all that is required for an effective initial computational screening of candidate materials and computer aided design. However, in order to make useful predictions of, say, active materials for OLED applications, one would very much like to be able to say the same for predictions of non-radiative decay rates. There has been considerable experimental progress in understanding the mechanisms of non-radiative decay in organometallic complexes (see Ref. [2] for a recent review). While some pioneering theoretical efforts have been made to understand some mechanisms of non-radiative decay [250-259], the prediction of non-radiative decay rates remains *the* big challenge for theory. We stress that an important part of this challenge is to develop simple models for non-radiative decay. This will allow for a deeper conceptual understanding of the factors that determine the relative importance of different mechanisms and thence to understand how to control these via materials design. New models may also lead to better progress with the computational approach.

Yersin *et al.* [2, 139] have emphasised that the non-radiative decay rate, $k_{NR}^m$, of the $m^{\text{th}}$ substate of $T_1$ in Ir(ppy)$_3$ (**2**) in PMMA is proportional to the radiative decay rate of that substate, $k_R^m$ – the non-radiative rate is an order of magnitude smaller than the radiative rate for all three substates. This means that the PLQY ($\phi_{PL}^m = k_R^m/(k_R^m + k_{NR}^m)$) is very similar, ~90 % in all three substates. This, they argue, is evidence that selection rules prevent non-radiative decay from (pure) triplet states and so the degree of mixing to singlets via SOC, which determines $k_R$, also determines $k_{NR}$. The same trend is also observed in Ir(biqa)$_3$ (**17**) [2]. It would be interesting to know from modelling, computation and experiment how generally this result holds as it may provide important information about the dominant non-radiative decay mechanisms.

### 7.1.5  Vibronic coupling

Early on Nozaki *et al.* [136] demonstrated that Franck-Condon effects on emission could be accurately reproduced by TDDFT calculations (although they did not include SOC in these calculations.) Furthermore, they showed that this could be combined with experiment to give important information about the localisation of excitations. Nevertheless, this kind of analysis has not been widely applied. Once SOC is included other effects may also appear, for example Herzberg-Teller coupling [260] may induce a finite lifetime for substate I of $T_1$ [2, 140]. However, one still expects this lifetime to be extremely long as the decay requires both SOC (to allow a spin-forbidden transition) and Herzberg-Teller coupling (to allow a dipole-forbidden transition). To date there have not been any theoretical attempts to quantify such effects that we are aware of.

## 7.2   Lessons Learnt

We conclude this review by returning the Zewail's challenge [61] and listing some of the key insights that have come from the large theoretical literature on pseudooctahedral cyclometalated $t_{2g}^6$ organometallic complexes.

- It is empirically observed (see, e.g., [2]) that pseudooctahedral complexes are better suited to OLED applications than, e.g., square planar complexes. This is a direct result of the fact that in an octahedral $t_{2g}^6$ complex the occupied d-orbitals remain degenerate. This degeneracy is only weakly lifted by the, say, trigonal ligand field in a real complex. In contrast in, say, a square planar complex the ligand field is such that the occupied d-orbitals have significantly different energies. This means that SOC has a much weaker effect in square planar complexes than in pseudooctahedral complexes.
- Simple pseudo-angular momentum models of the low-energy excited states can describe the phosphorescence and the nature of the $T_1$ excitation. In particular they explain why substate I has a radiative decay rate that is orders of magnitude smaller than those of II and III. These models are built from a recognition of the pseudooctahedral symmetry of the complex. Symmetry lowering terms from the trigonal crystal field can be included as corrections to this framework.
- Comparison of the pseudo-angular momentum with experiment demonstrate that in the complexes listed in Table 2 excitations are localised to a single ligand. First principles calculations also suggest this is the case. Thus organometallic complexes provide an interesting playground for exploring the issue of localised versus delocalised states in mixed valence compounds [137]. The role of polar solvents appears to be particularly interesting (see section 6.1, Figure 19 and Ref. [136]).
- The low-energy virtual states (LUMO, LUMO+1 and LUMO+2) have little metal character because they are odd in the (approximate) reflection through the plane perpendicular the ligand that bisects the chelate angle. This appears to hold even when this symmetry is quite a poor approximation.
- The highest energy occupied states (HOMO, HOMO-1 and HOMO-2) are superpositions of metal $t_{2g}$ and ligand orbitals that are even in the (approximate) reflection symmetry perpendicular to the plane of the ligand. Therefore, the ligand π- and metal d-orbitals can strongly mix if the chemistry allows it. Electronic correlations (configuration interactions) play a key role in determining how strongly such mixing occurs.

- Low energy excitations with ~50 % MLCT and ~50 % LC character are good for generating fast phosphorescence. This follows, in part from the previous two points, and from Equation (6), which shows us that we need strong MLCT and LC character to have both strong SOC and a large exchange interaction (i.e., singlet-triplet splitting). The former is required in order to get radiative excitations out of "triplet" states (cf. the case of Ir(bpy)$_3^{3+}$ (**1**) discussed in section 6.1, where the strong LC character of the excited state leads to slow radiative decay). The latter is required to ensure efficient triplet harvesting, i.e., the funnelling of excitations into the triplet state. (This assumes fast intersystem crossing, but that is guaranteed by the large SOC and in these complexes intersystem crossing occurs on timescales orders of magnitude faster than the excited state lifetime [56, 153].)

  Interestingly, new approaches have recently been discussed involving "singlet harvesting" [2]. This requires a weak exchange interaction to ensure that S$_1$ is at reasonably small energies above T$_1$. To date this has been achieved by moving away from $t_{2g}^6$ transition metals and away from pseudooctahedral complexes. Another approach might be to take a pseudooctahedral complex with a larger degree of MLCT (in particular, a larger $\theta$ in Equation (6), i.e. a HOMO that is very strongly concentrated on the metal).

- The pseudooctahedral and (pseudo)trigonal symmetries of the complexes discussed here determines which substates of T$_1$ may mix with (which) singlets under SOC. This determines both the relative energies and the radiative decay rates of the substates. In particular, symmetry considerations dictate that substate I has a radiative decay rate that is at least an order of magnitude slower than the others. If the trigonal symmetry is broken, the remaining pseudotrigonal symmetry nevertheless dictates that substate II has a slower rate of radiative decay than substate III. This is borne out quantitatively in TDDFT calculations with SOC.

- First principles calculations that include scalar relativistic effects[12] and SOC are accurate enough at predicting the ZFS and radiative decay rates of T$_1$ to be of practical help in designing new generations of complexes. These calculations are sufficiently fast on modern supercomputers to allow for the screening of large numbers of candidate materials in a short time. But such an enterprise would be infinitely more valuable if non-radiative processes were sufficiently well understood that non-radiative decay rates could also be calculated routinely.

# 8   Acknowledgements

I thank Anthony Jacko, Ross McKenzie and Claire Tonnelé for helpful comments on this review and my other collaborators in this area for many helpful discussions. I was supported by an Australian Research Council (ARC) Future Fellowship (FT130100161) while preparing this manuscript.

---

[12] Either explicitly or via pseudopotentials.

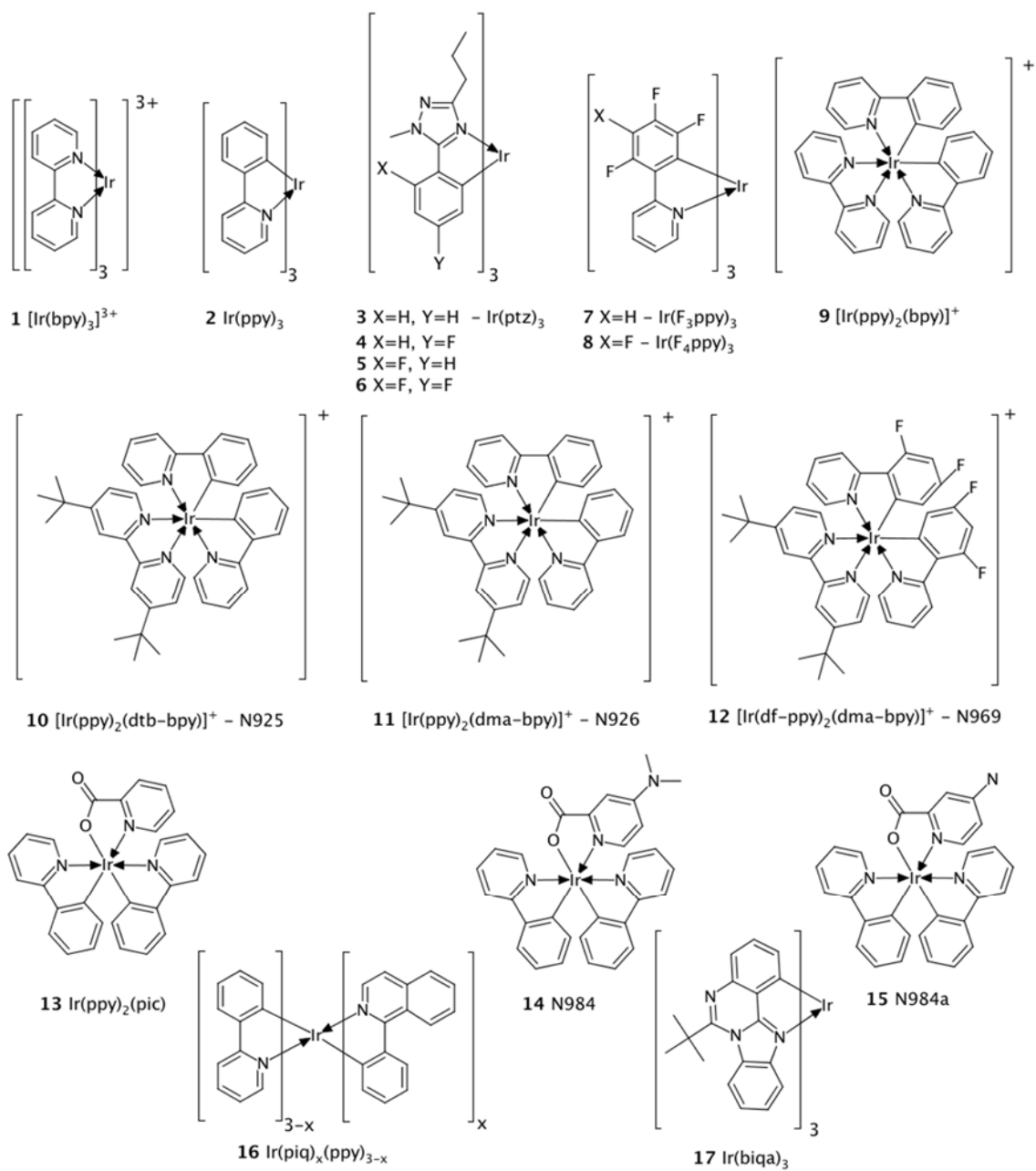

Figure 1. Some of the complexes discussed in this review. Here, and throughout this review, we consider only facial isomers and adopt the following abbreviations: bpy is 2,2'-bipyridine; ppy is 2-phenylpyridyl); ptz is 1-methyl-5-phenyl-3-*n*-propyl; $F_3$ppy is 2-(3',4',6'-trifluorophenyl)pyridine; $F_4$ppy is 2-(3',4',5',6'-tetrafluorophenyl)pyridine; dtb-bpy is 4,4'-di-tert-butyl-2,2'-dipyridyl; dma-bpy is 4,4'-dimethylamino-2,20-bipyridine; df-ppy is 2,4-difluorophenylpyridine; pic is picolinate; N984 is bis(2-phenylpyridine)(2-carboxy-4-dimethylaminopyridine)iridium(III); N984a is bis(2-phenylpyridine)(2-carboxy-4-aminopyridine)iridium(III); piq is 1-phenylisoquinoline; and biqa is 8-*tert*-butyl-benzimidazo(1,2-c)chinazolin.

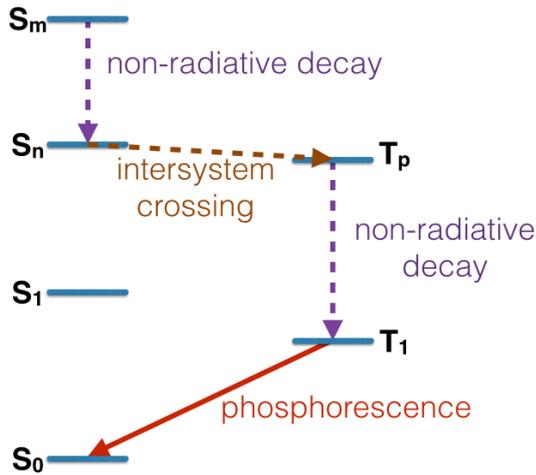

Figure 2. Sketch of a possible decay pathway in the active component of an OLED. An excited state is created when an electron and a hole meet at the molecule. Here we consider the formation of a singlet in state $S_m$, which decays non-radiatively to some lower energy state, $S_n$. SOC then allows intersystem crossing to a triplet state, $T_p$. This is particularly efficient when SOC is large and the energy difference between $S_n$ and $T_p$ is small. The state can decay non-radiatively to the lowest energy triplet, $T_1$. Finally, radiative decay (phosphorescence) occurs back to the ground state, $S_0$.

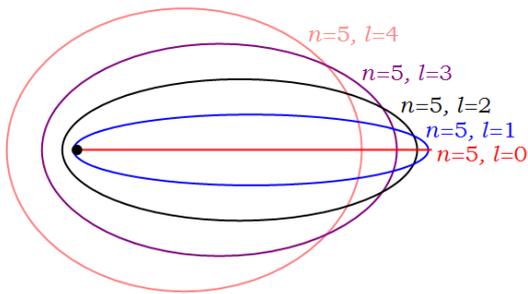

Figure 3. Classical orbits analogous to quantum states for principal quantum number $n = 5$ and quantum-mechanically allowed angular momenta, $l$. The black dot at the left-hand focus marks the nucleus. These are the solutions of the Bohr-Sommerfeld model [261] with the values of the angular momenta corrected to give the correct semi-classical limit. This corresponds to ellipses with semi-major axes of length $\frac{\hbar^2}{me^2} \frac{n^2}{Z}$ and semi-minor axes of length $\frac{\hbar^2}{me^2} \frac{nl}{Z}$, where $Z$ is the charge on the nucleus.

| atomic orbital | occupation | non-relativistic [eV] | scalar relativistic [eV] | ΔE (scalar – non-relativistic) [eV] |
|---|---|---|---|---|
| 1s | 2 | -69361.24 | -76024.44 | -6663.21 |
| 2s | 2 | -11512.87 | -13326.77 | -1813.90 |
| 2p | 6 | -11068.07 | -11581.42 | -513.35 |
| 3s | 2 | -2657.90 | -3127.09 | -469.19 |
| 3p | 6 | -2467.33 | -2615.01 | -147.68 |
| 3d | 10 | -2065.29 | -2048.96 | 16.33 |
| 4s | 2 | -546.82 | -669.96 | -123.14 |
| 4p | 6 | -466.21 | -501.70 | -35.49 |
| 4d | 10 | -297.20 | -293.55 | 3.65 |
| 5s | 2 | -73.88 | -96.50 | -22.61 |
| 4f | 14 | -72.53 | -61.56 | 10.97 |
| 5p | 6 | -47.54 | -52.73 | -5.19 |
| 5d | 9 | -4.82 | -4.54 | 0.28 |
| 6s | 0 | -2.22 | -3.87 | -1.65 |
| 6p | 0 | 2.52 | 1.84 | -0.67 |
| 7s | 0 | 3.91 | 2.75 | -1.15 |
| 6d | 0 | 8.26 | 8.25 | -0.01 |
| 8s | 0 | 25.92 | 20.70 | -5.22 |

Table 1. Calculated energies of the Kohn-Sham orbitals of an isolated Ir atom from a non-relativistic and a scalar relativistic DFT calculation [84]. We see that in general scalar relativistic effects stabilise s- and p- electrons and destabilise d- and f- electrons, which spend less time close to the nucleus, cf. Figure 3. The only exception here is the 6d orbital which has almost the same energy in both calculations.

| | $\Delta_{I,II}$ [cm$^{-1}$] | $\Delta_{II,III}$ [cm$^{-1}$] | $\tau_I$ ($1/k_R^I$) [µs] | $\tau_{II}$ ($1/k_R^{II}$) [µs] | $\tau_{III}$ ($1/k_R^{III}$) [µs] |
|---|---|---|---|---|---|
| Ir(biqa)$_3$ | 14 | 64 | 107 (**114**) | 5.6 (**5.7**) | 0.36 (**0.38**) |
| Ir(ppy)$_3$ (in PMMA) | 12.2 | 113 | 154 (**175**) | 15 (**17**) | 0.33 (**0.34**) |
| Ir(ppy)$_3$ (in CH$_2$Cl$_2$) | 19 | 151 | 116 | 6.4 | 0.2 |
| Ir(dm-2-piq)$_2$(acac) | 9.5–10 | 140–150 | 80–124 | 6.5–8.6 | 0.33–0.44 |
| [Os(phen)$_2$(dppm)]$^{2+}$ | 16 | 106 | 95 | 13 | 0.6 |
| [Os(phen)$_2$(dpae)]$^{2+}$ | 21 | 92 | 100 | 10 | 0.7 |
| Ir(piq)(ppy)$_2$ | 16 | 91 | 64 | 10.5 | 0.3 |
| Ir(4,6-dFppy)$_2$(acac) | 16 | 93 | 44 | 9 | 0.4 |
| Ir(pbt)$_2$(acac) | 6 | 97 | 82 | 25 | 0.4 |
| Ir(piq)$_2$(acac) | 9 | 87 | 47 | 8 | 0.3 |
| [Os(dpphen)$_2$(dpae)]$^{2+}$ | 19 | 75 | 92 | 9 | 0.7 |
| [Os(phen)$_2$(DPEphos)]$^{2+}$ | 16 | 68 | 104 | 14 | 0.9 |
| [Os(phen)$_2$(dppe)]$^{2+}$ | 19 | 55 | 107 | 12 | 0.9 |
| Ir(piq)$_2$(ppy) | 9 | 56 | 60 | 6.4 | 0.44 |
| [Os(phen)$_2$(dppene)]$^{2+}$ | 18 | 46 | 108 | 15 | 1.1 |
| [Ru(bpy)$_3$]$^{2+}$ | 8.7 | 52 | 230 | 8 | 0.9 |
| Ir(piq)$_3$ | 11 | 53 | 57 | 5.3 | 0.42 |
| Ir(4,6-dFppy)$_2$(pic) | 9 | 67 | 47 | 21 | 0.3 |
| Ir(thpy)$_2$(acac) | 3.5 | 31 | 113 | 35 | 1.5 |
| Ir(ppy)$_2$(ppy-NPH$_2$) | 6 | 21 | 188 | 19 | 1.8 |
| Ir(ppy-NPH$_2$)$_3$ | 6 | 20 | 177 | 15 | 1.4 |
| Ir(ppy)(ppy-NPH$_2$)$_2$ | 6 | 17 | 163 | 20 | 2 |
| Ir(btp)$_2$(acac) | 2.9 | 22 | 150 | 58 | 2 |
| Ir(btp)$_2$(acac) | 2.9 | 11.9 | 62 | 19 | 3 |
| Ir(s1-thpy)$_2$(acac) | 3 | 13 | 128 | 62 | 3 |
| Ir(ppy)$_2$(CO)(Cl) | <1 | <1 | 300 | 85 | 9 |
| [Rh(bpy)$_3$]$^{3+}$ | - | - | 4.5 × 10$^3$ | 1.35 × 10$^3$ | 650 |

Table 2. Key spectroscopic data for pseudo-octahedral $t_{2g}^6$-complexes. $\Delta_{I,II}$ is the energy gap between the two lowest energy substates of T$_1$, $\Delta_{II,III}$ is the energy gap between the two highest energy substates of T$_1$ and $\tau_m = (k_R^m + k_{NR}^m)^{-1}$, where $k_R^m$ and $k_{NR}^m$ and the radiative and non-radiative lifetimes of substate $m$, is the total lifetime of substate $m$. For Ir(ppy)$_3$ and Ir(biqa)$_3$ we also list $1/k_R^m$ (in bold) which,

unsurprisingly given the high photoluminescent quantum yields in these complexes, shows the same trend as $\tau_m$. We are not aware of measurements of $k_R^m$ in other relevant complexes. Note that in all complexes $\Delta_{I,II} < \Delta_{II,III}$ and $\tau_I > \tau_{II} > \tau_{III}$, which suggests that $k_R^I < k_R^{II} < k_R^{III}$. To avoid selection bias we have included all and only those pseudo-octahedral d⁶-complexes included in Table 2 of the recent review by Yersin *et al.* [2]. The two rows for Ir(btp)₂(acac) correspond to different sites.

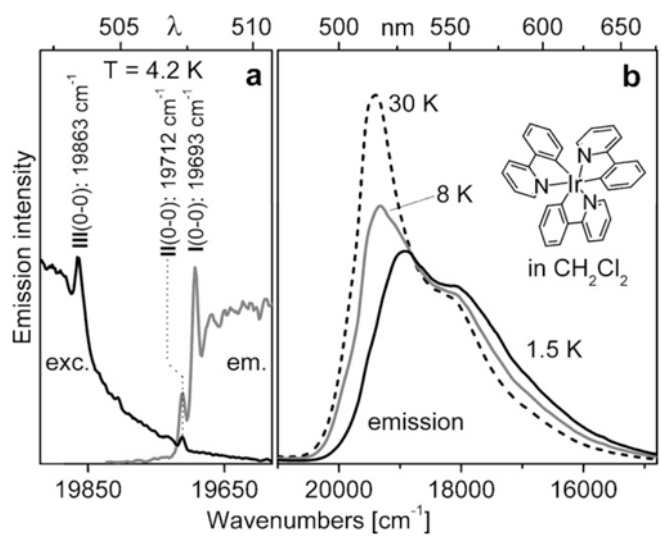

Figure 4. (a) The emission and excitation spectra of Ir(ppy)₃ (**2**) in CH₂Cl₂. (b) Temperature dependence of the emission spectra. From [2].

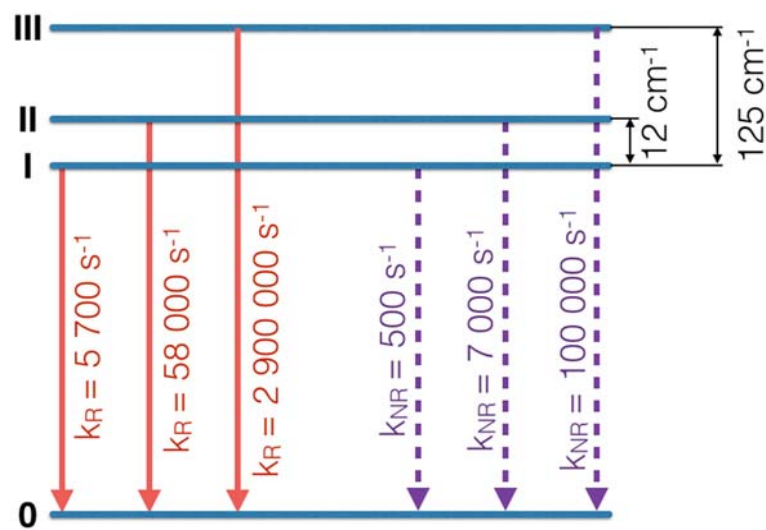

Figure 5. Energy level diagram of the lowest energy triplet excited state of Ir(ppy)₃ (**2**). Measured in ploy(methyl methacrylate) (PMMA) data from Refs. [2, 139]. Note that the lifetimes and zero-field splitting differ by up to a factor of two when measured in a different solvent, e.g., CH₂Cl₂ [2, 139].

| Complex | PL $\lambda_{max}$ (nm) | CIE (x, y) | $\Phi_{PL}$ | $\tau$ ($\mu$s) | $k_r$ ($\mu$s$^{-1}$) | $k_{nr}$ ($\mu$s$^{-1}$) |
|---|---|---|---|---|---|---|
| (**3**) X=H, Y=H | 449 | 0.158, 0.202 | 0.66±0.07 | 1.08±0.03 | 61±8 | 32±10 |
| (**4**) X=H, Y=F | 428 | 0.157, 0.127 | 0.27±0.05 | 1.25±0.30 | 22±09 | 58±28 |
| (**5**) X=F, Y=H | 443 | 0.155, 0.161 | 0.06 | 0.15 | 40 | 630 |
| (**6**) X=F, Y=F | 425 | 0.159, 0.117 | 0.03±0.01 | 0.15±0.07 | 20±16 | 650±330 |

Table 3. Key photophysical properties of Ir(ptz)$_3$ and its fluorinated analogues (**4-6**). Data from [54, 144]. Fluorine substitution shifts the emission from sky to deep blue [as evidenced by both the wavelength of the photoluminescence maximum (PL $\lambda_{max}$) and the Commission Internationale de l'Eclairage (CIE) coordinates], however the PLQY ($\Phi_{PL}$) falls off precipitously. Both the absolute values and the trend in the calculated radiative rates are similar to those determined from experiment. Errors have not been reported for complex (**5**). Nevertheless it is important to note that the radiative rates of complexes (**4**) and (**6**) are the same within experimental error and if there is a difference between the radiative rates of complexes (**3**) and (**5**) it has not yet been seen in these experiments. Likewise, the non-radiative rates of (**3**) and (**4**) are the same within error; as are those of (**5**) and (**6**).

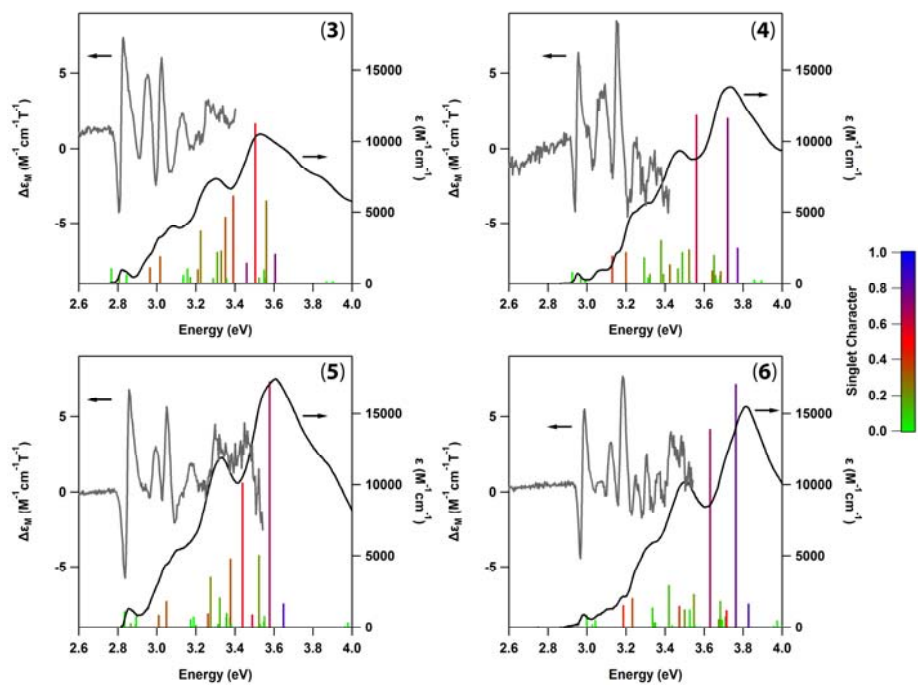

Figure 6. Low temperature absorption, MCD, and calculated relativistic TDDFT excitations (which include SOC perturbatively) for (**3**) Ir(ptz), X=Y=H; (**4**) X=H, Y=F; (**5**) X=F, Y=H; (**6**) X=Y=F; see Figure 1 for the definition of X and Y. $\varepsilon$ is the molar extinction coefficient while $\Delta\varepsilon_M$ is the MCD extinction coefficient scaled to the magnetic field strength. The height of the excitations represent the calculated oscillator strengths. Calculated excitations are color-coded according to the degree of singlet character. In all the complexes, a strong MCD A-term occurs around the first absorption band. Modified from [144].

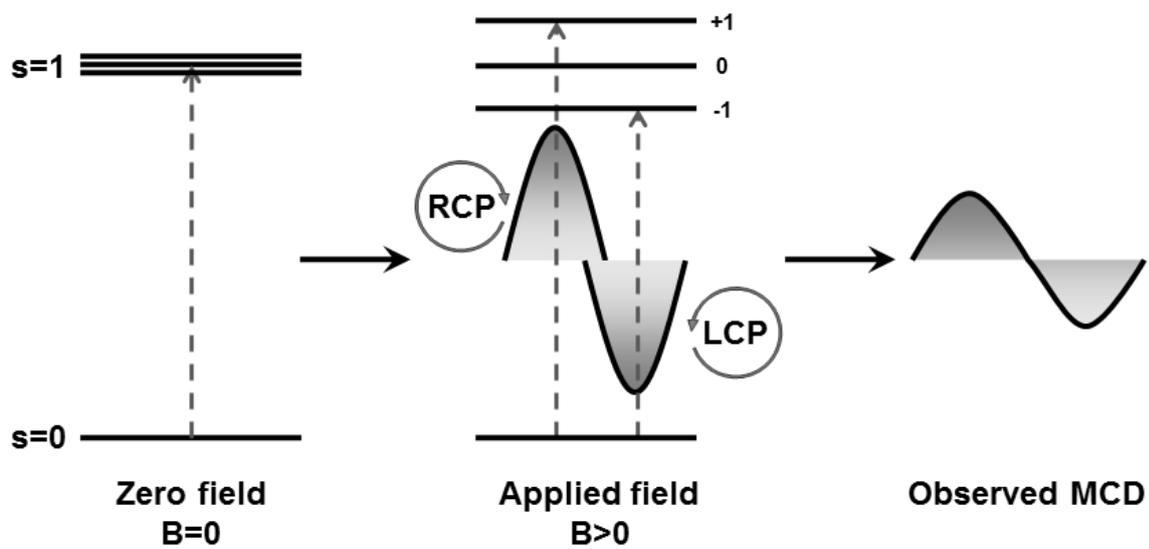

Figure 7. Schematic representation of an MCD A-term resulting from a transition into a degenerate excited state. In this example, a triplet spin state with no ZFS is split by an applied magnetic field. Selection rules dictate that right and left-hand polarised light are absorbed by different substates. From [144].

|  | irrep | $E$ | $C_3$ | $C_3^2$ |
|---|---|---|---|---|
| s, $p_z$ | A | 1 | 1 | 1 |
| $p_x + i p_y$ | $E^+$ | 1 | $\omega$ | $\omega^*$ |
| $p_x - i p_y$ | $E^-$ | 1 | $\omega^*$ | $\omega$ |

Table 4. The character table of C₃ after [98]. Here $\omega = e^{i2\pi/3}$.

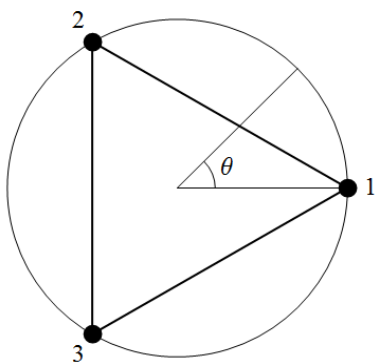

Figure 8. Coordinate system used to study the three site Hückel model.

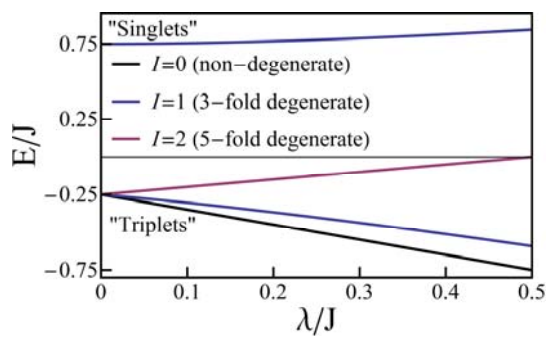

Figure 9. Energy eigenvalues of $H_{O_h}$ (Equation (5)) with $L_z = 0$. At $\lambda = 0$ the singlets have $E = 3J/4$ and the triplets have $E = -J/4$. For $\lambda > 0$ the labels 'singlet' and 'triplet' are no longer strictly defined (in their usual sense) nevertheless the relatively small energy shifts suggest that these labels retain some meaning, this claim is supported by directly examining the character of the eigenstates. It is interesting to note that, already in the octahedral problem the lowest energy (non-degenerate) state has no singlet contribution to its wavefunction for any value of $\lambda$, thus radiative transitions from this state are forbidden. The eigenstates with $L_z = \pm 1$ (not shown for clarity) simply increase the degeneracy of all states by a factor of 3. From [150].

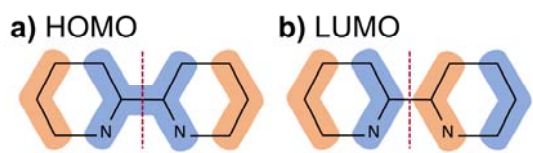

Figure 10. Sketches of π orbitals for a single bpy molecule that are (a) symmetric or (b) antisymmetric with respect to reflection through the plane perpendicular to the ligand that bisects the chelate angle (marked by the dashed line). From [150].

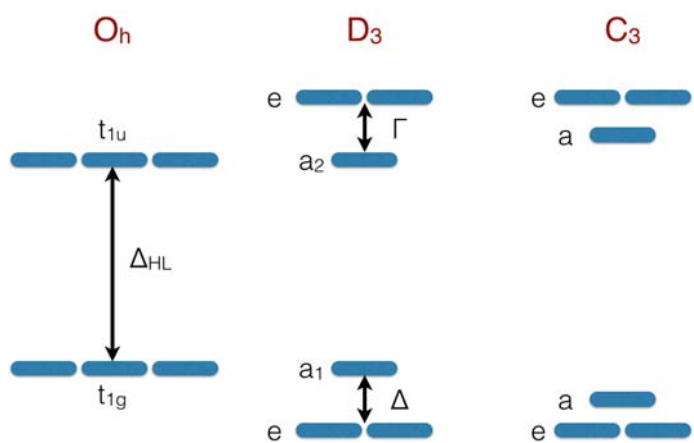

Figure 11. Schematic diagram of the frontier orbitals of complexes with $O_h$, $D_3$ and $C_3$ symmetry. The irreducible representations of orbtials are labelled as are the energy gaps: $\Delta_{HL}$, the HOMO-LUMO gap, $\Delta$ and $\Gamma$ the trigonal splitting of the HOMO and LUMO levels respectively once $O_h$ symmetry is broken. Note that in a $D_3$ symmetric complex the ($a_1$) HOMO and ($a_2$) LUMO are forbidden from mixing by symmetry. This restriction is removed in $C_3$ symmetric complexes as both the HOMO and LUMO transform according to the same representation (a). This stabilises the HOMO and destabilises the LUMO, reducing both $\Delta$ and $\Gamma$ - all other things being equal.

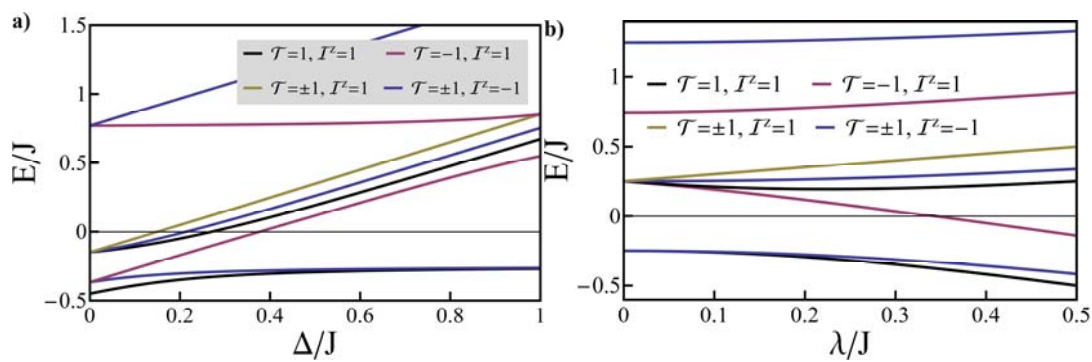

Figure 12. Solution of the pseudo-angular momentum model of a trigonal complex defined by Equation (8). (a) spectra for $\lambda = 0.2J$ and varying $\Delta/J$; (b) spectra for $\Delta = 0.2J$ and varying $\lambda/J$. Quantum numbers of the states are also reported. In both panels the states with quantum numbers labelled as $\mathcal{T} = \pm 1$ are two-fold degenerate. The eigenstates with $L_z = \pm 1$ (not shown for clarity) have the same properties except that their energies are increased by $\Gamma$ and all of the degeneracies are doubled corresponding to the two values of $L_z = \pm 1$. From [150].

| Name | Relation to eigenstates of $H_o$ for $\lambda = 0$, $|I, I^z, S\rangle$ | $\mathcal{T}$ | $\mathcal{I}^z$ | Wavefunctions in the form $|S_L S_H L_H\rangle$ | Singlets mixed with in full model |
|---|---|---|---|---|---|
| $|S_z\rangle$ | $|1,0,0\rangle$ | -1 | 1 | $\frac{1}{\sqrt{2}}[|\uparrow\downarrow\Rightarrow\rangle - |\downarrow\uparrow\Rightarrow\rangle]$ | - |
| $|S_x\rangle$ | $\frac{1}{\sqrt{2}}(|1,1,0\rangle - |1,-1,0\rangle)$ | 1 | -1 | $\frac{1}{2}[|\uparrow\downarrow\Uparrow\rangle - |\downarrow\uparrow\Uparrow\rangle - |\uparrow\downarrow\Downarrow\rangle + |\downarrow\uparrow\Downarrow\rangle]$ | - |
| $|S_y\rangle$ | $\frac{1}{\sqrt{2}}(|1,1,0\rangle + |1,-1,0\rangle)$ | -1 | -1 | $\frac{1}{2}[|\uparrow\downarrow\Uparrow\rangle - |\downarrow\uparrow\Uparrow\rangle + |\uparrow\downarrow\Downarrow\rangle - |\downarrow\uparrow\Downarrow\rangle]$ | - |
| $|T_{\mathbb{1}}\rangle$ | $|0,0,1\rangle$ | 1 | 1 | $\frac{1}{\sqrt{3}}(|\uparrow\uparrow\Downarrow\rangle + |\downarrow\downarrow\Uparrow\rangle) - \frac{1}{\sqrt{6}}(|\uparrow\downarrow\Rightarrow\rangle + |\downarrow\uparrow\Rightarrow\rangle)$ | None |
| $|T_z\rangle$ | $|1,0,1\rangle$ | -1 | 1 | $\frac{1}{\sqrt{2}}(|\uparrow\uparrow\Downarrow\rangle - |\downarrow\downarrow\Uparrow\rangle)$ | $|S_z\rangle$ |
| $|T_x\rangle$ | $\frac{1}{\sqrt{2}}(|1,1,1\rangle - |1,-1,1\rangle)$ | -1 | -1 | $\frac{1}{2}\left[|\uparrow\uparrow\Rightarrow\rangle - |\downarrow\downarrow\Rightarrow\rangle - \frac{1}{\sqrt{2}}(|\uparrow\downarrow\Downarrow\rangle + |\downarrow\uparrow\Uparrow\rangle - |\uparrow\downarrow\Uparrow\rangle - |\downarrow\uparrow\Downarrow\rangle)\right]$ | $|S_y\rangle$ |
| $|T_y\rangle$ | $\frac{1}{\sqrt{2}}(|1,1,1\rangle + |1,-1,1\rangle)$ | 1 | -1 | $\frac{1}{2}\left[|\uparrow\uparrow\Rightarrow\rangle + |\downarrow\downarrow\Rightarrow\rangle - \frac{1}{\sqrt{2}}(|\uparrow\downarrow\Uparrow\rangle + |\downarrow\uparrow\Uparrow\rangle + |\uparrow\downarrow\Downarrow\rangle + |\downarrow\uparrow\Downarrow\rangle)\right]$ | $|S_x\rangle$ |
| $|T_{z^2}\rangle$ | $|2,0,1\rangle$ | 1 | 1 | $\frac{1}{\sqrt{6}}(|\uparrow\uparrow\Downarrow\rangle + |\downarrow\downarrow\Uparrow\rangle) + \frac{1}{\sqrt{3}}(|\uparrow\downarrow\Rightarrow\rangle + |\downarrow\uparrow\Rightarrow\rangle)$ | None |
| $|T_{xz}\rangle$ | $\frac{1}{\sqrt{2}}(|2,1,1\rangle - |2,-1,1\rangle)$ | -1 | -1 | $\frac{1}{2}\left[|\uparrow\uparrow\Rightarrow\rangle - |\downarrow\downarrow\Rightarrow\rangle + \frac{1}{\sqrt{2}}(|\uparrow\downarrow\Uparrow\rangle + |\downarrow\uparrow\Uparrow\rangle - |\uparrow\downarrow\Downarrow\rangle - |\downarrow\uparrow\Downarrow\rangle)\right]$ | $|S_y\rangle$ |
| $|T_{yz}\rangle$ | $\frac{1}{\sqrt{2}}(|2,1,1\rangle + |2,-1,1\rangle)$ | 1 | -1 | $\frac{1}{2}\left[|\uparrow\uparrow\Rightarrow\rangle + |\downarrow\downarrow\Rightarrow\rangle + \frac{1}{\sqrt{2}}(|\uparrow\downarrow\Uparrow\rangle + |\downarrow\uparrow\Uparrow\rangle + |\uparrow\downarrow\Downarrow\rangle + |\downarrow\uparrow\Downarrow\rangle)\right]$ | $|S_x\rangle$ |
| $|T_{xy}\rangle$ | $\frac{1}{\sqrt{2}}(|2,2,1\rangle - |2,-2,1\rangle)$ | -1 | 1 | $\frac{1}{\sqrt{2}}[|\uparrow\uparrow\Uparrow\rangle - |\downarrow\downarrow\Downarrow\rangle]$ | $|S_z\rangle$ |
| $|T_{x^2-y^2}\rangle$ | $\frac{1}{\sqrt{2}}(|2,2,1\rangle + |2,-2,1\rangle)$ | 1 | 1 | $\frac{1}{\sqrt{2}}[|\uparrow\uparrow\Uparrow\rangle + |\downarrow\downarrow\Downarrow\rangle]$ | None |

Table 5. The basis set used to study the pseudo-angular momentum model. The wavefunctions are given in the form $|S_L S_H L_H\rangle$ with $\uparrow$ ($\downarrow$) indicating $S_v^z = +\frac{1}{2} (-\frac{1}{2})$ and $\Uparrow, \Rightarrow$ and $\Downarrow$ indicating $L_H^z = 1, 0$ and $-1$ respectively. $\mathcal{I}^z = (-1)^{I^z}$. In this table we list only the $L_L^z = 0$ [$\mathfrak{L}^z = (-1)^{L_L^z} = 1$] states. Each state has two partners with $L_L^z = \pm 1$ and hence $\mathfrak{L}^z = -1$. The latter two, but not the former, mix under the action of the full Hamiltonian.

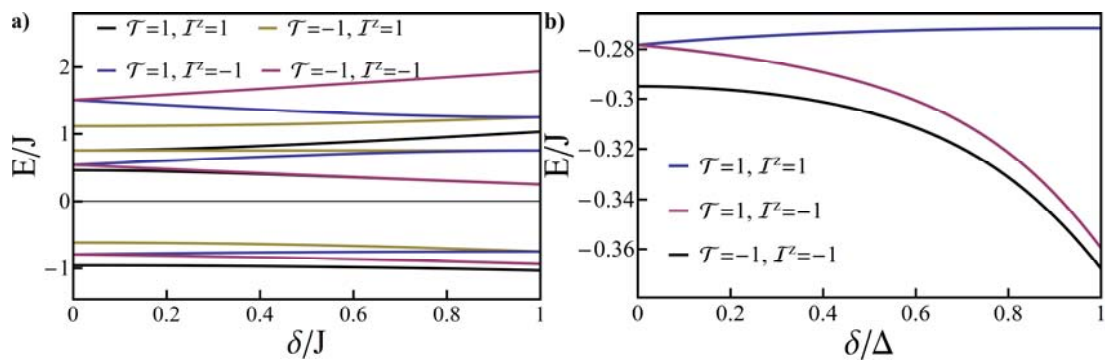

Figure 13. Solution of the pseudo-angular momentum model of a complex with broken trigonal symmetry - due either to chemical modification or excited state localisation. Panel (a) shows the full spectrum with $L_z = 0$. Panel (b) shows only the $T_1$ states, which are our primary concern. Here we take $\Delta = J/2$ and $\lambda = J/5$. From [150].

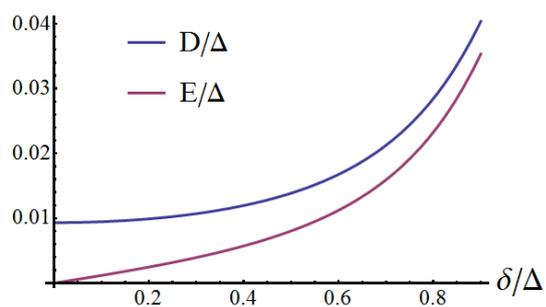

Figure 14. Calculated spin Hamiltonian parameters D and E (not to be confused with the energy, $E$) for the pseudo-angular momentum model of a complex with broken trigonal symmetry. Here we take $\Delta = \Gamma = J$ and $\gamma = \delta$.

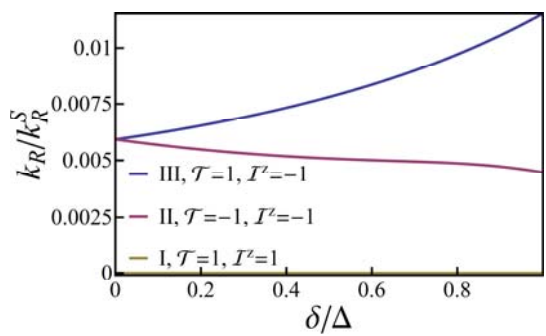

Figure 15. The radiative decay rates of the three substates of $T_1$. The conservation of $\mathcal{T}$ and $\mathcal{I}^z$ leads to the absence of radiative decay for state I. It can be seen that once the Jahn-Teller distortion becomes significant the radiative decay rate from state II is significantly smaller than that from state III, in good agreement with experiment (cf. Table 2). Here, as above, we take $\Delta = J/2$ and $\lambda = J/5$. Because of the underlying octahedral symmetry we take $\langle S_0 | \mu_\beta | S_n \rangle$ to be independent of $n$ and further we assume that the zero field splitting is small compared to the $S_0 \rightarrow T_1$ excitation energy, i.e., that $E_I \simeq E_{II} \simeq E_{III}$. It is convenient to define $k_R^S = \frac{4 m_e e^4 \alpha E_I^3}{9(4\pi\varepsilon_0)^2 h^3} \sum_{\beta \in \{x,y,z\}} \sum_{n \in \{x,y,z\}} \langle S_0 | \mu_\beta | S_n \rangle$, this corresponds to the radiative decay rate for a pure singlet with an excitation energy equal to the $T_1$ manifold. From [150].

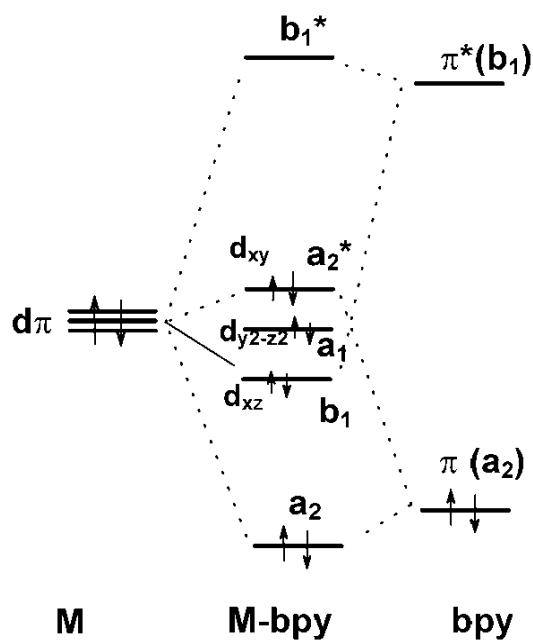

Figure 16. Correlation diagram for a $t_{2g}^6$ metal coupled to a single bpy ligand to form a complex with $C_{2v}$ symmetry. From [160].

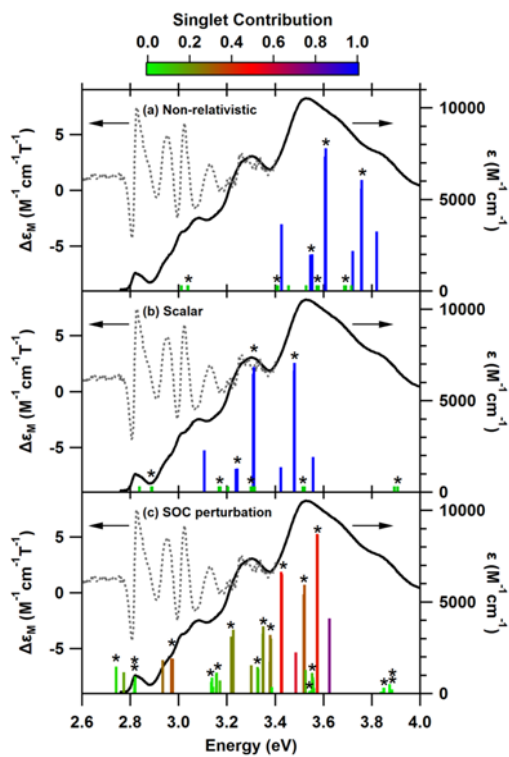

Figure 17. Excitations 1-8 calculated from one- and two-component methods for Ir(ppy)$_3$ (2) and Ir(ptz)$_3$ (3). Asterisks mark two-fold degenerate E states and states with significant oscillator strengths are labelled. Good agreement between the two methods is found for both complexes. These low lying excitations dominate the optical properties, and hence the technological applications, of these complexes. The excitation spectra of the two complexes are quite similar in these energy ranges. From [111].

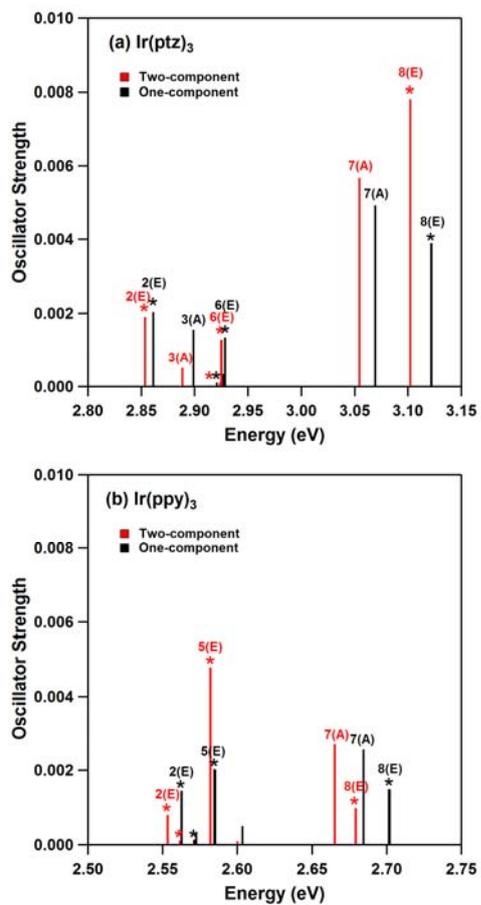

Figure 18. Comparison of the excitation spectra from two- and one-component relativistic TDDFT calculations with the measured absorption spectra for Ir(ppy)$_3$ (**2**) and Ir(ptz)$_3$ (**3**). Both calculations are clearly very similar. From [111].

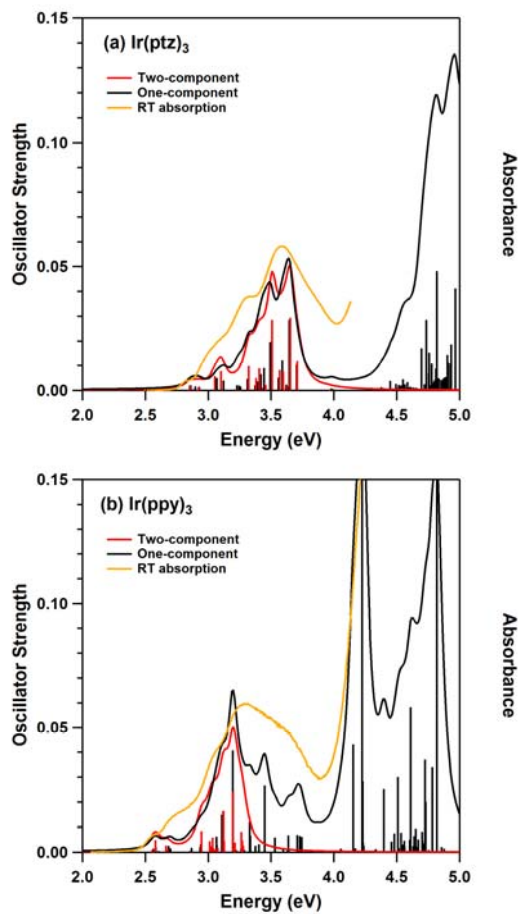

Figure 19. Calculated dipole moments, μ, of the $T_1$ states and reorganisation energies, $\lambda_{vib}$, for $T_1 \rightarrow S_1$ for (a) $[Ru(bpy)_3]^{2+}$ and (b) $[Os(bpy)_3]^{2+}$. The discontinuous changes in both properties result from the change from $D_3$ to $C_2$ conformations when the polarisability of the solvent is increased. The solvent is described by the Onsager model [175] and $g = [2(\epsilon - 1)]/[(2\epsilon + 1)r]$, where $\epsilon$ is the static dielectric constant of the solvent and $r$ is the radius of the cavity enclosing the organometallic complex, i.e., more polar solvents correspond to larger $g$. From [136].

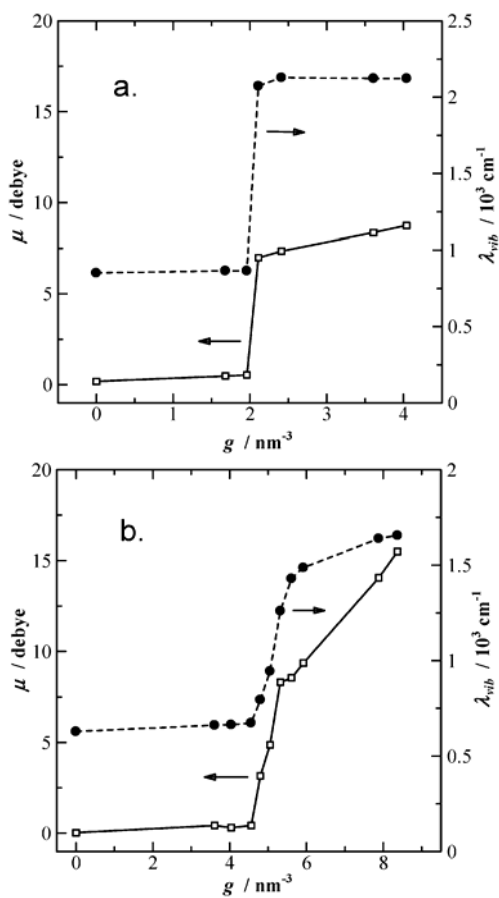

Figure 20. Frontier Kohn-Sham molecular orbitals and their energies in Ir(ppy)₃ (**2**) and the corresponding molecular orbital energies. From [111].

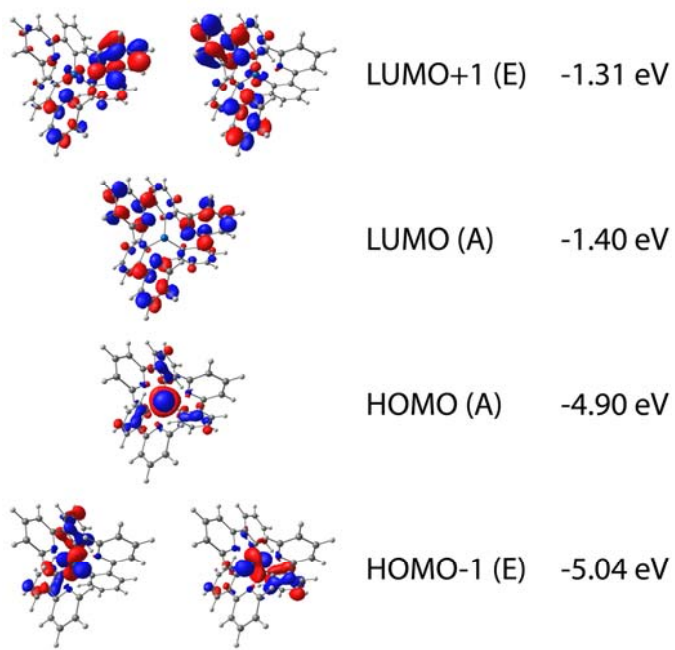

LUMO+1 (E)    -1.31 eV

LUMO (A)    -1.40 eV

HOMO (A)    -4.90 eV

HOMO-1 (E)    -5.04 eV

Figure 21. (a) HOMO and (b) LUMO calculated of Ir(ppy)₃ (**2**) in the $S_0$ geometry. From [163].

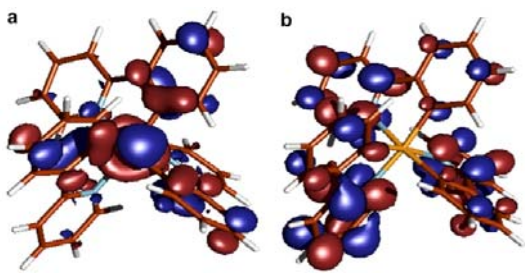

Figure 22. (a) Lowest and (b) highest energy occupied open shell molecular orbitals of Ir(ppy)$_3$ (**2**) in the T$_1$ geometry. From [163].

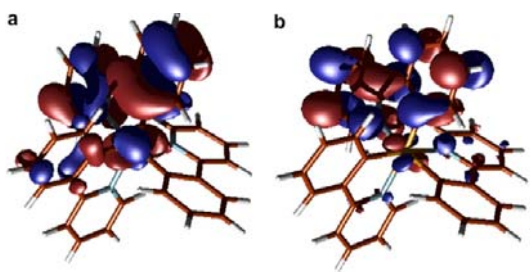

Figure 23. Frontier Kohn-Sham molecular orbitals and their energies in Ir(ptz)$_3$ (**3**) and the corresponding molecular orbital energies. From [111].

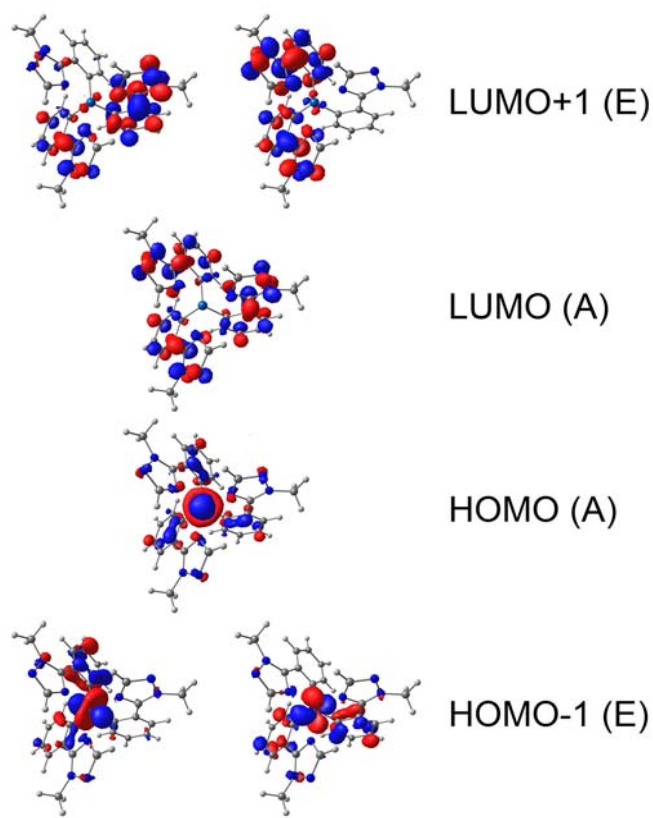

LUMO+1 (E)

LUMO (A)

HOMO (A)

HOMO-1 (E)

Figure 24. Absorption (solid line) and MCD (dashed line) spectra of Ir(ptz)$_3$ (**3**) collected at 10 K under an applied field of 5 T compared to the unbroadened absorption spectra calculated by (a) non-relativistic TDDFT, (b) scalar relativistic TDDFT and (c) perturbative SOC correction to the scalar relativistic calculation. All calculations were performed in a C$_3$ geometry. Degenerate (E) states are denoted by * (** marks two nearby E states). In the NR and SR calculations the (formally forbidden) triplet excitations are given small arbitrary oscillator strengths for clarity. Taken from [84].

| Orbital | Non-relativistic | Scalar relativistic |
|---------|------------------|---------------------|
| LUMO+1  | -0.94 eV         | -0.93 eV            |
| LUMO    | -1.09 eV         | -1.10 eV            |
| HOMO    | -5.21 eV         | -4.93 eV            |
| HOMO-1  | -5.36 eV         | -5.10 eV            |

Table 6. Comparison of the frontier molecular orbital energies of Ir(ptz)$_3$ (**3**) calculated from non-relativistic and scalar relativistic DFT. The indirect effect destabilises the Ir-5d orbital, which contributes significantly (~50 %) to the HOMO and HOMO-1 but negligibly to the LUMO and LUMO+1 (cf. Figure 23). This results in significant destabilisation of the HOMO and HOMO-1 once scalar relativistic effects are included and explains the negligible changes to the virtual orbitals from scalar relativity. Data from [84].

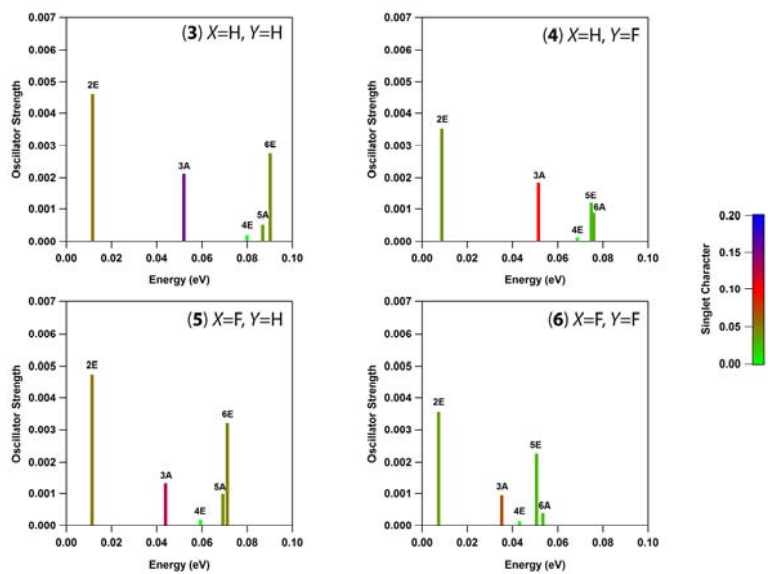

Figure 25. Lowest six excitations of Ir(ptz)$_3$ (**3**) and its fluorinated analogues (**4-6**) calculated from SOC perturbation TDDFT. Energies are plotted relative to the energy of the first excitation 1A [which has an extremely small (f < 10$^{-5}$ au) oscillator strength], the energy range between excitations 1 and 6 decreases with fluorine substitution. The ZFS of the T$_2$ manifold (excitations 3-6) is also reduced by fluorination. The colour coding indicates the singlet character of the excitations. Modified from [144].

| Complex | HOMO-1 | HOMO | LUMO | LUMO+1 | HOMO-HOMO-1 | LUMO+1-LUMO |
|---------|--------|------|------|--------|-------------|-------------|
| (**3**) X=H, Y=H | -5.154 | -4.980 | -1.166 | -0.989 | 0.174 | 0.177 |
| (**4**) X=H, Y=F | -5.583 | -5.393 | -1.405 | -1.229 | <span style="color:red">0.190</span> | 0.176 |
| (**5**) X=F, Y=H | -5.544 | -5.382 | -1.489 | -1.335 | 0.162 | <span style="color:red">0.154</span> |
| (**6**) X=F, Y=F | -5.969 | -5.781 | -1.701 | -1.557 | <span style="color:red">0.188</span> | <span style="color:red">0.144</span> |

Table 7. Calculated energies of frontier molecular orbitals in Ir(ptz)$_3$ (**3**) and its fluorinated analogues. All energies in eV; data from [144].

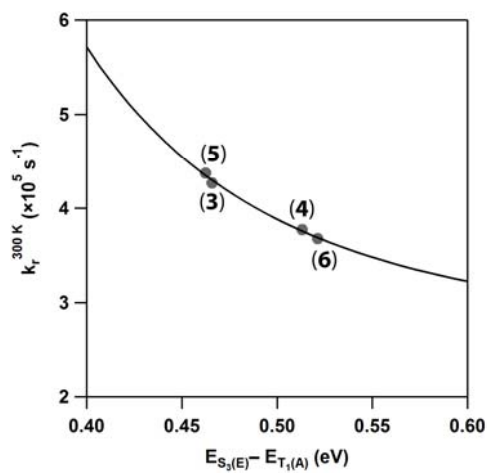

Figure 26. Calculated total radiative rate at 300 K for Ir(ptz)₃ (**3**) and its fluorinated analogues (**4-6**), plotted against the calculated energy gap between the scalar TDDFT excitations $S_3$ and $T_1$. The $S_3$-$T_1$ energy gap is found to be strongly dependent on the fluorination at the $Y$ position, whereas fluorination at $X$ does not change the relative energy separation significantly. The line is a best fit for the predicted dependence between the inverse fourth power of the radiative rate and the singlet-triplet energy gap [169]. However, we note that the limited variation in this data means that it does not strongly discriminate between different power laws. The calculated radiative rate is the same order of magnitude as the experimentally measured rate [54]. Modified from [144].

|  | (**3**) $X$=H, $Y$=H | (**4**) $X$=H, $Y$=F | (**5**) $X$=F, $Y$=H | (**6**) $X$=F, $Y$=F | Error in Eq. (10) |
|---|---|---|---|---|---|
| iridium | 0.4383 | 0.4417 | 0.4086 | 0.4130 | 0.0010 |
| triazolyl | 0.0368 | 0.0360 | 0.1391 | 0.1392 | 0.0009 |
| phenyl | -0.4705 | -0.4731 | -0.5461 | -0.5482 | 0.0005 |

Table 8. Partial charge per fragment based on Hirshfeld population analysis from scalar relativistic DFT. The total charge distribution changes with fluorination, as electron density is redistributed from the triazolyl to the phenyl ring. The column labelled Error in Eq. (10) is a test of the sum rule for fluorination where an entry of 0 indicates perfect agreement between the Hirshfeld population analysis and the predictions of Eq. (10). The observation that entries of this row are all zero to a very high accuracy indicates that redistributions of charge caused by fluorination at the $X$ and $Y$ positions are to a very high degree independent of one another. Data from [144].

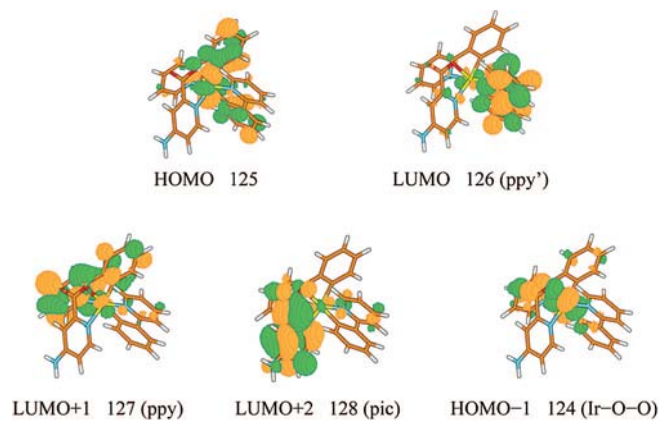

Figure 27. The molecular orbitals of N984a (**15**). From [180].

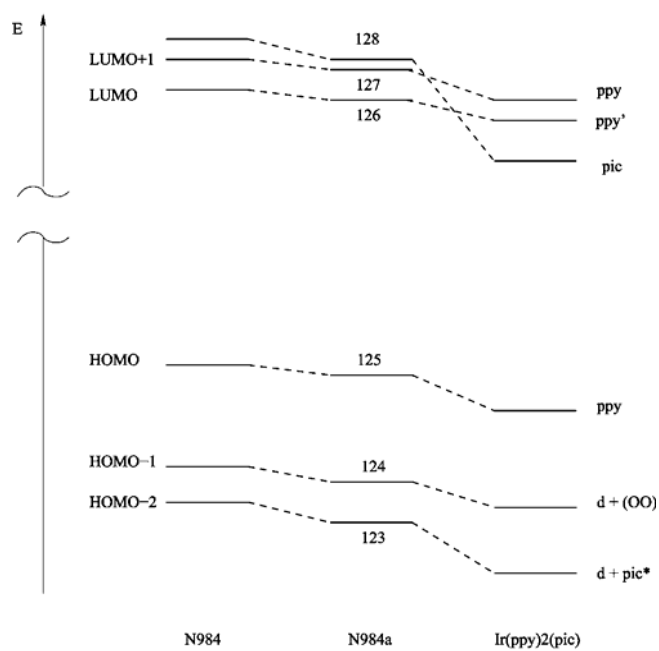

Figure 28. Changes in the molecular orbitals energies in the series Ir(ppy)$_2$(pic) (**13**), N984 (**14**) and N984a (**15**). From [180].

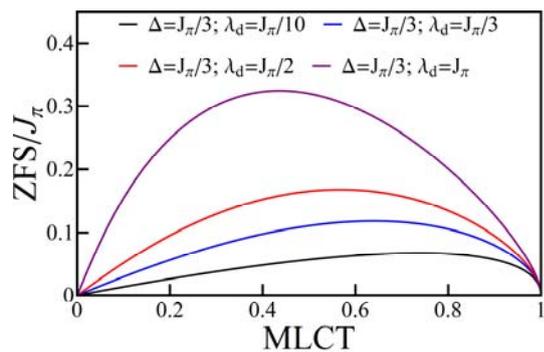

Figure 29. Zero-field splitting (ZFS) as a function of the degree of metal-to-ligand charge transfer (MLCT) of the low energy excitations for the trigonal pseudo-angular momentum model defined by Equation (8) for various parameters. Note that the ZFS is only proportional to the MLCT when $\lambda$ is small compared to both $\Delta$ and $J$. Note that $\theta$, cf. Equation (6), is varied parametrically in this plot.